\documentclass[usenatbib]{mnras}
\usepackage{newtxtext,newtxmath}
\usepackage[T1]{fontenc}

\DeclareRobustCommand{\VAN}[3]{#2}
\let\VANthebibliography\thebibliography
\def\thebibliography{\DeclareRobustCommand{\VAN}[3]{##3}\VANthebibliography}

\usepackage{amsmath}
\usepackage{graphicx}
\usepackage[usenames,dvipsnames]{xcolor}
\usepackage{mathtools}
\usepackage{gensymb}
\usepackage{tabularx}
\usepackage{subcaption}
\usepackage{physics}
\usepackage{subfiles}
\usepackage{multicol}
\usepackage[export]{adjustbox}

\newcommand{\dobib}{\bibliography{references}}

\title{Cloud Atlas: Navigating the Multiphase Landscape of Tempestuous Galactic Winds}

\author[B. Tan and D. B. Fielding] {
  \href{https://orcid.org/0000-0003-4805-6807}{Brent Tan}$^1$\thanks{E-mail: zunyibrent@ucsb.edu} \&
  \href{https://orcid.org/0000-0003-3806-8548}{Drummond B. Fielding}$^2$\\
    $^1$University of California - Santa Barbara,
    Department of Physics, CA 93106-9530, USA\\
    $^2$Center for Computational Astrophysics, Flatiron Institute, 162 5th Ave, New York, NY 10010, USA
}
\date{Accepted XXX. Received YYY; in original form ZZZ}
\pubyear{2023}

\begin{document}
\renewcommand{\dobib}{}
\label{firstpage}
\pagerange{\pageref{firstpage}--\pageref{lastpage}}
\maketitle

\begin{abstract}
Galaxies comprise intricate networks of interdependent processes which together govern their evolution. Central among these are the multiplicity of feedback channels, which remain incompletely understood.
One outstanding problem is the understanding and modeling of the multiphase nature of galactic winds, which play a crucial role in galaxy formation and evolution. We present the results of three dimensional magnetohydrodynamical simulations of tall-box interstellar medium patches with clustered supernova driven outflows. Dynamical fragmentation of the interstellar medium during superbubble breakout seeds the resulting hot outflow with a population of cool clouds. We focus on analyzing and modeling the origin and properties of these clouds. Their presence induces large scale turbulence, which in turn leads to complex cloud morphologies. Cloud sizes are well described by a power law distribution and mass growth rates can be modelled using turbulent radiative mixing layer theory. Turbulence provides significant pressure support in the clouds, while magnetic fields only play a minor role. We conclude that many of the physical insights and analytic scalings derived from idealized small scale simulations of turbulent radiative mixing layers and cloud-wind interactions are directly translatable and applicable to these larger scale cloud populations. This opens the door to developing effective subgrid recipes for their inclusion in global-scale galaxy models where they are unresolved.
\end{abstract}



\section{Introduction}  \label{sect:intro}

Understanding the inherently nonlinear, dynamical structures that underlie the complex ecosystems of galaxies is a challenging task. It is, however, essential towards making the most of our evergrowing body of increasingly detailed observations. A multitude of open problems still surround galaxy formation and evolution today. One that sits at the very heart is the challenge to understand the multiphase nature of not just the gas within galaxies and their surrounding environment (circumgalactic medium (CGM)), but also the flow of this material in and out of galaxies and how this regulates galaxy evolution \citep[often dubbed the Cosmic Baryon Cycle; see][]{tumlinson17}. Galactic outflows driven by feedback mechanisms carry material outwards \citep{veilleux05} while inflowing gas provides fuel for new star formation \citep[e.g. ][]{ keres05, dekel06, fraternali15}. This cycling connects processes on stellar ($\sim$\,pc) scales to galactic ($\sim$\,kpc) and intergalactic (IGM; $\sim$\,Mpc) scales, weaving them into a tightly interdependent and highly multiscale tapestry.

Galactic winds, which play a central role in regulating the baryon cycle, have garnered significant attention in recent years due to the clear evidence of their influence on galaxy evolution. They are thought to be primarily driven by feedback processes originating from either massive stars (for lower mass haloes) or active galactic nuclei (AGN; at the higher end of the mass spectrum), and are necessary ingredients for any realistic model of galaxy evolution \citep[see][for detailed reviews]{somerville15,naab17}. The result is the formation of large-scale outflows that transport material out of the galaxy while also shaping the CGM \citep{tumlinson17}.

Observations of galactic winds reveal them to be ubiquitous across star forming galaxies \citep{martin99,rubin14}. They also exhibit a complex, multiphase structure consisting of cold, cool, warm, and hot gas components (see Table~\ref{tab:phases} for the temperature ranges we use for each in this work), with the various phases spanning a range of different properties and dynamics \citep{veilleux05, strickland09, steidel10, rubin14, heckman17, Bolatto:2021}. A better understanding of the formation, survival, and growth of cool gas clouds which make up the most readily observable phase via emission and absorption lines, as well as their interactions with the other gas phases, is needed in order to provide vital insights into the mass transport mechanisms and overall energetics of these outflows. This in turn is essential for constructing a comprehensive picture of the multiphase nature of galactic winds and their impact on galaxy evolution.

In parallel, theoretical modeling and simulations of galactic winds have become increasingly sophisticated, aiming to reproduce and understand the multiphase nature and complex dynamics observed in these outflows. \citet{chevalier85} introduced an early analytic model where mass and energy injection into a spherically symmetric region powered a hot outflowing wind. Building on this simple model, recent theoretical work has included radiative cooling and gravity \citep{thompson16}, non-spherical expansion \citep{nguyen21}, more realistic non-uniform injection \citep{bustard16, nguyen23}, and frameworks for coupled multiphase evolution \citep{huang20, fielding22}. In particular, the multiphase models describe analytic prescriptions for evolving unresolved clouds in the cool phase and their coupling with the local hot background wind.

\begin{table}
    \centering
        \begin{tabular}{c|c}
            \hline 
            \hline
            Phase & Temperature Range \\
            \hline
            Cold & $T < 5 \times 10^3$\,K  \\
            Cool & $5 \times 10^3$\,K~$< T < 2 \times 10^4$\,K \\
            Warm & $2 \times 10^4$\,K~$< T < 5 \times 10^5$\,K \\
            Hot & $T > 5 \times 10^5$\,K \\
            \hline
        \end{tabular}
    \caption{Temperature ranges defined for the four phases of interest referred to throughout this work.}
    \label{tab:phases}
\end{table}
In this work, we focus on winds driven by stellar feedback, primarily in the form of energy released by core-collapse supernovae (SNe). While large scale cosmological simulations are now approaching high enough resolutions where multiphase structures (on scales of order 100 pc) are seen to develop self consistently in the CGM \citep{nelson20,ramesh23} and in galactic winds (e.g., in TNG50 \citealt{nelson19} and FIRE \citealt{angles17,pandya21}), the detailed properties of multiphase outflows are still most clearly seen in high resolution galaxy scale simulations \citep[e.g.,][]{schneider20, steinwandel22, rey23}, or interstellar medium (ISM) patch simulations (e.g., TIGRESS \citealt{kim17b,kim18}; SILCC \citealt{Gatto:2017, Rathjen:2023}). The $\sim$\,pc-scale resolution of these simulations makes it possible to resolve the energy injection by SNe and the resulting interactions with the surrounding ISM \citep[e.g.,][]{deAvillez00,joung06,hill12}. Such simulations have also revealed a trove of complications surrounding the launching of these winds. For example, the efficacy of this process is sensitive to the spatial distribution of SNe \citep{creasey13,martizzi15,li17,smith21} as well as spatiotemporal clustering \citep{kim17,fielding18} and self-consistency of treatments of star formation and feedback \citep{kim17b,kim18}.

The survival and growth of cool gas clouds entrained in these winds has also been a hotbed of analytic and numerical studies in recent years. Theoretically, this picture was initially problematic as these cool clouds should be quickly destroyed by hydrodynamic instabilities during this process \citep{klein94}. The timescale for the acceleration of an adiabatic cool cloud is a factor of ${\gtrsim}10$ times longer than the timescale for its destruction via Kelvin-Helmholtz (KH) and Rayleigh-Taylor (RT) instabilities \citep{zhang17}. This was also verified in simulations \citep{cooper09,scannapieco15,schneider17}. However, it was recently shown that under the right conditions, these cool clouds can not only survive but even grow within the hot wind \citep{marinacci10, armillotta17,gronke18,gronke20}. This is possible when the cooling time of mixed gas is shorter than the destruction timescale. Subsequent studies have shown that exact characterization of this parameter space is complex \citep{li20,sparre20,kanjilal21,farber22,abruzzo22}. The mechanism that drives this growth is the formation of a long tail structure of cool gas where mixing and cooling takes place efficiently in turbulent radiative mixing layers at the interface of the two gas phases \citep{ji18,fielding20,tan21}. Crucially, this requires long simulation domains that were not captured in earlier simulations, highlighting the importance of capturing the appropriate scales and boundary conditions that define such problems. 

\begin{figure}
    \centering
    \includegraphics[width=\columnwidth]{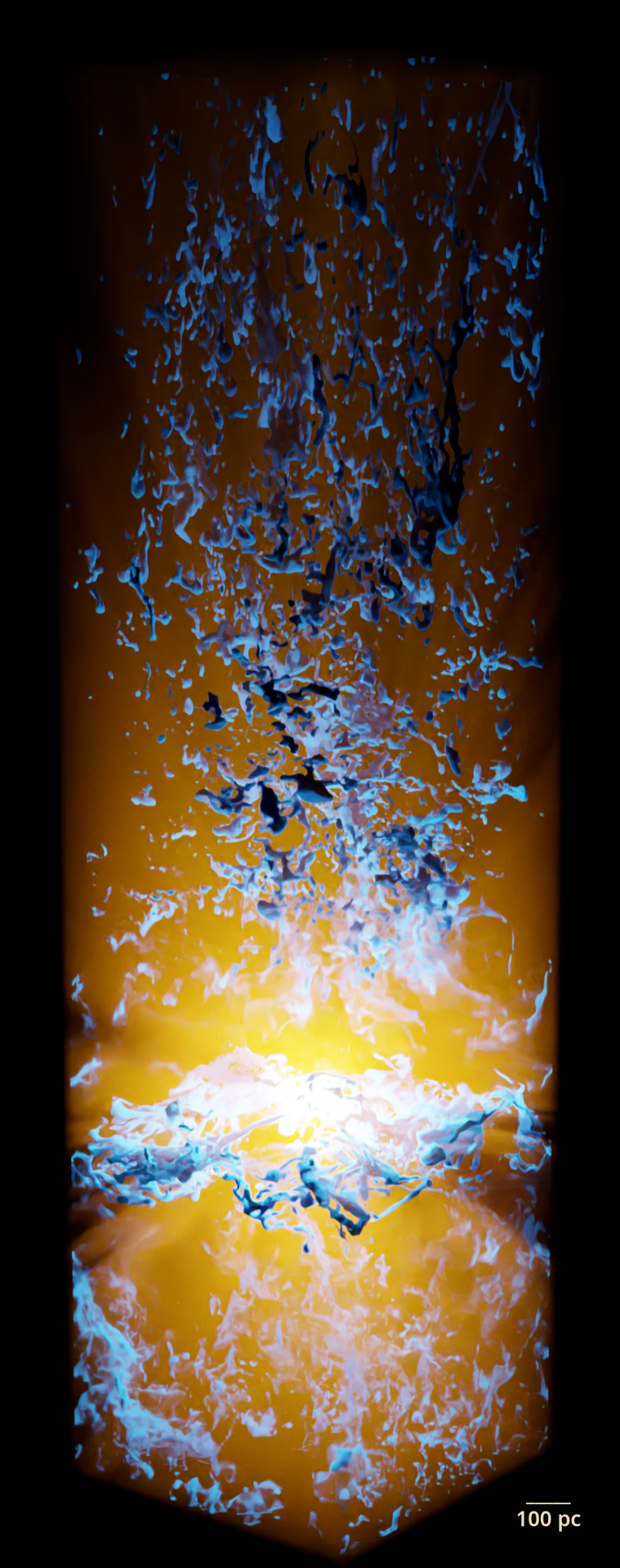}
    \vspace{2pt}
    \caption{Volume rendering of the main simulation used in this work that demonstrates the presence of numerous cool clouds (shown in opaque blue) embedded within the hot outflow (shown in transparent orange) powered by the SNe (shown as the light source) exploding within the disc mid-plane.}
    \label{fig:render}
\end{figure}
The interplay between the cool and hot gas phases and the role of various physical processes are hence crucial in determining the fate of cool gas clouds and their impact on the overall dynamics of galactic winds \citep{fielding22}. Aside from radiative cooling, other important processes at play include turbulence \citep{schneider17, banda19}, magnetic fields \citep{gregori99,mccourt15,gronnow18,hidalgo23}, and thermal conduction \citep{kooij21, bruggen23}. Despite these advances, many aspects of galactic wind simulations still rely on simplified models and assumptions, such as idealized cloud geometries or time-constant wind properties. The next step forward is to study the problem under more realistic conditions, incorporating the diverse range of physical processes and interactions that govern the evolution of galactic winds and their multiphase structures.

To that end, in this work we connect results from small scale idealized cloud crushing simulations to large scale multiphase galactic winds by studying how clouds form, evolve and interact with the wind. This mesoscale approach bridges the microscale phenomena controlling individual clouds to the macroscale phenomena controlling galactic scale evolution by simultaneously resolving the lifespan of thousands of individual clouds as well as the driving of the highly turbulent, temporally variable, hot wind that they interact with. The basis of our study builds upon a series of simulations following a similar design to those presented in \cite{fielding18}, which capture the dynamics in a vertically stratified (512 pc)$^2$ patch of a galactic disc. In these simulations, clustered SNe lead to the formation of superbubbles which propagate quickly enough to break out of the galactic disc, providing a pathway for the energy and momentum released to vent into the halo in the form of a galactic wind. The shredding and entrainment of cool dense clumps (which span a wide range of scales) is seen in these winds, and is expected to be important in influencing the structure and dynamics of the outflow. We extend the box asymmetrically so as to follow the clouds in the outflow out to a larger height, as well as include additional physics such as self-developed magnetic fields in the turbulent ISM. Figure~\ref{fig:render} shows a rendering from our simulation of cool structures (in blue) in the hot outflowing wind (in orange). Our main focus is on analyzing the clouds in these winds. More specifically, we investigate how they are formed during superbubble breakout and their properties relative to predictions from idealized small scale simulations of turbulent mixing layers and individual clouds.

The outline of this paper is as follows. In Section~\ref{sect:methods}, we provide a detailed description of our simulation setup. In Sections~\ref{sect:results_wind} and \ref{sect:results_clouds}, we present simulation results, with the former focusing on bulk properties of the outflow and the the latter on analysis of cloud properties including size distribution and growth. Lastly, we discuss some implications and complications in Section~\ref{sect:discussion} before summarizing and concluding in Section~\ref{sect:conclusions}. 

\dobib

\section{Methods}  \label{sect:methods}
In this work, we carry out 3D magnetohydrodynamical (MHD) simulations using the publicly available code Athena\verb!++! \citep{stone20} aimed at capturing the multiphase dynamics in a vertically stratified patch of a galactic disc. SNe corresponding to a star cluster are seeded by hand at the center of the midplane of the disc, which leads to the formation of a superbubble that propagates rapidly enough to break out of the disc and release energy and momentum into the halo in the form of a galactic wind. This builds upon the design of a similar suite of simulations presented in \cite{fielding18} where additional details and discussion can be found. In our analyses in the following sections, we focus on studying the formation and dynamics of a resulting population of cool material that gets sown into the wind. Before that, we describe in this section our simulation setup, implementations of various physics, and the initial conditions we adopt.

\subsection{Setup}
All simulations are run in three dimensions on Cartesian grids using the HLLD approximate Riemann solver and the third-order accurate Runge-Kutta time integrator. Our simulation setup consists of a rectangular box with dimensions $512 \times 512 \times 2048$\,pc and a fiducial resolution of 2\,pc on a uniform grid (we also discuss in this work a simulation run with a higher resolution of 1\,pc but which fails to drive a successful outflow). The box is asymmetric---the mid-plane is located a quarter box length from the bottom---so as to follow the outflow on the upper side out to a greater height. We use periodic boundary conditions on the sides of the box, along the $x$ and $y$ directions. We adopt outflowing boundary conditions (zero gradient with inflow explicitly disallowed) along the $z$ direction at the top and bottom. We also implement a density floor of $n = 10^{-8} {\rm ~cm}^{-3}$. 

\subsection{Source Terms}
We include a background gravitational profile, driven turbulence, optically thin radiative cooling, photoelectric heating, and SNe injection. Here, we provide more details about the implementation of each of these in turn. We do not include any explicit viscosity or thermal conduction. A $\gamma=5/3$ equation of state is adopted throughout. 

\subsubsection{Gravity}
We first consider the density profile of a thin isothermal disc in hydrostatic equilibrium. We begin by assuming that the self-gravity of the gas is a subdominant component of the total gravitational potential and is hence unimportant (i.e., $\sim 2 \pi G \Sigma_{\rm gas} z$ is small, which is true for $\Sigma_{\rm gas} \ll 1000 M_{\odot}/{\rm pc}^2$). We hence neglect self gravity in our simulations. The spherical potential at a vertical height $z$ above the disc and a radial distance $R$ from the center is
\begin{align}
    \Phi(R,z) = -\frac{GM}{r} = -\frac{GM}{(R^2 + z^2)^{1/2}},
\end{align}
and hence the gravitational acceleration is 
\begin{align}
    g(R,z) = -\frac{\partial \Phi}{\partial z} = -\frac{GM}{(R^2 + z^2)^{3/2}}.
\end{align}
For a thin disc, $z << R$,
\begin{align}
    g(R,z) \approx - \frac{GM}{R^3} z = -\Omega^2 z,
\end{align}
where $\Omega = v_{\rm cir}/R$ is the angular velocity. Our equation of hydrostatic equilibrium is thus
\begin{align}
    \frac{dP}{dz} = \rho g z = -\rho \Omega^2 z.
\end{align}
With $P = c_i^2 \rho$ ($c_i$ is the isothermal sound speed, related to the adiabatic sound speed by $c_s^2 = \gamma c_i^2$), we can easily solve this for the density profile:
\begin{align}
    \rho(z) = \rho_0\exp\left[ - \frac{z^2}{2H^2} \right],
\end{align}
where $H \equiv c_i/\Omega$ is the scale height and $\rho_0$ is the mid-plane density at $z=0$.

We can however drop the thin disc assumption and solve for the profile out to larger $z$. We now have
\begin{align}
     g(R,z) = -\frac{GM}{R^3} \frac{R^2}{(R^2 + z^2)^{3/2}} z = -\Omega^2 \frac{R^2}{(R^2 + z^2)^{3/2}} z.
\end{align}
Solving this gives us:
\begin{align}
    \rho = \rho_0 \exp \left[ \frac{R^2}{H^2} \left ( \frac{R^2}{(R^2+z^2)^{1/2}} -1 \right) \right].
\end{align}
These are the profiles we adopt in our simulations. We test systems in hydrostatic equilibrium with this gravitational profile in Appendix~\ref{sect:apdx_c}. We note that there are caveats to the simplified approach we have taken of assuming a simple external spherical potential which would affect the profiles above. A more sophisticated treatment should account for both disc and stellar spheroid components \citep[e.g.,][]{strickland00}, turbulent support \citep[e.g.,][]{cooper08}, a dark matter halo component \citep[e.g.,][]{li17} and self-gravity. These are particularly important if one is going to distances $\gg 1$\,kpc \citep{li17}. However, our simplified assumption is sufficient for our box size.

\begin{figure}
    \centering
    \includegraphics[width=\columnwidth]{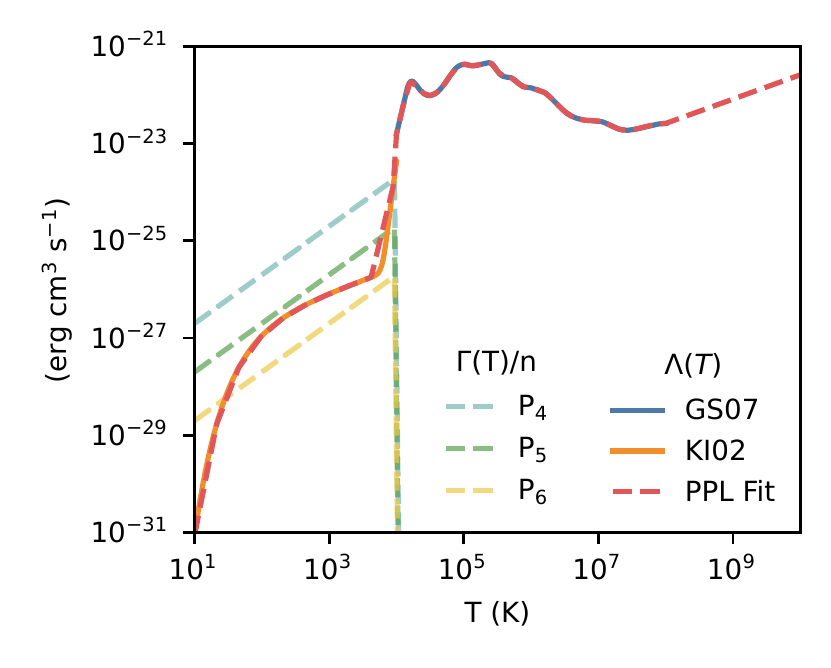}
    \caption{Cooling curves along with our fit and heating curves for $\langle n_H \rangle = 100 {\rm ~cm}^{-3}$ at several different pressures (e.g., $P_4$ is $P = 10^4 k_{\rm B } {\rm ~cm}^{-3}$~K).}
    \label{fig:cooling_curve}
\end{figure}

\subsubsection{Stratified Turbulence}
We drive mixed turbulence on large scales so that the mass-weighted velocity dispersion $\delta v \equiv \langle v^2 \rangle^{1/2} \approx 10$\,km/s, which is roughly the sound speed of $10^4$\,K gas and consistent with observed velocity dispersions in the ISM. Physically, this turbulence is driven by either star formation feedback or gravitational instability in ISM gas. The turbulent kinetic energy injection rate is thus $\Dot{E}_{\rm turb} \approx \rho \delta v^3 L_{\rm box}^2$ where $L_{\rm box}^2$ is the horizontal box length. The turbulence is driven on the scale of the disc scale height with power distributed evenly between wave numbers $k=2$ to $k=3$. The turbulence is constrained to follow a Gaussian profile with scale height $H \sim 80$\,pc (we test this implementation of constrained turbulence in Appendix~\ref{sect:apdx_d}). We drive the turbulence every $5 \times 10^{-3}$ crossing times $t_{\rm cross}$. The driven turbulence in Athena\verb!++! adopts an Ornstein-Uhlenbeck process to smoothly vary the velocity perturbations over some correlation time \citep{lynn12}, which we set to 8~Myr. Turbulence along with heating and cooling leads to the formation of a turbulent multiphase ISM which we allow to form for 60 Myrs (roughly a turnover time) prior to any supernova explosions. 

\subsubsection{Cooling/Heating}
The net cooling rate per unit volume is formulated as
\begin{equation}
    \rho \mathcal{L} = n^2 \Lambda - n\Gamma ,
\label{eq:cooling_eq}
\end{equation}
where $\Lambda$ is the cooling function and $\Gamma$ is the heating rate. Our cooling function $\Lambda(T)$ combines the collisionally ionized equilibrium (CIE) cooling curve in \citet{Gnat2007} for $T \geq 10^4$\,K with the cooling function for $T \leq 10^4$\,K in \citet{koyama02}. We obtain our cooling curve by performing a piece-wise power law fit over $\sim 50$ logarithmically spaced temperature bins, starting from a temperature floor of $10$\,K up to a maximum temperature of $10^{10}$\,K. We also include a photoelectric heating (PEH) rate $\Gamma = 10^{-26} (\langle n_H \rangle /{\rm ~cm}^{-3})$\,erg s$^{-1}$, where we have scaled the solar value by the average mid-plane density to approximate the scaling of PEH with higher star formation rates in denser regions. This approach is similar to those used in \citet{kim17} and \citet{fielding18}, which run similar setups. We assume solar metallicity ($X=0.7$, $Z=0.02$). We also use a fixed mean molecular mass $\mu \sim 0.62$ for a fully ionized plasma. This has a $\leq2$ factor of error in the temperature of neutral/partially ionized gas below $10^4$\,K, but should should not affect our overall conclusions. Figure~\ref{fig:cooling_curve} shows the aforementioned cooling curves along with our fit and heating curves for $\langle n_H \rangle = 100 {\rm ~cm}^{-3}$ at several different pressures (e.g., $P_4$ is $P = 10^4 k_{\rm B } {\rm ~cm}^{-3}$~K).

We also adopt an additional constraint on the simulation timestep over the standard CFL constraint where we require that that the timestep is less than or equal to one quarter of the shortest single-point cooling time $t_{\rm cool}$ across the whole domain. This is to ensure good coupling between cooling and the hydrodynamical evolution.

Finally, we develop and implement an extended version of the fast and robust exact cooling algorithm described in \citet{townsend} to include heating (See Appendix~\ref{sect:apdx_b} for full details).

\begin{figure}
    \centering
    \includegraphics[width=\columnwidth]{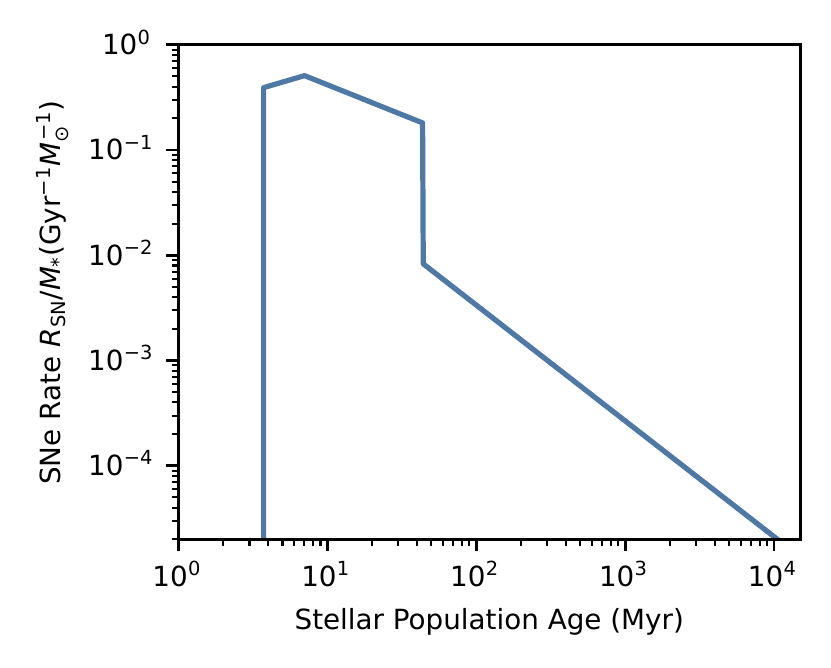}
    \caption{The SNe rate as as function of the stellar population age. This consists of the core-collapse SNe rate ($<44$~Myr) and the Ia rate ($>44$~Myr).}
    \label{fig:fig1}
\end{figure}

\subsubsection{Supernova Injection}
We include both core-collapse supernova and Ia rates using the piece-wise power law fits given in \citet{fire3}. The combined SNe rate (shown in Figure~\ref{fig:fig1}) is
\begin{align}
    \frac{R_{\rm SN}/M_{*}}{{\rm Gyr}^{-1} M_{\odot}^{-1}} =
    \begin{cases}
         \: 0                 & (t < t_1) \\
         \: a_1 (t/t_1)^{s_1} & (t_1 \leq t \leq t_2) \\
         \: a_2 (t/t_2)^{s_2} & (t_2 \leq t \leq t_3) \\
         \: a_3 (t/t_3)^{s_3} & (t > t_3),
    \end{cases}
\end{align}
where $s_1 \equiv \ln{(a_2/a_1)}/\ln{(t_2/t_1)}$, $s_2 \equiv \ln{(a'_2/a_2)}/\ln{(t_3/t_2)}$, $s_3 = -1.1$, $(a_1,a_2,a'_2,a_3) = (0.39,0.51,0.18,0.0083)$ and $(t_1,t_2,t_3) = (3.7,7.0,44)$~Myr. The time of the first Ia is set to be after the time of the last core-collapse SNe ($t = t_3$). For $M \equiv M_{*}/M_{\odot}$, the SNe number per initial solar mass $n_{\rm SN} \equiv N_{\rm SN}/M$ is thus given by
\begin{align}
    n_{\rm SN}(t) =
    \begin{cases}
         \: 0                                                            & (t < t_1) \\
         \: \frac{a_1}{s_1+1} \frac{t_1}{1 {\rm Gyr}} \left[ (t/t_1)^{s_1+1} -1 \right] & (t_1 \leq t \leq t_2) \\
         \: n_{\rm SN}(t_2) +  \frac{a_2}{s_2+1} \frac{t_2}{1 {\rm Gyr}} \left[(t/t_2)^{s_2+1} -1 \right] & (t_2 \leq t \leq t_3) \\
         \: n_{\rm SN}(t_3) +  \frac{a_3}{s_3+1} \frac{t_3}{1 {\rm Gyr}} \left[(t/t_3)^{s_3+1} -1 \right] & (t > t_3).
    \end{cases}
\end{align}
The time where $n$ supernovae have gone off is hence
\begin{align}
    t_{\rm SN}(M, n) =
    \begin{cases}
         \: t_1 \left[ \frac{n(s_1+1)}{Ma_1(t_1/1 {\rm Gyr})} \right]^{1/(s_1+1)} & (n \leq n_1) \\
         \: t_2 \left[ \frac{(n-n_1)(s_2+1)}{Ma_2(t_2/1 {\rm Gyr})} \right]^{1/(s_2+1)} & (n_1 \leq n \leq n_2) \\
         \: t_3 \left[ \frac{(n-n_2)(s_3+1)}{Ma_3(t_3/1 {\rm Gyr})} \right]^{1/(s_3+1)} & (n \geq n_2).
    \end{cases}
\end{align}

We deposit $10^{51}$~ergs per SN as pure thermal bombs. While one could determine the amount of thermal and kinetic energy injected by each SN (see \citet{martizzi15}, who calibrated their model to high resolution simulations of individual SN remnants),
this would matter only for the first couple of SNe. Since the SNe are clustered tightly both in space and time, most of them barring the first few occur in the hot and low density remnant of the previous SNs, assuming that $\Delta t_{\rm SN}< t_{\rm PE}$ ($t_{\rm PE}$ being the timescale over which the SNR reaches pressure equilibrium or mixes with the ambient ISM) so that a coherent bubble can be driven \citep{fielding18}. Hence their cooling radii $r_{\rm cool}$ are around an order of magnitude larger than the injection radius $r_{\rm inj}$, which means that most of the energy from the SN is transferred to the surrounding gas.

We assume a spherical geometry for the injection site with radius $r_{\rm inj} = 20$\,pc. We assume a stellar ejecta mass of $8.72 M_{\odot}$ for core-collapse SN and $1.4 M_{\odot}$ for Ia. We also assume that all SN go off at the origin, since the cluster radius $R_{\rm cl} = 10$\,pc is relatively small. 

While out of the scope of this work, we note that in reality, stars far from the disc center could have a disproportionate effect in driving winds. For instance, Ia at late times from stars that settle high above the disc or OB (binary) runaways \citep[e.g.,][]{Steinwandel:2022}. This is mainly due to the dual effects of a larger $r_{\rm cool}$ in the lower density region along with the lower scale height contributing to a large $r_{\rm cool}/H$ ratio. 

For each cell with center within the sphere radius, we then inject energy from the SN uniformly. Note that there will be some error due to resolving the volume of the sphere, but the error is small as long as we resolve the injection radius by more than a few cells. 

\subsection{Initial Conditions}
We adopt a circular velocity $v_{\rm cir} = 175$\,km/s and galactic radial distance $R=1$\,kpc, which corresponds to a scale height $H=66$\,pc. The box is initially filled with $T = 10^{4}$\,K gas with a mid-plane density $n = 100$\,cm$^{-3}$ at $z=0$. The average density within one scale height is $\langle n \rangle = 86 {\rm \,cm}^{-3}$.
Using $ \langle n \rangle \equiv \Sigma_{\rm g} /(2H \mu m_{\rm p})$, this corresponds to a gas surface density $ \Sigma_{\rm g} \approx 175 \, M_{\odot} \, {\rm pc}^{-2}$. For a a star formation efficiency $\epsilon_* \equiv M_{\rm cl}/\pi h^2 \Sigma_g \sim 0.1$ (as found in \citet{grudic18}), this corresponds to a star cluster mass $M_{\rm cl} \sim 2.5 \times 10^5 \,M_{\odot}$. For comparison, the simulation from the suite of TIGRESS simulations under the SMAUG project with the highest star formation rate (R2) has $ \Sigma_{\rm g} \approx 74 \, M_{\odot} \, {\rm pc}^{-2}$ \citep{kim20a}. We initialize a magnetic field aligned along the x-direction with plasma beta $\beta \equiv P_{\rm gas}/P_{\rm B} = 100$ everywhere. However, through cooling and turbulent dynamo amplification, $\beta \sim 1$ is rapidly achieved in the ISM.  

\begin{figure*}
    \centering
	\includegraphics{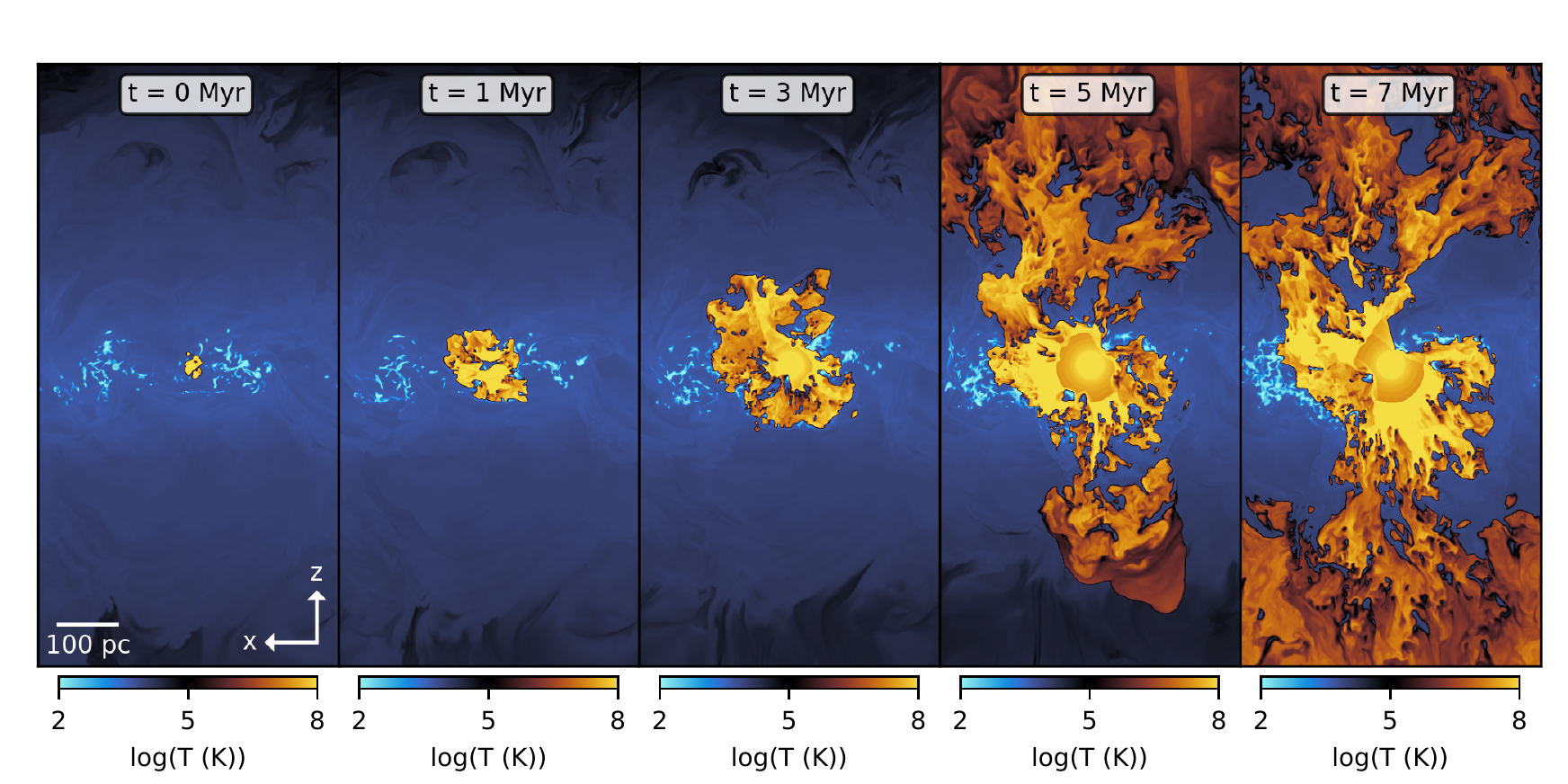}
	\caption{The various stages of a superbubble breakout, driven by clustered SNe. As the first SNe go off, a hot bubble expands outwards, propagating through the multiphase ISM, leading to a complex fractal structure. As the superbubble eventually breaks out of the disc, the ISM fragments and breaks up, thus seeding the hot outflow with cool clouds.}
	\label{fig:slices_progression_early}
\end{figure*}

\dobib

\section{Results : Breakout and Wind Outflow} \label{sect:results_wind}
In this section, we provide context for the winds that contain our clouds, setting the stage for our main focus---a closer look at cloud properties---in the next section. There is already a large body of existing work studying these SN driven outflows---we concentrate mainly on analysis meant to highlight the most salient points that are relevant to these clouds.

\subsection{Superbubble Breakout}
Supernovae (SNe) are powerful energetic events that release a tremendous amount of energy ($\sim10^{51}$\,ergs) and mass into their surrounding environments. However, individual SNe are ineffective at driving galactic winds because most of this energy is radiated away before it can break out of the disc. Spatiotemporal clustering of SNe resolves this problem, and is motivated by the fact that stars form primarily in clusters. When multiple supernova explosions occur in the same region of space and over a relatively short period of time, they can overlap and form superbubbles. These superbubbles are able to drive the expansion of the material outwards from the cluster, breaking out of the disc before the cluster runs out of SNe. After superbubble breakout the hot material inside is able to vent its energy into the halo and power an outflowing galactic wind. As the superbubble propagates through the clumpy multiphase environment, it sweeps around higher-density regions, resulting in a surface that is fractal in nature. This complex interplay leads to the fragmentation and break up of the ISM, seeding the hot wind with a population of cool clouds. Figure~\ref{fig:slices_progression_early} illustrates this by showing temperature slices at various stages of the process described above.

\begin{figure}
    \centering
    \includegraphics[width=\columnwidth]{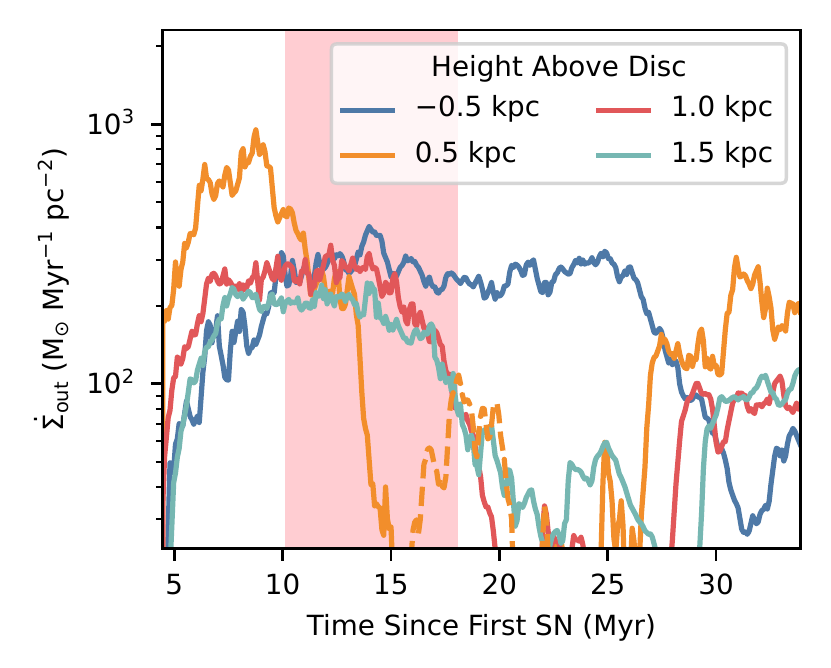}
    \caption{The outwards mass flux over time at four different heights showing the initial breakout and outflow above the disc. This then reverses sides twice over the timespan of the SNe. Dashed line represents negative values, while the red shaded region highlights the time window after the SNe rate peaks in which we analyze wind embedded clouds in the next section.}
    \label{fig:mass_flux_time_profile}
\end{figure}

\begin{figure*}
    \centering
    \includegraphics[width=\textwidth, valign=c]{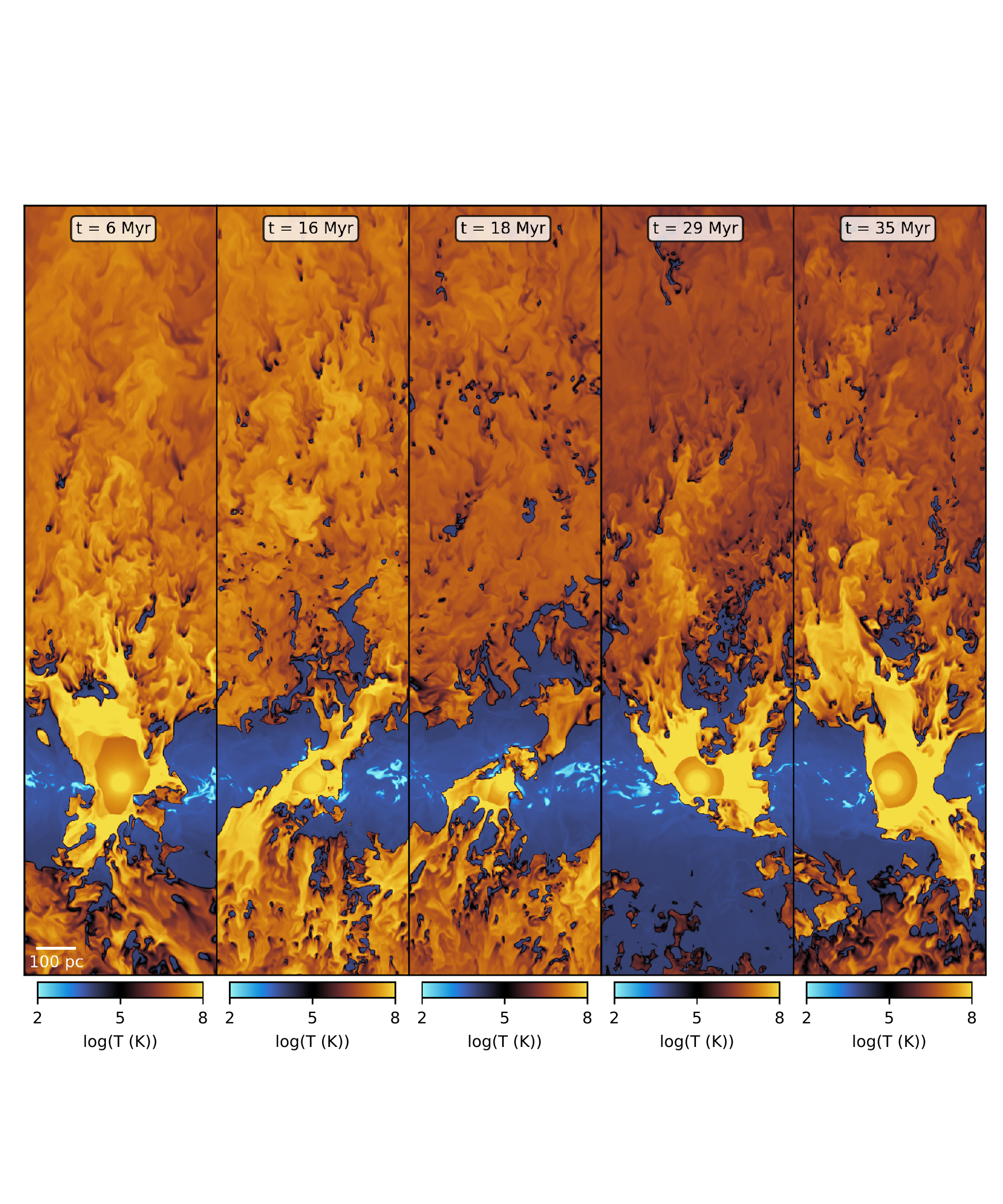}
    \caption{Slices of temperature at various times from our fiducial simulation where the breakout successfully launches a wind above the disc, despite the outflow being interrupted in the middle.}
    \label{fig:slices_progress_256}
\end{figure*}

\subsection{Asymmetric Outflows}
By virtue of having to make its way through a multiphase ISM, the expansion and eventual break out of the superbubble can exhibit great asymmetry in terms of driving outflows above and below the disc. This can be self-reinforcing---whichever side breaks through first is then free to vent into the low density halo, creating a channel of lower resistance for the hot gas to funnel through and leading to a weak or non-existent outflow on the opposite side. In addition, turbulent motions in the ISM can exacerbate or reverse this asymmetry. For example, if cold dense material in the ISM moves over the cluster region, it `caps' and inhibits the path of the hot expanding wind, much like turning off a valve, resulting in a redirection towards the other side of the disc. In this manner, an outflowing wind can be quenched at the base on one side midway during the lifetime of the cluster SNe. While we did not investigate this in further detail, we observed that this was a common occurrence in our simulations---due to the nature of the setup, a wind that breaks out on one side pushes the ISM radially outwards in the plane of the disc. Because of the periodic boundary conditions, this leads to a build up of ISM material on the sides which then falls back towards the cluster. This sloshing motion often eventually halts the wind and drives it to the opposing side. In a more realistic larger scale setup, this indicates that dynamical interactions between different star clusters could complicate wind driving, either reinforcing or disrupting outflows depending on their spatiotemporal separations. In Appendix~\ref{sect:apdx_f}, we show and analyze a case where the wind is quenched.

In Figure~\ref{fig:mass_flux_time_profile}, we show the outwards mass flux per unit area at four different heights in our box as a function of time for our fiducial simulation. The blue and orange lines show the mass fluxes below and above the disc respectively at 0.5\,kpc, while the red and teal lines are at 1 and 1.5\,kpc above the disc. The dashed orange line shows when the mass flux becomes negative, i.e., material is falling back to the disc. We can see that the breakout is initially on the top side which launches an outflowing wind. As the outflow is interrupted at the base, the wind loses its power source, and begins to slow and dissipate. Meanwhile, the outflow is directed to the bottom side of the disc. This eventually reverses again nearing the end of the cluster SN period. The red region highlighted in Figure~\ref{fig:mass_flux_time_profile} shows the time range in which we analyze clouds in Section~\ref{sect:results_clouds}. This choice corresponds to a time period right after the peak of the SNe rate (see Figure~\ref{fig:fig1}). To illustrate this more clearly, Figure~\ref{fig:slices_progress_256} shows temperature slices at various times in the same simulation. We see that the wind initially breaks out on the topside and that fragments of the ISM become entrained in the hot outflowing wind, whereas it fails to do the same on the bottom side, which is apparent when looking at the rightmost two panels at later times.

\subsection{Turbulent Winds}
\begin{figure*}
    \centering
    \vspace*{\fill}
    \includegraphics[width=\textwidth, valign=c]{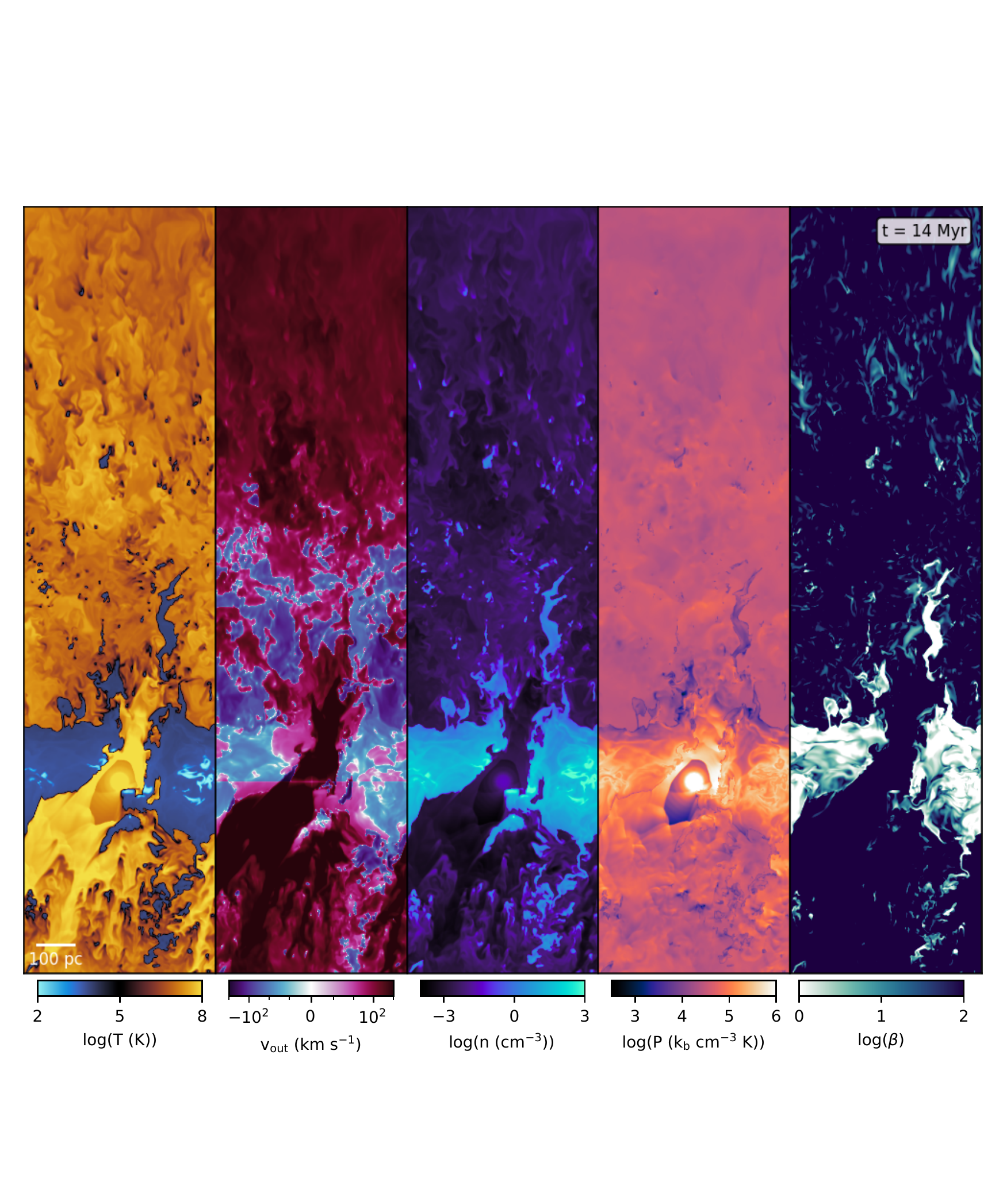}
    \vspace*{\fill}
    \caption{Slices of temperature, outflow velocity, number density, pressure and magnetic plasma beta for a single time snapshot 14 Myr after the first SN. Clustered SNe drive a hot outflowing wind which contains a population of cool clouds in this wind.}
    \label{fig:slices_prim}
\end{figure*}
\begin{figure*}
    \centering
    \vspace*{\fill}
    \includegraphics[width=\textwidth, valign=c]{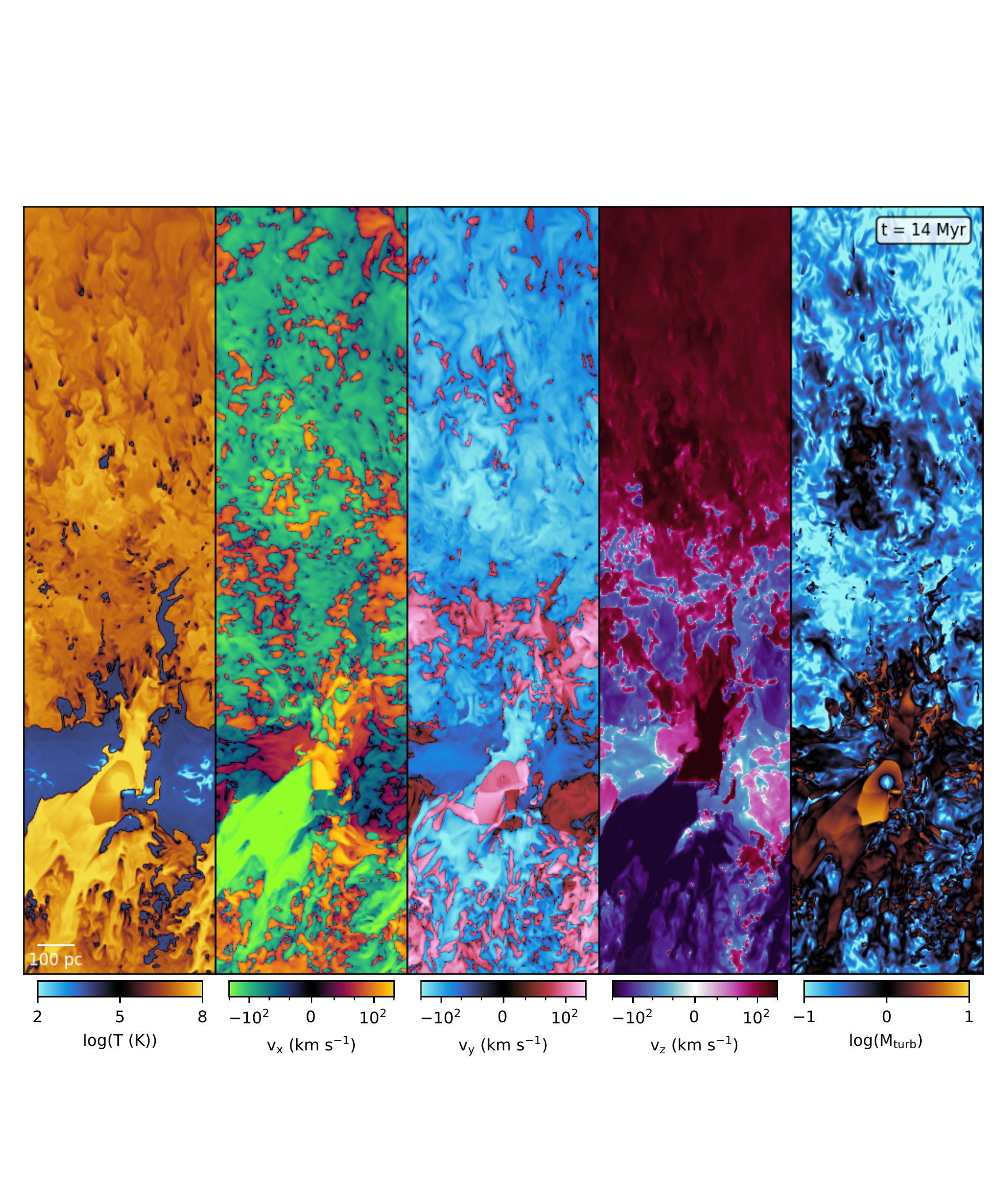}
    \vspace*{\fill}
    \caption{Slices of the velocities in the three axis-aligned directions. The vertical velocities $v_{\rm z}$ reflect the bulk outflow velocity, while the velocities $v_{\rm x}$ and $v_{\rm y}$ parallel to the disc plane reflect the strong turbulence in the wind.}
    \label{fig:slices_turb}
\end{figure*}
\begin{figure*}
    \centering
    \vspace*{\fill}
    \includegraphics[width=\textwidth, valign=c]{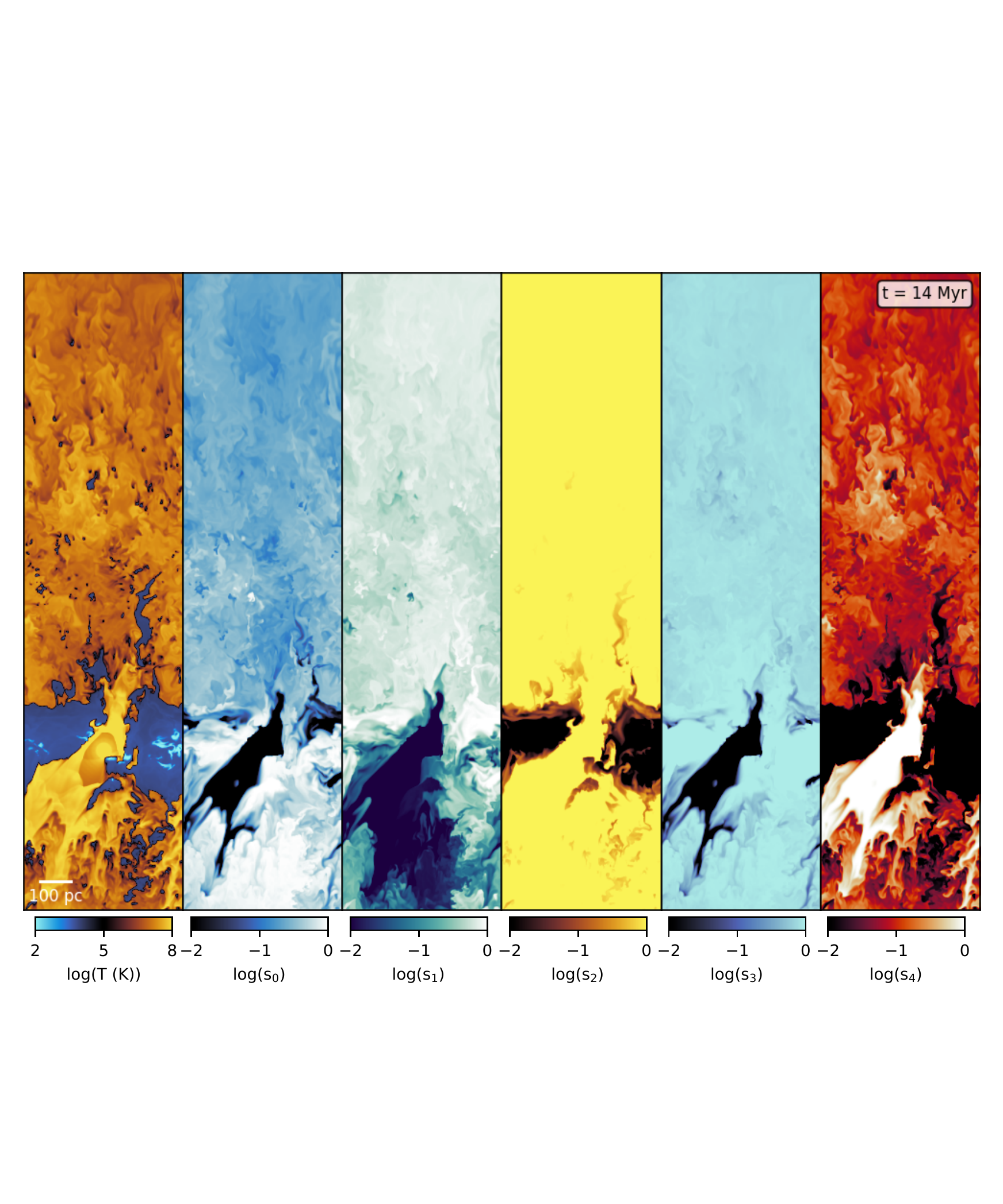}
    \vspace*{-40pt}
    \caption{Slices of the passive scalars we have used in this simulation. We use a total of 5 passive scalars. The first two are introduced right before the first SN explosion. All cold gas ($T < 5 \times 10^3$\,K) is dyed with passive scalar $s_0$, and all cool gas ($5 \times 10^3$\,K < $T$ < $2\times10^4$\,K) is dyed with passive scalar $s_1$. Passive scalars $s_2$ and $s_3$ track how much of the gas in a cell was {\it ever} hot or cold. Lastly, passive scalar $s_4$ tracks all SN injected gas. Beyond the disc and the injection site, $s_2$ and $s_3$ are close to unity everywhere, indicating that there is strong mixing between phases.}
    \vspace*{\fill}
    \label{fig:slices_scalars}
\end{figure*}
\begin{figure}
    \centering
    \includegraphics{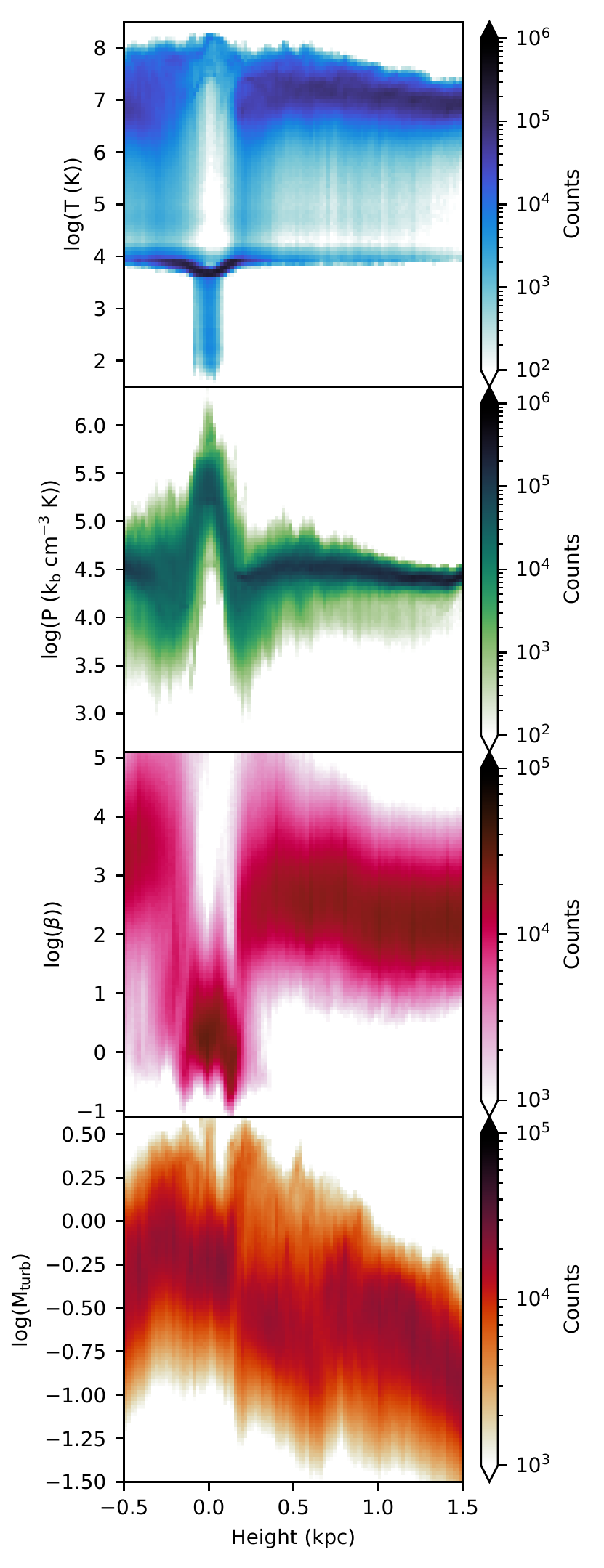}
    \caption{2D volume weighted histograms of temperature, pressure, magnetic plasma beta, and turbulent Mach number with height 14 Myr after the first SN. These reflect wind properties: $T \sim 10^7$\,K, $P/k_b \sim 10^{4.5}$~cm$^{-3}$~K, $\beta \sim 10^2 - 10^3$, and $M_{\rm turb} \sim 0.25$.}
    \label{fig:distribution_heights}
\end{figure}
Figures~\ref{fig:slices_prim}, \ref{fig:slices_turb}, and \ref{fig:slices_scalars} show the range of quantities we track in our simulations, and allow us to make some general statements about the outflowing winds. Figure~\ref{fig:slices_prim} shows slices of temperature, outflow velocity, number density, pressure and magnetic plasma beta for a single time snapshot 14 Myr after the first SN, in the middle of the red shaded region shown in Figure~\ref{fig:mass_flux_time_profile}. This shows the general picture of clustered SNe driving a hot outflowing wind which contains a population of cool clouds, which we focus our analysis on in the next section. Figure~\ref{fig:distribution_heights} is composed of 2D histograms that show how these properties vary with height. The histograms are volume weighted and hence largely reflect wind properties, especially at heights above 0.5 kpc. The top-most panel shows that the wind has a temperature of $T \sim 10^7$\,K, along with cool clouds that have a temperature of $T \sim 10^4$\,K. In the second panel from the top, it can be seen that pressure drops off rapidly away from the disc. The pressure in the wind is roughly $P/k_b \sim 10^{4.5}$~cm$^{-3}$~K . In terms of non-thermal components in the wind, both magnetic fields and turbulence are small compared to thermal pressure. We can see from the third panel that although the disc is strongly magnetized with plasma beta $\beta \sim 1$, the wind itself has a very weak magnetic field with $\beta \sim 10^2 - 10^3$. Lastly, the bottom-most panel shows that above 0.5 kpc, the turbulent Mach number in the wind $M_{\rm turb} \sim 0.25$. This brings us to a discussion of the nature of turbulence and mixing in the wind, where we will also define how $M_{\rm turb}$ is computed.

Dispersed multiphase flow systems such as particles or droplets in liquid or gaseous flows are common in various engineering fields \citep{balachandar10}. When the mass fraction of the dispersed phase is comparable to that of the carrier phase, the back reaction is non-negligible and the system is said to have two-way coupling. The nature of this multiphase turbulence is still considered an open problem due to the involved complexities. For example, it is unclear how particles can affect turbulence in the carrier phase---studies have shown generally that small particles can attenuate turbulence by dissipating energy for turbulent eddies, while larger particles can amplify turbulence through mechanisms such as vortex shedding \citep{gai20}. Similarly, it is apparent in our simulations that the presence of cloud fragments induces turbulence on the wind scale. It has also been widely shown that instabilities on cloud scales such as KH, RT, and Richtmyer-Meshkov not only disrupt clouds but mass load the wind with turbulent gas \citep{klein94,maclow94,pittard16,banda20}. This arises firstly during the breakout phase as the superbubbles propagates through the multiphase ISM. Once this reaches the outflow phase, the flow of the hot wind around the population of clouds similarly induces a large scale turbulent field on the scale of the disc scale height. Figure~\ref{fig:slices_turb} shows slices of the velocities in the three axis-aligned directions. The vertical velocities $v_{\rm z}$ reflect the bulk outflow velocity, while the velocities $v_{\rm x}$ and $v_{\rm y}$ parallel to the disc plane reflect the turbulence in the wind.

\begin{figure*}
\centering
\begin{minipage}{.45\textwidth}
    \centering
    \includegraphics{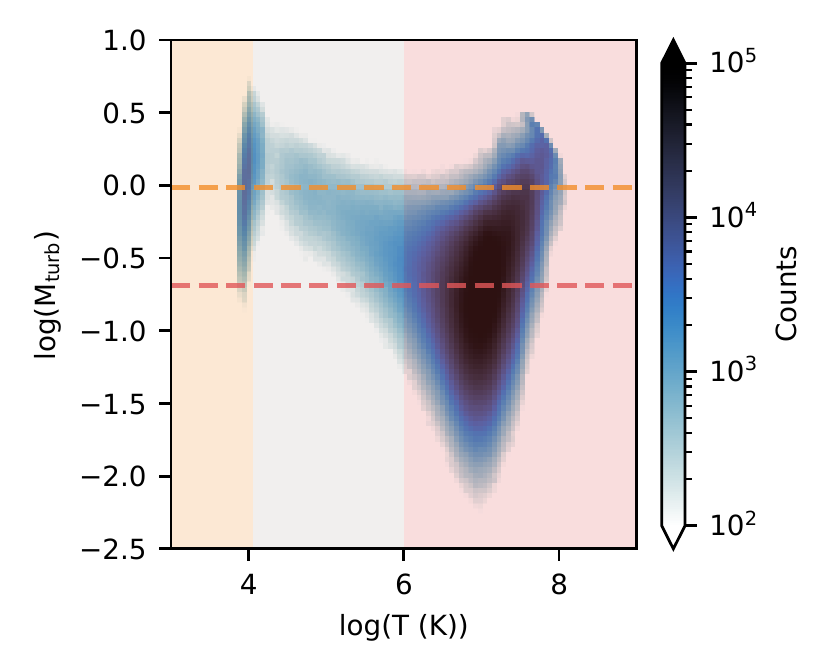}
    \caption{2D histogram of turbulent Mach number $M_{\rm turb}$ at heights above 0.5 kpc estimated using a filtering method to account for bulk flows and temperature $T$. The orange dashed line shows the average $M_{\rm turb}$ ($\sim 1$) within the orange shaded region while the red dashed line shows the average $M_{\rm turb}$ ($\sim 0.25$) in the red shaded region. These probe the cool clouds and the hot wind respectively.}
    \label{fig:phase_mach_t}
\end{minipage}\hspace{20pt}%
\begin{minipage}{.45\textwidth}
    \centering
    \includegraphics{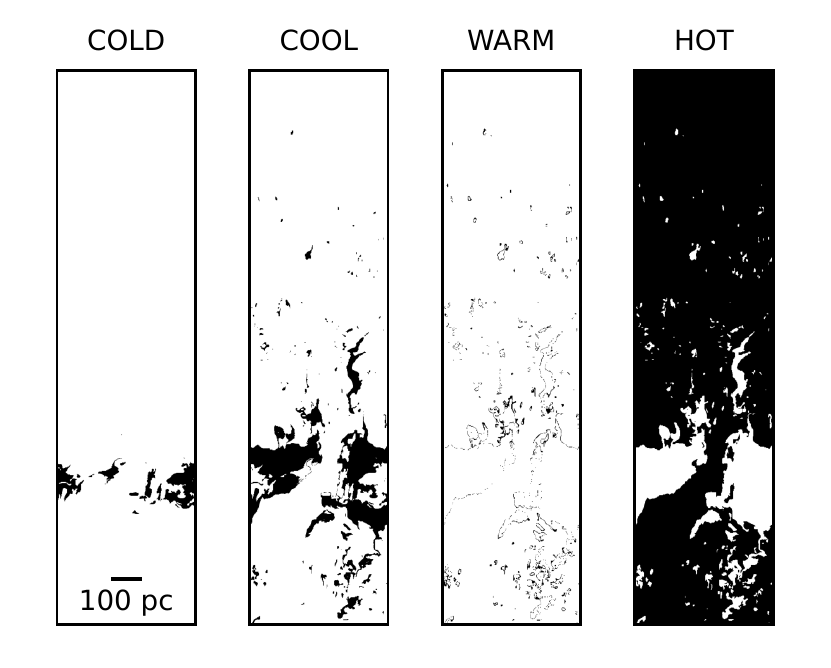}
    \caption{2D slices at 14 Myr showing the distribution of the following gas phases: hot gas ($T > 5 \times 10^5$\,K), warm gas ($2 \times 10^4$\,K~$< T < 5 \times 10^5$\,K), cool gas($5 \times 10^3$\,K~$< T < 2 \times 10^4$\,K) and cold gas ($T < 5 \times 10^3$\,K). Cold gas is located within the ISM while cool gas extends to clouds that are contained in the wind, which is made up of the volume filling hot phase. Warm gas exists mainly in interfacial mixing layers.}
    \label{fig:gas_phases}
\end{minipage}
\end{figure*}
\begin{figure*}
    \centering
    \includegraphics{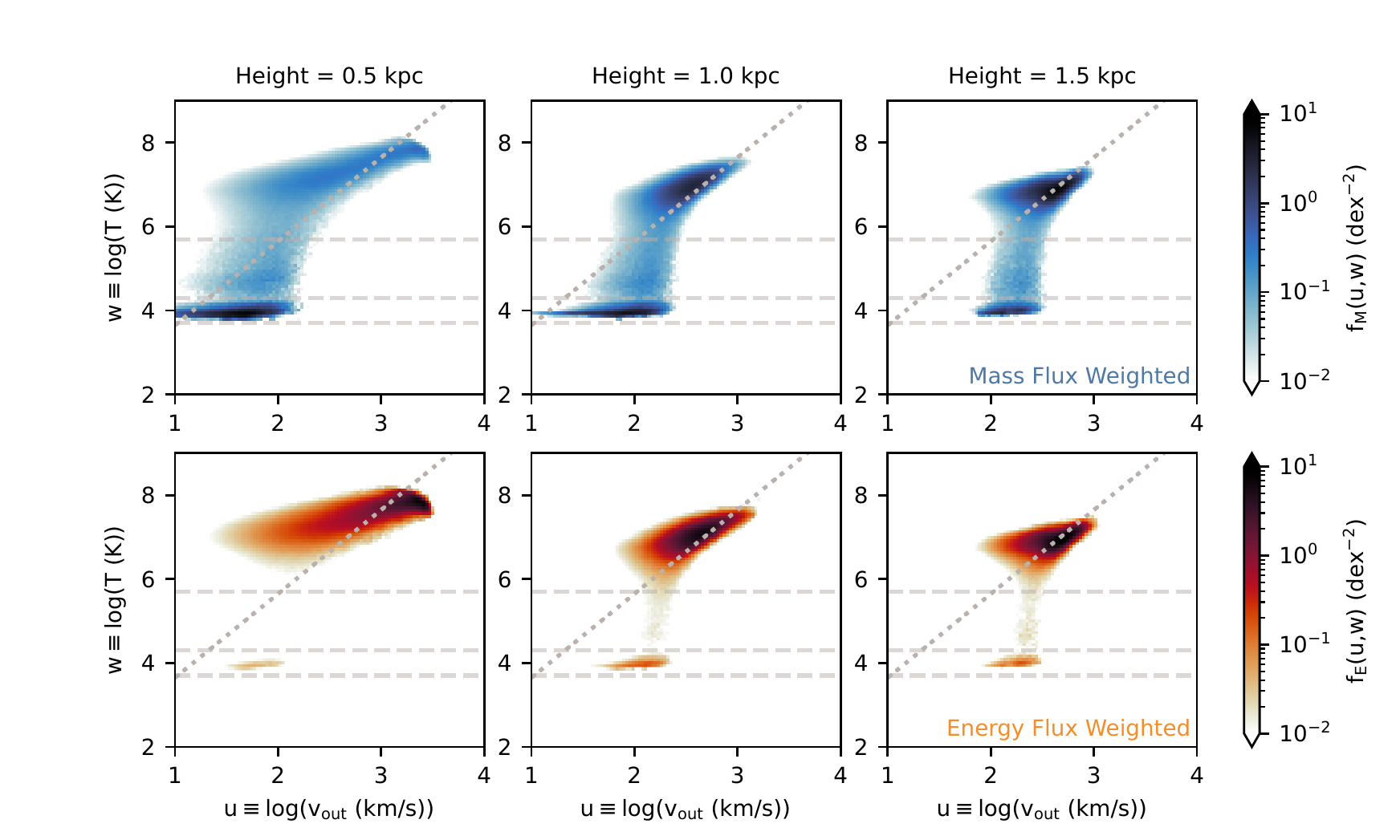}
    \caption{Temperature-velocity joint PDFs at different heights above the disc, weighted by mass flux in the top row and energy flux in the bottom row. The horizontal grey dashed lines demarcate the boundaries between the phases defined above. The hot phase dominates the energy transport while there is significant mass contained in the cool phase. The grey dotted line shows the sound speed of the gas at each temperature---the wind is transonic with a Mach number close to unity. Cool material gets more entrained at increasing heights.}
    \label{fig:histogram_fluxes}
\end{figure*}

To examine the turbulence more closely, we apply a Gaussian filter to estimate the turbulent velocity $v_{\rm turb}$. This method involves approximately removing large scale variations related to the bulk flow. This is done by convolving each velocity component with a Gaussian kernel to estimate the bulk velocity along that component. Subtracting this thus leaves the turbulent velocity of that component. While exact results depend on the kernel size, qualitative conclusions remain unchanged \citep{abruzzo22}. We choose to use 16 pc, which sits comfortably within the range of sizes spanned by the cool clouds. The $i$-th component of the turbulent velocity is hence defined here as
\begin{equation}
    v_{i, {\rm turb}} (\boldsymbol{x}) = v_i(\boldsymbol{x}) - 
    \frac{\iiint f_{\sigma}(\boldsymbol{x}-\boldsymbol{r})\rho(\boldsymbol{r} v_i(\boldsymbol{r})) d^3 \boldsymbol{r}}{\iiint f_{\sigma}(\boldsymbol{x}-\boldsymbol{r})\rho(\boldsymbol{r}) d^3 \boldsymbol{r}},
\end{equation}
where $f_{\sigma}(\boldsymbol{x})$ is the three-dimensional Gaussian filter. We weight the filter by density so that the velocities within the cloud are not dominated by the that of the hot gas. The right most panel of Figure~\ref{fig:slices_turb} shows how the turbulent Mach number using the above definition for turbulent velocity ($M_{\rm turb} \equiv v_{\rm turb}/c_{\rm s}$) varies in the slices. In general, the turbulent Mach number in the wind is high only close to the disc, and quickly weakens away from it. On the other hand, the cool clouds have transonic internal turbulence. This is shown more quantitatively in Figure~\ref{fig:phase_mach_t}, which shows a 2D histogram of $M_{\rm turb}$ and temperature $T$ at the same time snapshot as Figure~\ref{fig:distribution_heights}, but only considering heights above 0.5 kpc. The orange dashed line shows the average $M_{\rm turb}$ ($\sim 1$) within the orange shaded region which represents the cool clouds, while the red dashed line similarly shows the average $M_{\rm turb}$ ($\sim 0.25$) in the red shaded region which represents the hot wind. 
In all, there is significant turbulence both within the clouds and in the outflowing wind. This means that these environments are more similar to the turbulent boxes studied in \citep{gronke22} rather than idealized laminar wind tunnel setups. This is critical because it is turbulence that drives mixing between the phases and hence their coupling. 

Finally, Figure~\ref{fig:slices_scalars} shows slices of the passive scalars we have used in this simulation, and also illustrates the amount of mixing between phases happening in these winds. We use a total of 5 passive scalars. These passive scalars track gas with certain properties. The first two are introduced right before the first SN explosion. All cold gas ($T < 5 \times 10^3$\,K) is dyed with passive scalar $s_0$, and all cool gas ($5 \times 10^3$\,K < $T$ < $2\times10^4$\,K) is dyed with passive scalar $s_1$. This is to track how much cloud material was {\it originally} cold and cool ISM material. Passive scalars $s_2$ and $s_3$ track how much of the gas in a cell was {\it ever} hot or cold respectively, similarly starting from right before the first SN, so as to quantify the amount of mixing. Lastly, passive scalar $s_4$ tracks all SN injected gas, which similar to $s_0$ and $s_1$, tracks how much cloud material was from the SN. In the following section, we use $s_0$, $s_1$, and $s_4$ to quantify the original temperatures of cloud material. Here, we will just point out how beyond the disc and the base of the jet, both $s_2$ and $s_3$ are close to unity everywhere. This means that all gas not close to the injection point or part of the ISM has at some point been cold or hot.

\subsection{Mass and Energy Flux Phase Distribution and Cloud Entrainment}
Finally, we can look at the fraction of the wind’s mass and energy flux carried by the different phases as defined in Table~\ref{tab:phases}. Figure~\ref{fig:gas_phases} shows 2D slices of the general location or each of these phases at 14 Myr. Cold gas is only located within the ISM. Cool gas is found in the ISM and the clouds that are contained in the wind, which is made up of the volume filling hot phase beyond the disc. Warm gas is found in the interfacial mixing layers.

Figure~\ref{fig:histogram_fluxes} shows the temperature velocity distributions at different heights above the disc over the time period identified in Figure~\ref{fig:mass_flux_time_profile}, weighted by mass flux in the top row and energy flux in the bottom row. The PDFs $f_{\rm M}(u,w)$ and $f_{\rm E}(u,w)$ are weighted by mass and energy flux respectively, where $u\equiv \log(v_{\rm out})$ and $w\equiv \log(T)$ in units of km/s and K. The PDFs $f_{\rm q}$ for each quantity $q$ is defined as $\frac{1}{\langle q \rangle}\frac{d^2q}{dudw}$ where $\langle q \rangle$ is the average over both time and the horizontal slice, such that $\int f_{\rm q}\,dudw = 1$. We define the mass flux to be $\rho v_{\rm z}$ and the energy flux to be $E_{\rm z}v_{\rm z}$ where $E_{\rm z} = \rho (\frac{1}{2} v_{\rm z}^2 + e_{\rm th})$ and $e_{\rm th}$ is the specific internal energy. The horizontal grey dashed lines demarcate the boundaries between the phases defined above. Consistent with previous work, the hot phase dominates the energy loading while there is significant mass loading by the cool phase \citep{fielding18,kim20a,kim20b,Rathjen:2023}. Mass and energy loading factors are similar to those found in more global simulations such as \citet{schneider20}, who also observe cold phase mass loading in a volume filling hot wind. This is in line with previous studies of multiphase flows such as the more idealized setups of \citet{banda21}, where the hot phase occupies $\ge 90\%$ of the volume but only contains $\sim 10\%$ of the mass. The grey dotted line shows the sound speed of the gas at each temperature---the outflow velocity of the wind is transonic with a Mach number close to unity. We also see that the cool material gets more entrained at increasing heights, indicating that there is momentum transport from the hot to the cool phase, which has been found to be driven by mixing rather than ram pressure acceleration \citep{Melso:2019, schneider20, tonnesen21}.

\dobib

\section{Results : Clouds} \label{sect:results_clouds}
In this section we focus our attention on analysis of the clouds embedded within the hot wind and their properties. We first discuss how we define and identify clouds, followed by how we measure related cloud properties. We then analyze size distributions and various scaling relations associated with cloud growth rates. Finally, we explore other interesting aspects of cloud properties relating to non-thermal support, material origins, and wind alignment. 

\begin{figure}
    \centering
    \includegraphics[width=\columnwidth]{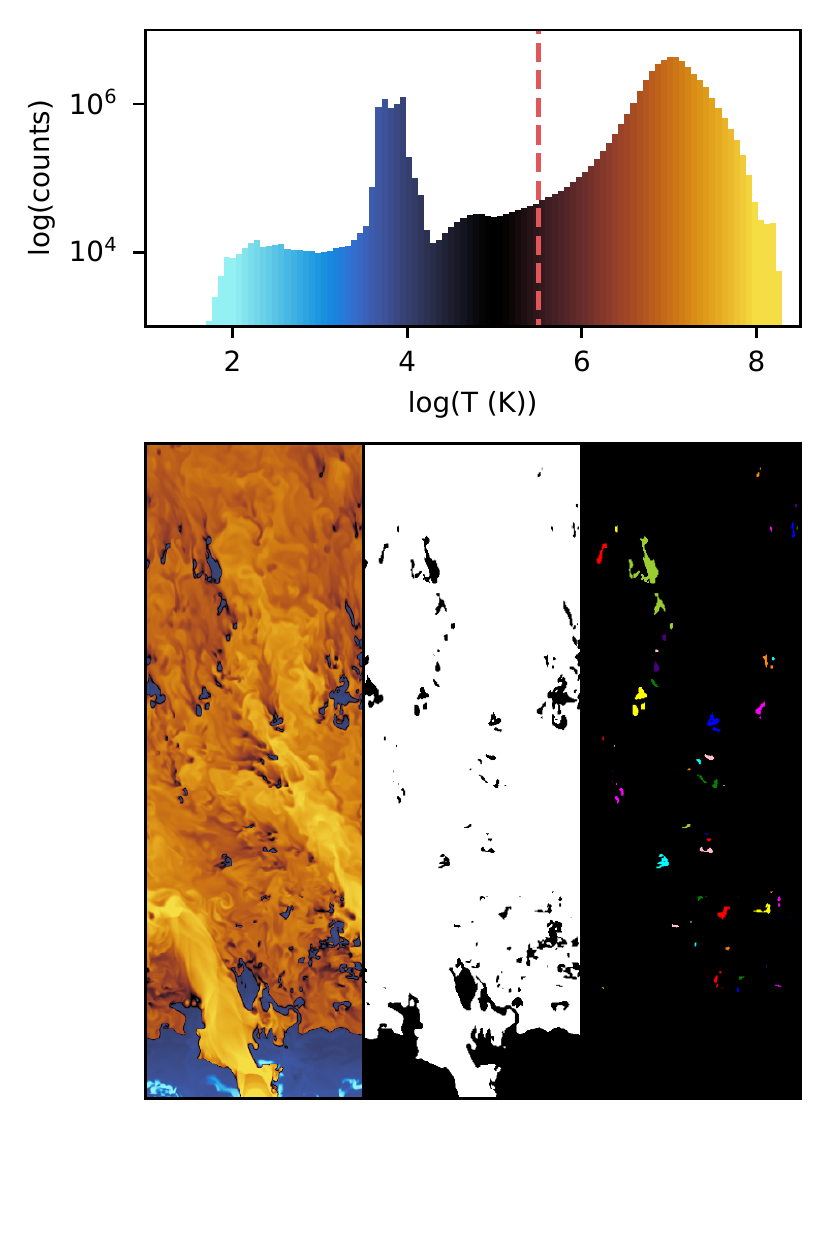}
    \vspace{-30pt}
    \caption{{\it Top panel:} Histogram of gas cell temperatures from a single snapshot. Most of the gas is in the volume filling hot wind, with a peak at $T = 10^4$\,K corresponding to cool clouds. The red dashed line represents our threshold value for identifying cloud material. {\it Bottom Panel:} From left to right: slices plots of temperature, cloud material, and identified clouds.}
    \label{fig:cloud_finder}
\end{figure}

\subsection{Cloud Identification}
In order to map out and catalogue the properties of clouds embedded in the hot outflowing wind, we must first define them. We identify clouds by applying a threshold at $T = 10^{5.5}$\,K, represented by the red dashed line the upper panel of Figure~\ref{fig:cloud_finder}. Any gas below this threshold is thus considered `cloud material'. The lower center panel of Figure~\ref{fig:cloud_finder} shows this cool material identified from the corresponding temperature slice. While we use a fixed threshold, we find that cloud identification in our simulations is generally insensitive to the exact value used.\footnote[2]{Instead of arbitrarily choosing a a threshold value, we also explored using a non parametric approach such as Otsu's image segmentation algorithm \citep{otsu79}, which determines a threshold for separating two classes by maximizing(minimizing) the variance between(within) them, in our case for the distribution of $\log(T)$. This approach gives a similar value close to $\log(T (K)) = 5.5$, but the exact value varies with different time slices.} We can understand this physically---most cloud gas is at a temperature $\sim  T = 10^4$\,K, while most of the wind is at $T > 10^6$\,K. Gas at intermediate temperatures mainly exists only at the thin interfaces between these two phases in turbulent radiative mixing layers. This can be seen visually in the temperature slice in the lower left panel of Figure~\ref{fig:cloud_finder} (also see the `warm' panel in Figure~\ref{fig:gas_phases}). It is hence not surprising that the identification of clouds is understandably robust to the choice of threshold provided that this threshold temperature lies in the mixing region. The simple approach of using a fixed threshold is thus as effective as it is because of the strong biphasic nature of the wind and the clouds. This can be seen in the upper panel of Figure~\ref{fig:cloud_finder} which shows the temperature distribution in the wind at a single time snapshot, along with the threshold value used. Using a constant threshold also allows for more consistency over multiple time slices. We can thus define a single cloud as a collection of cells that fall below the threshold and that are interconnected (where connected is defined as sharing at least one corner). Except for the analysis on cloud size distributions, where we account for clouds that intersect the periodic dimensions and hence wrap around, we ignore clouds that are touching the borders of our simulation. This means we naturally exclude disc material. We also only consider clouds that have a volume greater than 16 cells. The bottom right panel of Figure~\ref{fig:cloud_finder} shows an example of these clouds arbitrarily colored by their cloud identification number. We also compile our catalog over multiple time snapshots (a total of 10 spanning 8 Myr).

\subsection{Measuring Cloud Properties}
\begin{figure*}
    \centering
    \includegraphics[width=\linewidth]{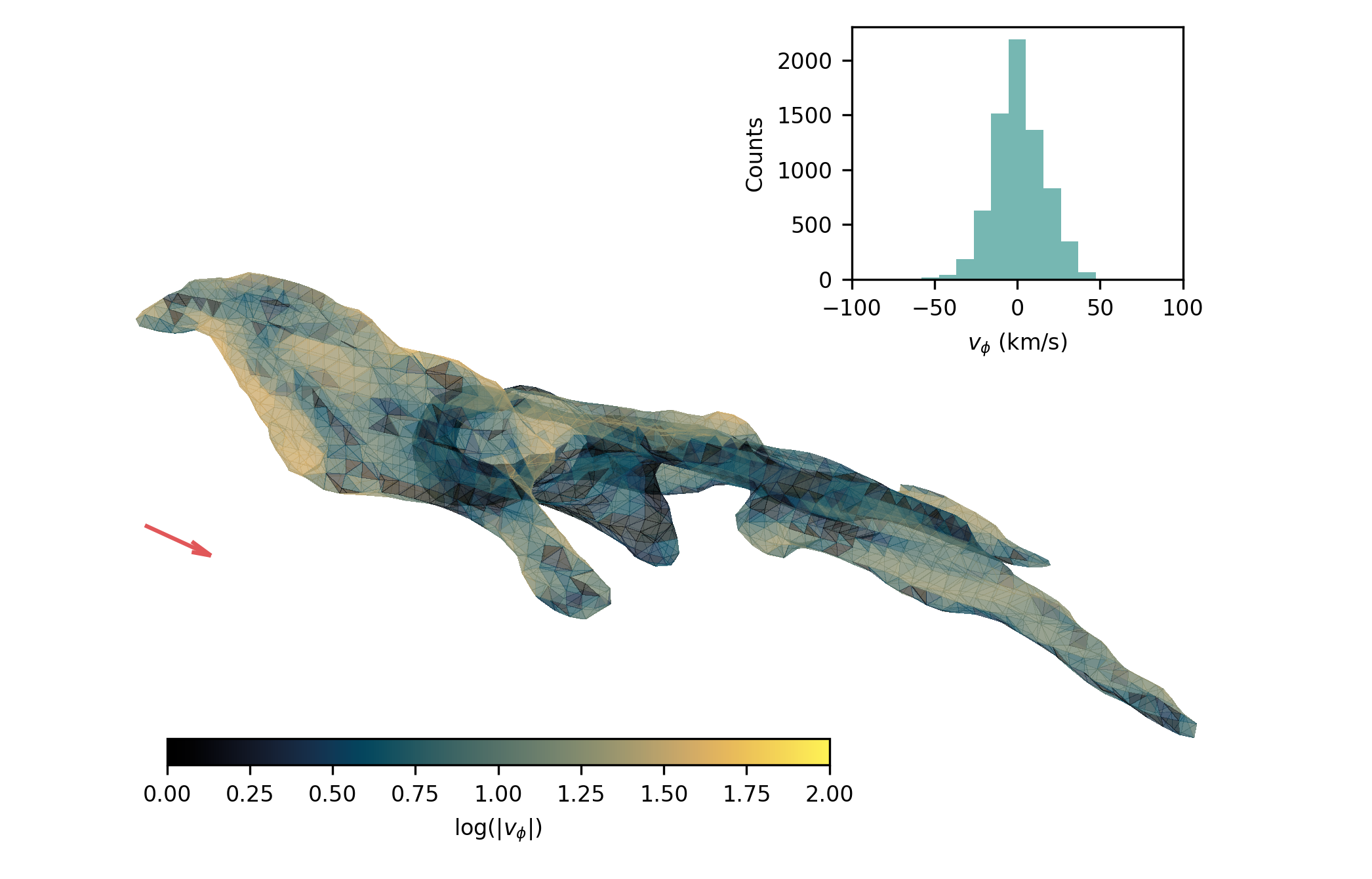}
    \includegraphics{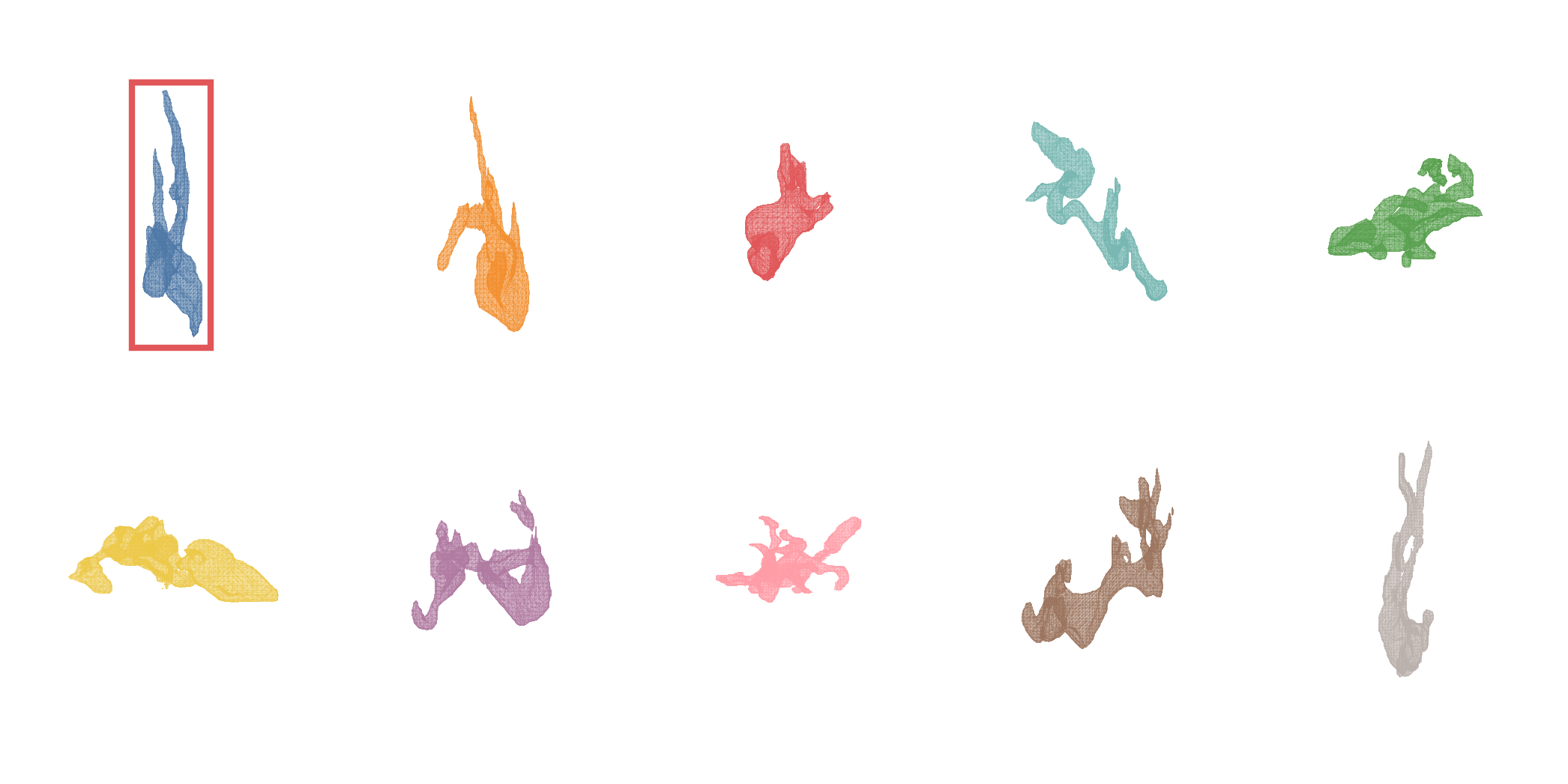}
    \caption{{\it Bottom Panel:} Meshes representing the temperature isosurface of a range of identified clouds. The cloud in the top left is shown in greater detail in the top panel. {\it Top panel:} Zoom-in of a a cloud surface colored by $\textbf{v}_{\phi}$. The wind direction relative to the cloud is denoted by a red arrow. The inset shows the surface area weighted histogram of $\textbf{v}_{\phi}$. $v_{\rm turb}$ is the standard deviation of this distribution. }
    \label{fig:clouds_3D}
\end{figure*}
For each identified cloud, we measure a number of properties as follows. The position of the cloud is computed as the volume weighted centroid of cloud material. Individual cloud properties such as velocity $\textbf{v}_{\rm cl}$, density $\rho_{\rm cl}$, thermal and magnetic pressures $P_{\rm cl}$ and $P_{\rm B, cl}$ are volume averaged over cool ($<1.01 \times 10^4$\,K) gas in the cloud. Temperature is density weighted, so that $P_{\rm cl} = \rho_{\rm cl} T_{\rm cl}$. Passive scalars are also density weighted. We define the local background environment of a cloud as any hot ($>10^6$\,K) wind gas within a cube centered on the cloud that has a length on each side equal to twice that of the longest axis of the tight bounding box that contains the cloud, and also require that there are at least 16 such cells in this volume. Corresponding wind properties such as the background wind velocity $\textbf{v}_{\rm w}$ are averaged in the same way as cloud properties. The turbulence in the wind $\textbf{v}_{\rm turb, w}$ is taken to be the rms velocity in the wind frame in this environment.

We adopt a geometric approach towards characterizing the turbulence on the surface of a cloud, which involves measuring quantities on a temperature isosurface constructed using the marching cubes algorithm described in \citet{lewiner2003}. This isosurface is represented by a mesh of triangular faces. Figure~\ref{fig:clouds_3D} shows examples of this mesh in the lower panel for various clouds. The cloud in the upper left is shown in greater detail in the upper panel. 

To measure the turbulence in the mixing layer on this surface, we use the method described in \citet{abruzzo22}, which they demonstrate provides a measure of turbulence that is comparatively robust to gradients in the laminar component of the background wind as compared to the other approaches of characterizing turbulence that they explored (filtering methods and velocity structure functions). The outline of the method is as follows: Consider a point on the temperature isosurface that is at the centroid of one of the triangular faces which constitute the mesh. Working in the cloud frame, we linearly interpolate the velocity $\textbf{v}$ and the logarithmic density $\rho$ at that point. We define a set of axes $\hat{\textbf{n}}, \hat{\textbf{w}}$ and $\boldsymbol{\hat{\phi}}$, where $\hat{\textbf{n}}$ is the inwards direction normal to the surface, $\hat{\textbf{w}}$ is the direction of the component of the background wind $v_{\rm wind}$ in the surface plane ($\hat{\textbf{w}} \equiv \textbf{v}_{w}/|\textbf{v}_{w}|$;$\textbf{v}_{w} \equiv \textbf{v}_{\rm wind} - (\textbf{v}_{\rm wind} \cdot \hat{\textbf{n}})\hat{\textbf{n}}$), and $\boldsymbol{\hat{\phi}}$ is the direction on the surface orthogonal to both $\hat{\textbf{n}}$ and $\hat{\textbf{w}}$, i.e., $\boldsymbol{\hat{\phi}} \equiv \hat{\textbf{n}} \cross \hat{\textbf{w}}$. These definitions are physically motivated---the normal direction captures turbulent radiative mixing layer driven accretion, while the wind direction captures the coherent flow of the background wind. Picking out the $\boldsymbol{\hat{\phi}}$ direction thus allows us to disentangle the turbulence from these flows. The velocity component is hence simply $\textbf{v}_{\phi} \equiv \textbf{v} \cdot \boldsymbol{\hat{\phi}}$. We thus estimate $v_{\rm turb}$ to be the area weighted standard deviation of $\textbf{v}_{\phi}$. The upper panel of Figure \ref{fig:clouds_3D} shows how $\textbf{v}_{\phi}$ varies over the surface of a cloud. The top right inset shows a surface area weighted histogram of $\textbf{v}_{\phi}$ on this surface. As expected, the distribution is normal and centered on zero. $v_{\rm turb}$ for this cloud is simply the standard deviation of this distribution. The inflow velocity through this surface is similarly given by $\textbf{v}_{\rm in} \equiv \textbf{v} \cdot \hat{\textbf{n}}$.

In Appendix~\ref{sect:apdx_e}, we detail how we estimate cloud growth rates $\Dot{m}$, comparing the effectiveness of various methods of doing so. Ultimately, we use the net inwards mass flux on the temperature isosurface ($\Dot{m}_{\rm surface} \equiv \sum_{\rm surface}\rho v_{\rm in} \Delta A$) with a scaling factor calibrated to match the total cooling luminosity, so $\Dot{m} \sim f_{\rm scale} \Dot{m}_{\rm surface}$. Appendix~\ref{sect:apdx_e} also presents an important caveat that these methods fail when the cloud is not actually growing, and thus do not reflect the transition to the cloud destruction regime. Instead, we show that we can estimate this scale from the cloud size distribution.

Now that we have discussed how we identify clouds and measure their properties, we are ready to further analyze these measured properties. We first present the distribution of cloud sizes in the simulation and compare to analytic formulations. We then compare estimated cloud growth rates with predicted scalings from models for the surface area and associated mass flux as given in equations~\eqref{eq:pld} -- \eqref{eq:vturb}.

\begin{figure}
    \centering
    \includegraphics[width=\columnwidth]{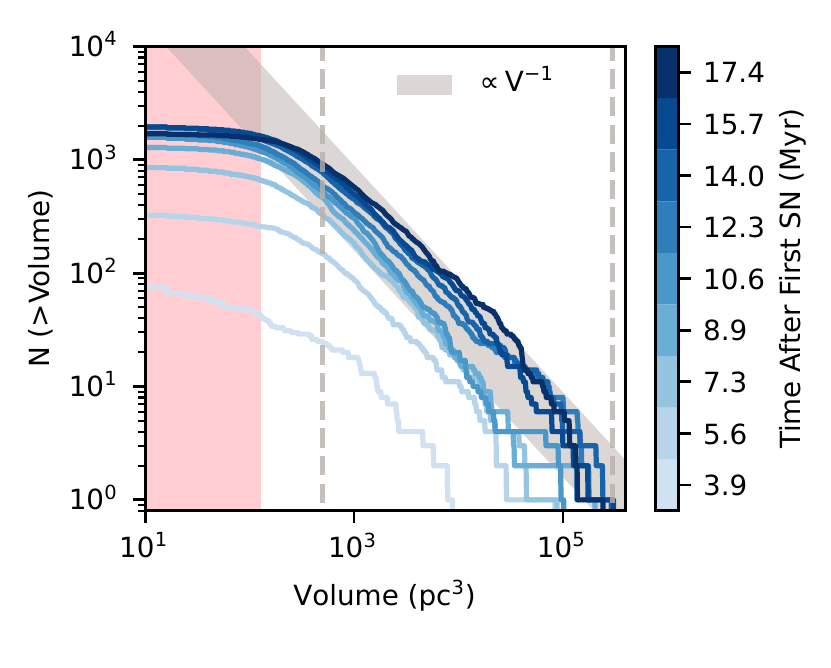}
    \caption{Cumulative cloud volume distribution at various time slices. The pink region indicates where clouds are resolved by less than 16 cells. The grey region shows the expected power law scaling of $-1$. Grey dashed lines show expected lower (for survival) and upper bounds on cloud sizes.}
    \label{fig:cloud_hist}
\end{figure}

\subsection{Distribution of Cloud Sizes}
The snapshots shown in Figures~\ref{fig:slices_progression_early} and \ref{fig:slices_progress_256} present an intriguing picture of an initial population of cold clouds seeded in the wind via direct fragmentation of ISM material. Here, we look at the implications of this process on the cloud size distribution.

How are these clouds generated? Unlike setups where cloudlets are formed via gravitational collapse or thermal instabilities, the process of superbubble breakout here itself leads to fragmentation of the ISM. While a detailed study would be needed to fully understand this fragmentation process, we believe there are three key pieces at play. Firstly, the bubble expands through a turbulent multiphase ISM, which was itself generated through a combination of driven turbulence and thermal instability. Both the turbulent field and the multiphase landscape distort the bubble geometry on a large range of scales and can lead to the formation of disconnected cloudlets. In this case, the nature of the turbulence and the initial distribution of cold ISM material might play important roles. Secondly, the thin shells of the expanding bubble interfaces are susceptible to a host of instabilities such as Kelvin-Helmholtz, Rayleigh-Taylor and Richtmyer–Meshkov, and have been shown to themselves be fractal in nature \citep{fielding20,lancaster21}. Lastly, the expansion of the bubble through this multiphase inhomogenous medium leads to large scale turbulence (independent of the initial driven turbulence in the ISM). This can also drive turbulent fragmentation \citep[e.g. Kolmogorov-Hinze theory for turbulent fragmentation of droplets;][]{kolmogorov49,hinze55}. The detailed influence of each of the above, along with their interplay, will be an important focus for future work.

Regardless, the simulations suggest that it is the fragmentation of the initial ISM that determines the distribution of cloud sizes, at least when the outflowing wind first breaks out. The initial cloud mass population in the hot wind can thus be viewed as being seeded by the fragmentation of the ISM, driven by the process of superbubble breakout. We will show below that our simulations support this picture of cloud genesis via ISM fragmentation during superbubble breakout, and leads to a distribution that scales as:
\begin{equation}
    \frac{dN(m)}{dm} \propto m^{-2}.
    \label{eq:pld}
\end{equation}
where $N(m)$ is the number of clouds with mass $m$. Inverse power law distributions with a -2 exponent, which indicates that there is a constant contribution from each logarithmic bin, are commonly observed. We refer the reader to Section~\ref{sect:discussion} for further discussion.

What sets the lower and upper limits on cloud sizes? Since the clouds are formed during the breakup of the ISM, the largest cloud size is simply constrained by the initial scale of the ISM, i.e., the scale height $H$ of the disc. The scale of the smallest clouds is a more complex question. While fragmentation can lead to arbitrarily small clouds, they should be rapidly destroyed in the hot turbulent wind below some critical size, where growth due to cooling becomes ineffective. This survival threshold $r_{\rm crit}$ presents a natural choice for the lower limit. However, the problem of what determines $r_{\rm crit}$ is a thorny one, and there has been no shortage of recent investigations into the matter. We will briefly review some of them.

For a cloud of size $r_{\rm cl}$ with overdensity $\chi$ relative to its background and temperature $T_{\rm cl}$ at rest embedded in a hot wind of temperature $T_{\rm hot}$ moving at a velocity $v_{\rm wind}$, \citet{gronke18} identified a characterize size $r_{\rm crit,cc}$ above which the clouds survive till entrainment and grow. This size corresponds to where the cloud crushing time $t_{\rm cc} \sim \sqrt{\chi}r_{\rm cl}/v_{\rm wind}$, the timescale on which a cloud is destroyed by hydrodynamic instabilities \citep{klein94}, becomes longer than the cooling time of mixed gas $t_{\rm cool,mix} \equiv t_{\rm cool}(T_{\rm mix})$, where $T_{\rm mix} \sim \sqrt{T_{\rm cl}T_{\rm hot}}$ \citep{begelman90}. From their turbulent box simulations, \citet{gronke22} found that there is also an additional empirically determined Mach number dependence of the functional form $f(M_{\rm turb}) \sim 10^{0.6 M_{\rm turb}}$. They attribute this additional Mach dependence to increased turbulent disruption, in contrast to the increased survival times seen in cloud crushing simulations with high Mach winds which stems from cloud compression \citep{scannapieco15,bustard22}.
The critical size for survival is thus given by
\begin{align}
    r_{\rm crit,cc} = 2\mathrm{~pc~} 
    \frac{T^{5/2}_{\rm cl,4} M_{\rm wind}}{P_3 \Lambda_{\rm mix,-21.4}}
    \frac{\chi}{100}
    f(M_{\rm turb}),
\end{align}
where $T_{\rm cl,4} \equiv (T_{\rm cl}/10^4\mathrm{~K})$, $P_3 \equiv nT/(10^3\mathrm{~cm}^{-3}\mathrm{~K})$, $\Lambda_{\rm mix,-21.4} \equiv \Lambda(T_{\rm mix})/(10^{-21.4}\mathrm{~erg~cm}^3\mathrm{~s}^{-1})$, and $M_{\rm wind}$ is the Mach number of the wind. 

Alternatively, \citet{li20} and \citet{sparre20} lay out a survival criterion $t_{\rm cool,wind} < t_{\rm life}$, where $t_{\rm cool,wind}$ is the cooling time of the {\it hot} background wind and $t_{\rm life} = \Bar{f}t_{\rm cc}$ is the predicted cloud lifetime. $\Bar{f}$ is an empirically calibrated function, given by 
\begin{equation}
    \Bar{f} = 10 \left(\frac{r_{\rm cl}}{1\mathrm{~pc}}\right)^{0.3}
        \left(\frac{n_{\rm hot}}{0.01\mathrm{~cm}^{-3}}\right)^{0.3}
        \left(\frac{T_{\rm hot}}{10^6\mathrm{~K}}\right)
        \left(\frac{v_{\rm wind}}{100\mathrm{~km~s}^{-1}}\right)^{0.6},
\end{equation}
where $n_{\rm hot}$ is the density of the hot gas. At $\chi = 100$, the difference between the two criteria is small, and when comparing the two, attributed to different ways of defining cloud survival \citep{kanjilal21}. At larger $\chi$ however, these criteria differs by orders of magnitude.
\citet{abruzzo22} find that at $\chi > 100$, the Li/Sparre criterion agrees much better with simulation results, but also point out that $t_{\rm cool,wind}$ cannot be physically important since their results are unchanged when they artificially shut off wind cooling. They instead propose a new criterion $t_{\rm cool,mix} < t_{\rm shear}$ which captures the empirically derived $\chi$ scaling of $r_{\rm crit}$ in a physically motivated model \citep{abruzzo23}. Here $t_{\rm shear} \sim r_{\rm cl}/v_{\rm wind}$ is the wind crossing time over the cloud. The physical argument is as follows. As hot phase fluid elements travel past the cloud, they mix with the cooler fluid elements from the cloud to produce intermediate temperature gas at the interface. Clouds will only grow if a parcel of gas is able to mix and cool before being advected past the cloud. This requires that the cooling timescale of the mixed gas is shorter than the wind crossing time past the cloud. This criterion gives a cloud survival size 
\begin{equation}
    r_{\rm crit, shear} \equiv \sqrt{\chi} r_{\rm crit,cc}.
    \label{eq:rcritshear}
\end{equation}

Figure~\ref{fig:cloud_hist} shows the cumulative size distribution of clouds at different times in the simulation. The distribution of $dN/dm \propto m^{-2}$ is equivalent to a cumulative distribution that scales as $N(>m) \propto m^{-1}$. Instead of a mass scaling, we consider instead the volume scaling $dN(>V)/dV$ since the fragmentation process is not driven by a density dependent process, although the difference made by this choice does not affect the results. The pink region in Figure~\ref{fig:cloud_hist} indicates where clouds are resolved by less than 16 cells (and hence not included in the catalog of cloud properties). The grey region show the $-1$ power law scaling for comparison.

The upper panel shows distributions that start from when the superbubble first breaks out from the disc. The number of clouds in the wind increases over time as the disc continues to fragment. We note that the $N(>V) \propto V^{-1}$ power law distribution is seen even at early times, supporting the argument that this is a result of the disc fragmentation process during the process of superbubble breakout. We have checked that the the distributions does not show any consistent trends with height in these simulations.

\begin{figure}
    \centering
    \includegraphics[width=\columnwidth]{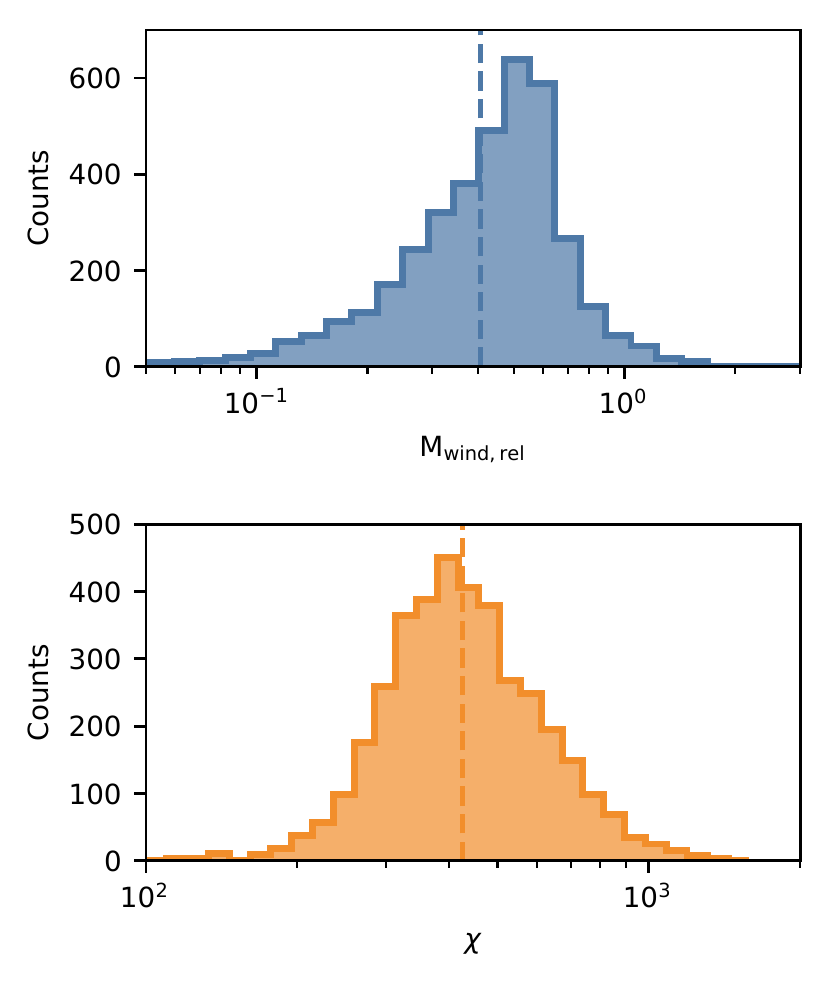}
    \caption{Distributions of the Mach number of the wind relative to the cloud (top) and cloud overdensity relative to the wind (bottom). Dashed lines show population means.}
    \label{fig:cloud_mach_chi}
\end{figure}

The distribution exhibits a flattening at small scales somewhat above the resolution limit, which is expected for clouds below the critical radii for sustained growth. To estimate these critical radii we take a typical set of parameters for these winds. For example, in our fiducial simulation, these are $M_{\rm wind,rel} \sim 0.4$, $M_{\rm turb,wind} \sim 0.5$, $P_3 \sim 30$ and $\chi \sim 400$. Figure~\ref{fig:cloud_mach_chi} shows the distributions of $M_{\rm wind,rel}$ and $\chi$ along with their means. This gives a cloud survival size of $r_{\rm crit,shear} \sim 5$\,pc.

This nicely match the scales below which we see the distributions flatten. The vertical grey dashed lines in Figure~\ref{fig:cloud_hist} show the upper and lower bound volume estimates corresponding to $r_{\rm crit,shear}$ and the disc scale height $H \sim 66$\,pc respectively. Ideally, the smallest surviving clouds should be resolved by multiple cells to achieve first order convergence in terms of total cloud mass. With a resolution of 2\,pc, we resolve $r_{\rm crit,shear}$ by a factor of $\sim 2$. In Appendix~\ref{sect:apdx_f}, we perform the same analysis for a simulation which fails to sustain an outflowing wind.

\subsection{Model for Cloud Growth}
\subsubsection{Theoretical Model}
Assuming that the conditions are right for a cloud to survive and grow in a wind, we can write its mass growth rate as 
\begin{equation}
    \Dot{m} \sim \rho_{\rm wind} v_{\rm in} A_{\rm cloud},
    \label{eq:mdot}
\end{equation}
where $\rho_{\rm wind}$ is the density of the hot wind, $v_{\rm in}$ is the inflow velocity corresponding to the mass flux from the hot background onto the cloud, and $A_{\rm cloud}$ is the effective surface area of the cloud.

The first ingredient in this cloud growth model is $A_{\rm cloud}$. It is important here to distinguish this effective surface area from the non-convergent surface area of the cool gas \citep{fielding20,abruzzo22}, which corresponds more closely to the temperature isosurfaces we measure in our simulations below. Instead, a more accurate characterization of $A_{\rm cloud}$ is an effective surface area corresponding to some smooth envelope around the cloud \citep{gronke20}. In cloud crushing simulations, surface area has an initial rapid growth stage associated with tail formation, followed by slower isotropic growth \citep{gronke20,abruzzo22}. This corresponds to $A_{\rm cloud}$ going from initially scaling as $\propto V$ (linearly with volume), to scaling as $\propto V^{2/3}$. The initial exponential growth can be attributed to a tail formation phase where the high shear mainly leads to mass growth in a growing tail downstream of the cloud. Eventually, the cloud becomes entrained and the surface area scales much as a monolithic cloud would. However, it was found in \citet{gronke22b} and \citet{tan23} that for clouds which never get entrained (either because of large scale turbulence or gravitational acceleration), and are hence subject to continuous shear, the cloud surface area instead scales as $A_{\rm cloud} \propto V^{5/6}$. This scaling can arise from the fractal nature of the mixing surface, which has been measured in mixing layer simulations to have a fractal dimension $D \sim 2.5$ \citep{fielding20}, which would mean $A \propto V^{D/3} \sim V^{5/6}$. If we assume that the smallest surviving clouds are the most spherical, we can model the effective cloud surface area as 
\begin{equation}
    A_{\rm cloud} = A_{\rm crit} \left(\frac{V}{V_{\rm crit}}\right)^{5/6},
    \label{eq:surface_area}
\end{equation}
where $A_{\rm crit}$ and $V_{\rm crit}$ are the spherical area and volumes respectively corresponding to the critical cloud survival size $r_{\rm crit}$, which we discuss below.

Next we discuss the second ingredient in the model, the inflow velocity  $v_{\rm in}$. In plane parallel simulations of radiative turbulent mixing layers \citep{tan21,fielding20}, $v_{\rm in}$ is found to scale as 
\begin{equation}
    v_{\rm in} \sim v_{\rm turb}^{3/4} \left( \frac{L_{\rm turb}}{t_{\rm cool,min}} \right)^{1/4},
    \label{eq:vin1}
\end{equation}
where $v_{\rm turb}$ is the turbulent velocity in the mixing layer, $L_{\rm turb}$ is the largest mixing scale (typically the outer scale of the turbulence), and $t_{\rm cool,min}$ is the minimum cooling time in the simulation, which scales inversely with pressure ($t_{\rm cool,min} \propto P^{-1}$). Additionally, these simulations find that the turbulent velocity scales with the relative shear velocity in these setups, such that $v_{\rm turb} \sim f_{\rm rel} v_{\rm shear}$, where $f_{\rm rel}$ is some constant of proportionality (\citet{tan21} report $f_{\rm rel} \sim 0.3$ and \citet{fielding20} find $f_{\rm rel} \sim 0.1-0.2$) that varies with the geometry of the setup (e.g., \citet{mandelkar19} find $f_{\rm rel} \sim 0.2-0.3$ in supersonic streams). However, in cloud crushing setups, clouds eventually entrain, i.e., $v_{\rm shear}$ goes to zero, yet these clouds continue to grow \citep{gronke20}, suggesting that this relationship breaks down at lower values of $v_{\rm shear}$. This is likely due to continued driving of mixing due to cooling, either via induced pulsations \citep{gronke22b} or is self-sustained as the cooling velocity drives ongoing turbulence after entrainment \citep{jennings21,abruzzo22}. To account for this saturation at low shear velocities, we can hence express the turbulent velocity as 
\begin{equation}
    v_{\rm turb} \approx \max(f_{\rm rel} v_{\rm shear}, c_{\rm s,cold}).
    \label{eq:vturb}
\end{equation}
In our simulations, we find that $f_{\rm rel} \sim 1/15$.

For strong cooling regimes, \citet{tan21} calibrate the equations above to simulations of turbulent mixing layers and find that 
\begin{equation}
    v_{\rm in} \approx 10\mathrm{~km~s}^{-1}
        \left(\frac{v_{\rm turb}}{50\mathrm{~km~s}^{-1}}\right)^{3/4}
        \left(\frac{L_{\rm turb}}{100\mathrm{~pc}}\right)^{1/4}
        \left(\frac{t_{\rm cool,min}}{0.03\mathrm{~Myr}^{-1}}\right)^{-1/4}.
    \label{eq:vin2}
\end{equation}
The above scalings assume that we are in the subsonic to transonic regime where the relative velocity between the hot wind and the cloud does not exceed the hot gas sound speed. The turbulent mixing velocity saturates as clouds becomes supersonic \citep{yang23}, which would change the scalings above. We also assume that we are in the rapid cooling regime, where the cooling time is much shorter than the turbulent mixing time $L/v_{\rm turb}$, i.e., the outer scale eddy turn over time \citep[Damk{\"o}hler number $Da \equiv L/(v_{\rm turb}t_{\rm cool,mix}) > 1$;][]{tan21}.

In the following sections, we compare the model outlined above to results from the clouds in our simulations.

\subsubsection{Cloud Surface Area}
\begin{figure}
    \centering
    \includegraphics[width=\columnwidth]{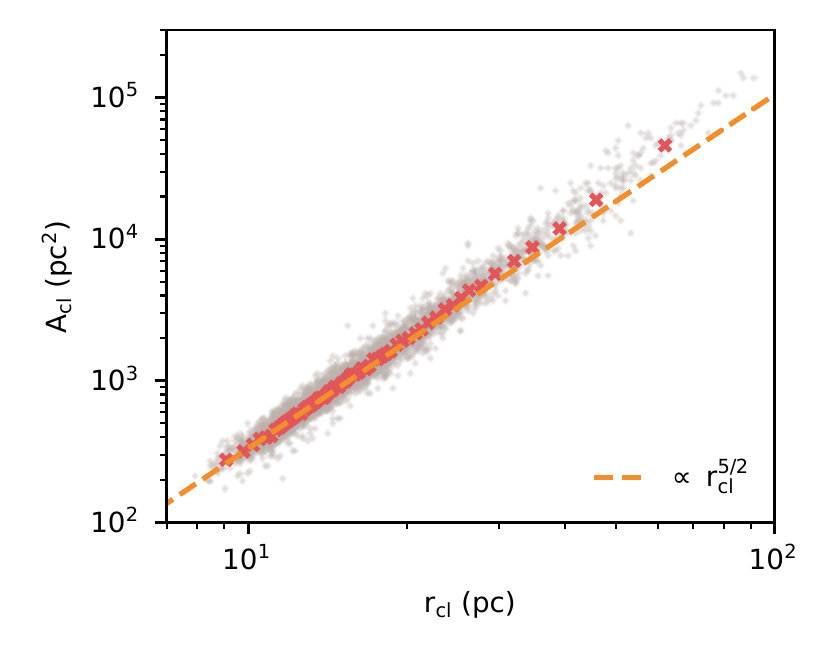}
    \caption{Surface area of clouds plotted against their size. Most clouds follow a $V^{5/6}$ or $r_{\rm cl}^{5/2}$ scaling denoted by the dashed orange line, although this relation appears to steepen for a handful of the largest clouds.}
    \label{fig:surfacetovolume}
\end{figure}
The isosurface method detailed previously allows us to compare the scaling of the cloud surface area as a function of cloud size to the model scaling in equation~\eqref{eq:surface_area}. We take the cloud surface area $A_{\rm cl}$ to be the area of the isosurface. We define this to be with respect to a measurement scale of 2 pc when constructing the isosurface, since the isosurface area varies with this choice of scale \citep{fielding20}. We do not expect this area to be the same as $A_{\rm cloud}$ from equation~\eqref{eq:surface_area}, since its magnitude in part depends on the choice of scale and temperature of the isosurface, but we do expect the scaling with size to remain the same. Figure~\ref{fig:surfacetovolume} shows a 2D histogram of $A_{\rm cl}$ and cloud radius $r_{\rm cl}$ (the radius of a sphere with equivalent volume), along with the $A_{\rm cl} \propto V^{5/6} \propto r_{\rm cl}^{5/2} $ scaling from equation~\eqref{eq:surface_area}. We find that most of the clouds agree with this scaling, except for the largest clouds which appear to have a steeper slope closer to $A_{\rm cl} \propto r_{\rm cl}^{11/4}$ or even $r_{\rm cl}^{3}$. We discuss this further below.

\subsubsection{Cloud Growth}
\begin{figure}
    \centering
    \includegraphics[width=\columnwidth]{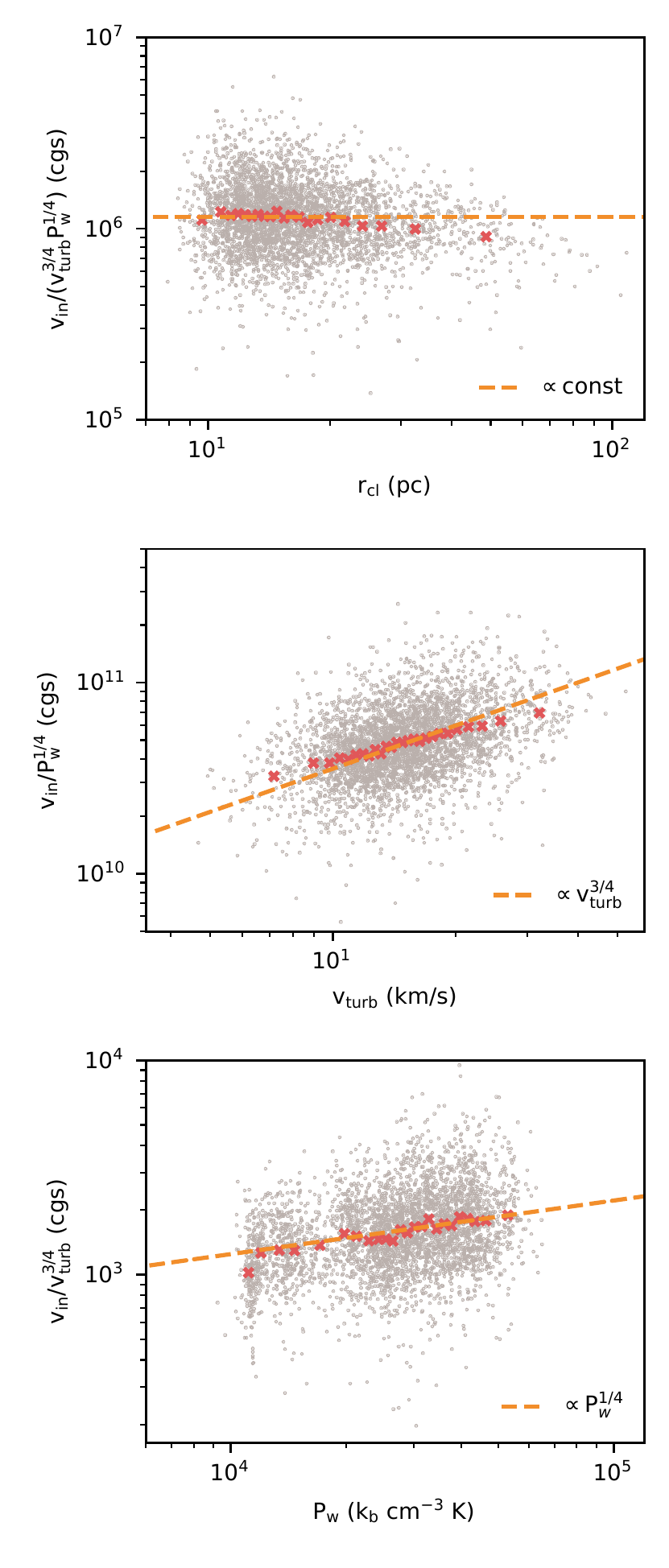}
    \caption{Scalings of the inflow velocity $v_{\rm in}$ with various parameters---the cloud size, turbulent velocity, and pressure (as a proxy for the cooling time). Observed scalings appear to match theoretical expectations.}
    \label{fig:cloud_scalings}
\end{figure}
We can also see how well the scalings in equations~\eqref{eq:vin1} and \eqref{eq:vin2} hold up. For reference, we expect the inflow velocity $v_{\rm in}$ to scale as 
\begin{equation}
    v_{\rm in} \propto v_{\rm turb}^{3/4} \left( \frac{L_{\rm turb}}{t_{\rm cool}} \right)^{1/4}
\end{equation}
in the strong cooling regime where the mixing time $L_{\rm turb}/v_{\rm turb}$ is longer than the cooling time of mixed gas $t_{\rm cool, mix}$ \citep{tan21}. For a typical cloud, $T_{\rm mix} \sim 10^{5.5}$\,K and $P_3 \sim 30$, giving $t_{\rm cool, mix} \sim 0.15$~Myr. In the mixing layers, $v_{\rm turb} \sim c_{\rm s,cool} \sim 15$\,km/s. This suggests that we are in the strong cooling regime if $L_{\rm turb} > 2$\,pc. We discuss this in further detail shortly, but $L_{\rm turb}$ is usually taken to be the outer scale of the turbulence, where mixing is the most effective, hence we expect our clouds to comfortably fall within this strong cooling regime. In fact, this criterion from \citet{tan21} is likely oversimplified since the cooling curve drops off sharply above $10^5$\,K---it is reasonable to expect that the requirement of $L_{\rm turb}$ might be even smaller.

To compare cloud properties with these scalings, we define 
\begin{equation}
v_{\rm in} \equiv \Dot{m}/(\rho_{\rm wind}A_{\rm cl})    
\end{equation}
following equation~\eqref{eq:mdot} but using the isosurface area and the associated mass flux through it. Figure~\ref{fig:cloud_scalings} demonstrates the $v_{\rm in}$ scaling with, $r_{\rm cl}$, $v_{\rm turb}$, and $P_{\rm w}$ respectively. Here we use $P_{\rm w}$ as proxy for the characteristic mixing layer cooling time $t_{\rm cool}$ since it scales inversely with $P_{\rm w}$, and we will discuss how $r_{\rm cl}$ relates to $L_{\rm turb}$. More specifically, each panel show the scaling of $v_{\rm in}$ divided by the expected scalings of the other variables so as to remove any cross-dependencies. Each grey point in the panels represents individual clouds, while red crosses mark a corresponding binned scatter plot to more easily visualize trends in the data \citep{binscatter}. Binned scatter plots partition the $x$ variable into fewer bins and display the average outcome for observations within that bin. Orange lines show the scalings we expect from our model. 

\begin{figure}
    \centering
    \includegraphics[width=\columnwidth]{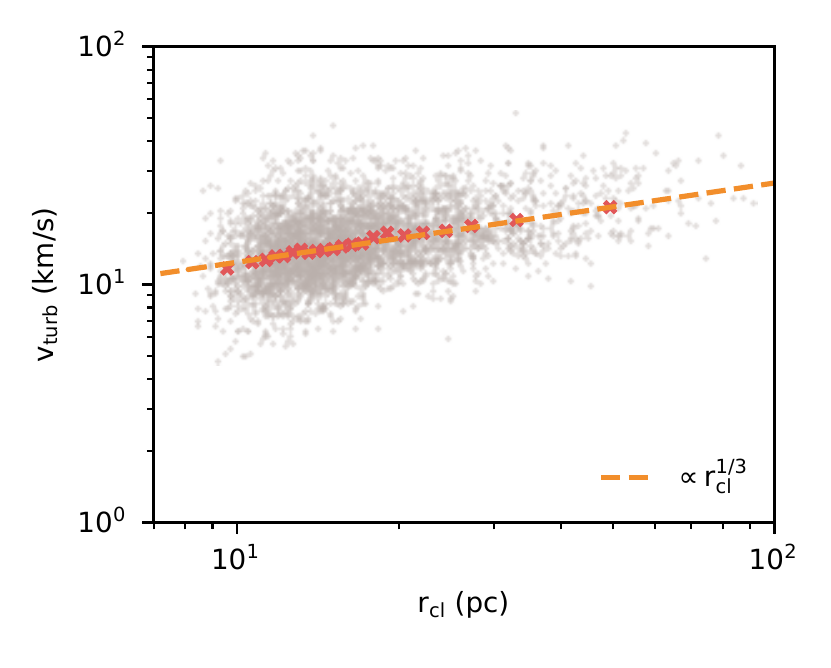}
    \caption{The turbulent velocity measured on the cloud surface scales with the cloud size in a manner that is consistent with subsonic Kolmogorov turbulence.}
    \label{fig:cloud_scalings_kol}
\end{figure}

The top panel of Figure~\ref{fig:cloud_scalings} shows $v_{\rm in}/(v_{\rm turb}^{3/4}P_w^{1/4})$ plotted against $r_{\rm cl}$. In previous work, it has been common to set the outer turbulent length scale $L_{\rm turb}$ to be either the box size (in the case of turbulent mixing layers) or the cloud size (for cloud simulations). In the latter, $L_{\rm turb}$ is in fact set to the {\it initial} cloud size and fixed for the rest of the simulation, even if the cloud grows in mass and volume by over an order of magnitude or fragments into many pieces. 
Here, we expect that if $L_{\rm turb} \sim r_{\rm cl}$, then we should see that $v_{\rm in}/(v_{\rm turb}^{3/4}P_w^{1/4}) \propto r_{\rm cl}^{1/4}$.
However we see in Figure~\ref{fig:cloud_scalings} that there is no scaling with $r_{\rm cl}$, save for the slight drop-off at large cloud sizes due to the steeper scaling of the surface area there. 

Despite this, we find that $v_{\rm in}$ still scales with cloud size, but this is because in our measurements, $v_{\rm in} \propto v_{\rm turb}^{3/4} \propto r_{\rm cl}^{1/4}$. Figure~\ref{fig:cloud_scalings_kol} shows this scaling, which is consistent with subsonic Kolmogorov turbulence. 
There are two possible interpretations here. The first is that our method of measuring $v_{\rm turb}$ is sensitive to the scale of the cloud, and that the scaling comes from measuring velocity differences at the surface on cloud scale separations. The second, which we lean towards, is that that the turbulent length scale here is not set independently by each individual cloud, but is instead a common scale over all the clouds. This is motivated by the observation that all the clouds are evolving in the same common turbulent wind, where the turbulent velocity field has been driven on the outer scale of the entire system of clouds ($\sim$ the disc scale height $H$). The idea that $L_{\rm turb}$ is set by the outer turbulent scale and not the size of each individual cloud is also consistent with previous work. For example, \citet{gronke22b} use the initial cloud size as $L_{\rm turb}$ for a turbulent box. They note that this choice becomes ambiguous in their simulations due to mass growth and cloud fragmentation/coagulation---in principle, at late times, $L_{\rm turb}$ approaches the driving scale of turbulence in their simulation $L_{\rm box}$. \citet{abruzzo22} directly measure velocity structure functions around wind tunnel clouds and also find that $L_{\rm turb} \sim$ the initial cloud size even as their clouds grow. 

The remaining panels in Figure~\ref{fig:cloud_scalings} hence assume that there is no explicit dependence of $v_{\rm in}$ on $r_{\rm cl}$ besides that from $v_{\rm turb}$. The middle panel shows the scaling of $v_{\rm in}/P_{\rm w}^{1/4}$, which is consistent with the expectation of scaling as $v_{\rm turb}^{3/4}$. Again, there is a drop-off at higher values of $v_{\rm turb}$, but this is similarly attributed to the higher surface area for the largest clouds that we divide by to get $v_{\rm in}$ (leading to a slope shallower by a power of $-1/4$ since $v_{\rm turb}$ scales with cloud size). The lowest panel of Figure~\ref{fig:cloud_scalings} shows the scaling of $v_{\rm in}/v_{\rm turb}^{3/4}$ with $P_{\rm w}$, where we see the 1/4 scaling with cooling time characteristic of turbulent mixing layer growth in the strongly cooling regime. 

\begin{figure}
    \centering
    \includegraphics[width=\columnwidth]{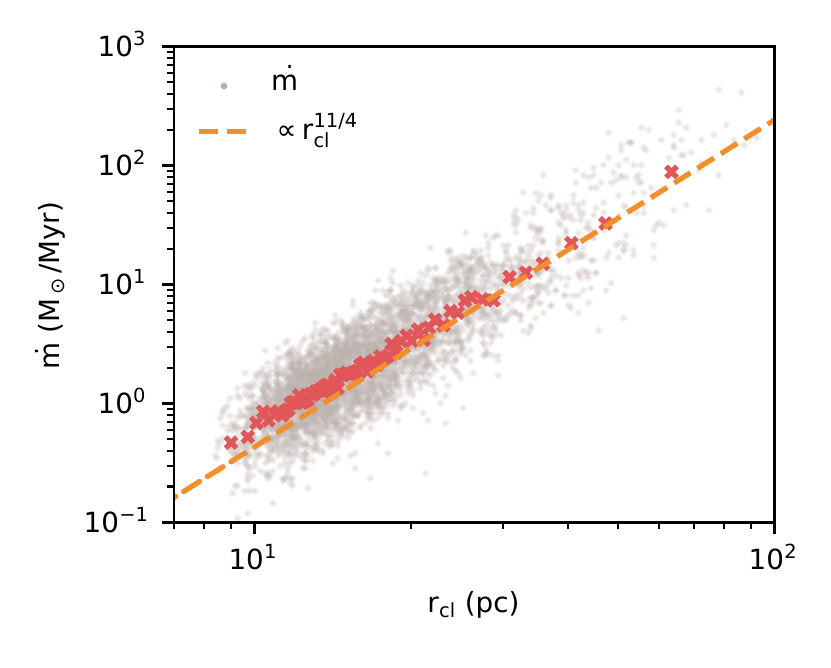}
    \caption{Mass growth rates compared to the predicted scaling from the analytic model (orange dashed line), show good agreement.}
    \label{fig:mdotscaling}
\end{figure}

Finally, combining all of the above, we can put in some numbers and directly compare to the model presented above for the mass growth rate of clouds (equations~\eqref{eq:mdot}, \eqref{eq:surface_area} and \eqref{eq:vin2}). The orange dashed line in Figure~\ref{fig:mdotscaling} shows our model with the typical values in our wind: $P_3 \sim 30$, $T \sim 10^7$\,K, $L_{\rm turb} \sim H \sim 66$\,pc, and $r_{\rm crit,shear} \sim 5$\,pc. For $v_{\rm turb}$, we use the orange dashed line from Figure~\ref{fig:cloud_scalings_kol}, so $v_{\rm turb} \sim 10 (r_{\rm cl}/10{\rm ~pc})^{1/3}$\,km/s. Note there are no scaling factors here. We find that the model agrees extremely well with the estimated mass growth rates of the clouds in the simulation. 
Here, it is important to highlight some complications in the simulation data here that are not immediately obvious. Our model predicts an overall scaling of $\Dot{m} \propto r_{\rm cl}^{11/4}$, arising from $A_{\rm cloud} \propto r_{\rm cl}^{5/2}$ and $v_{\rm in} \propto r_{\rm cl}^{1/4}$. However, for large clouds, we have seen that the entirety of this $r_{\rm cl}$ scaling is encapsulated by the surface area scaling. The model still working well here suggests that either the clouds are growing slower than expected given their surface area, or that our measurement of the surface area of these largest clouds overestimates the effective surface area. In addition, we have assumed that $\rho_{\rm wind}$ is independent of $r_{\rm cl}$, but we find that there is a selection effect stemming from clouds in higher pressure environment being found in patches of the wind which are being compressed. Since these patches are smaller in scale, corresponding clouds are likewise smaller. This artificially makes the slope of $\Dot{m}$ in Figure~\ref{fig:mdotscaling} visibly shallower at small cloud sizes.

\begin{figure}
    \centering
    \includegraphics[width=\columnwidth]{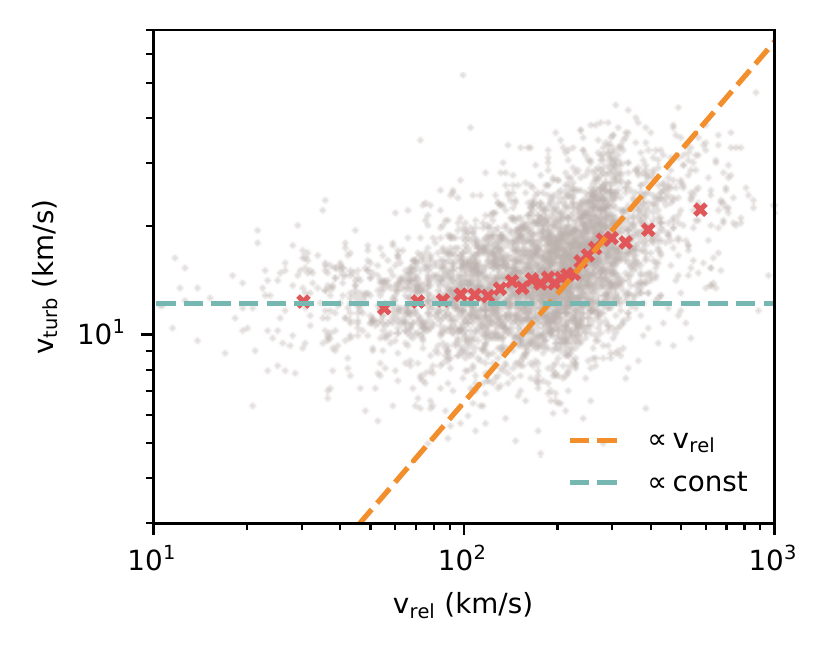}
    \caption{The turbulent velocity on the cloud surface scales with the relative velocity between the cloud and the wind when shear is high, but levels off as clouds get entrained to the sound speed of the cool gas.}
    \label{fig:turbvshear}
\end{figure}
\begin{figure*}
    \centering
    \includegraphics[width=\textwidth]{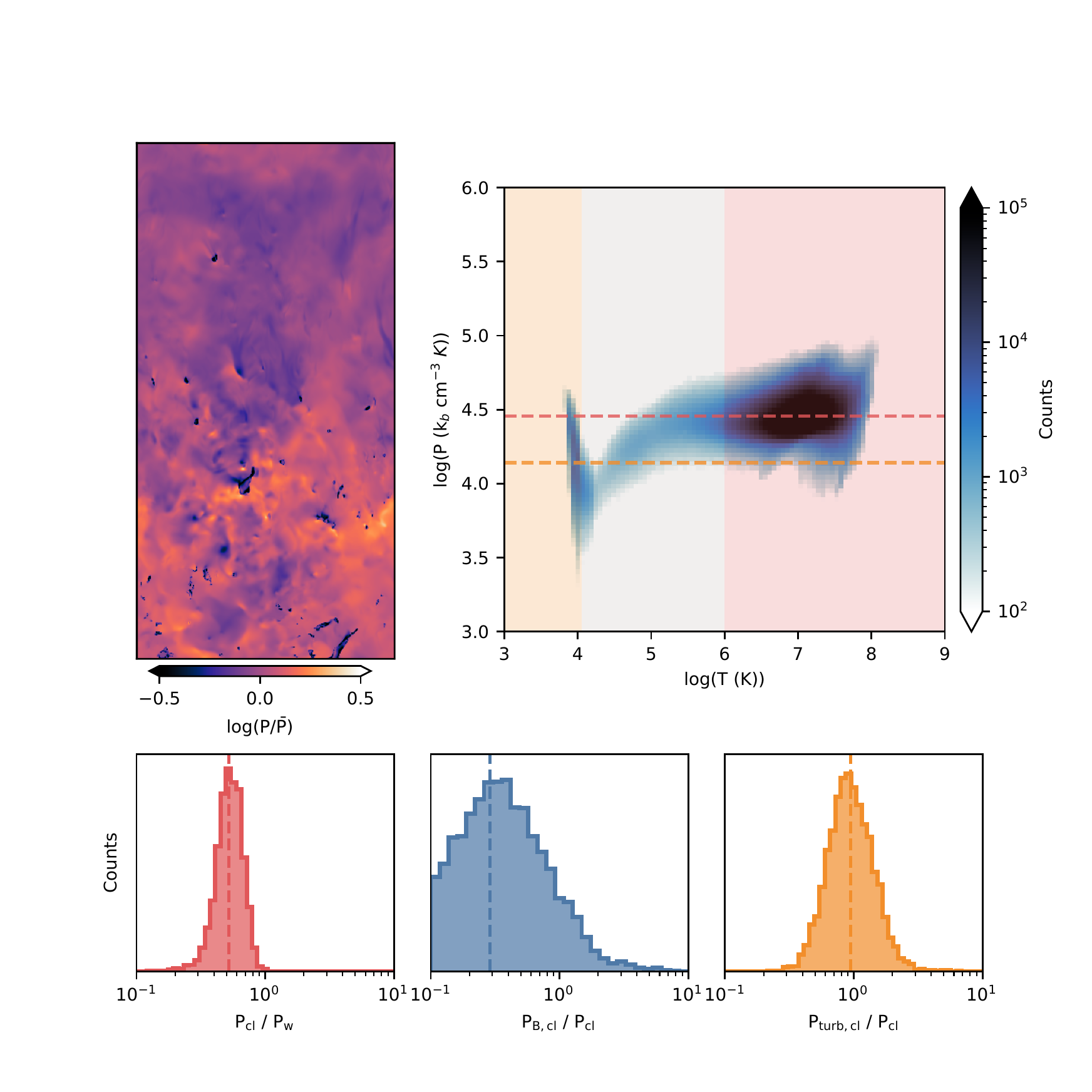}
    \caption{{\it Top Row:} On the left, we show a slice of thermal pressure normalized by the mean at 14 Myr. We see that that clouds appear generally under-pressured as compared to their surroundings. This is shown more clearly on the right, where the orange dashed line shows the mean thermal pressure of cool gas in the orange shaded region, and the red dashed line represents the mean thermal pressure in the hot gas marked by the red shaded region. {\it Bottom Row:} From left to right: histograms of the ratios of (i) the ratio of thermal pressures in the cloud and the surrounding wind, (ii) the ratio of magnetic pressure and thermal pressure in the cloud, and (iii) the ratio of turbulent pressure and thermal pressure in the cloud. These histograms are constructed from all the clouds in the catalog.}
    \label{fig:pressure_balance}
\end{figure*}

\subsubsection{Dependence of Turbulent Mixing on Wind Shear}
Lastly, we look at the scaling between the turbulent velocity $v_{\rm turb}$ in the mixing layer on the cloud surface with the relative shear velocity between the cloud and the surrounding wind $v_{\rm rel}$. As discussed above, plane parallel turbulent mixing layer simulations find that $v_{\rm turb} \propto v_{\rm rel}$, close to linear. However, cloud crushing simulations observe that turbulence persists even when the cloud is entrained. We find results consistent with both findings and well represented by equation~\eqref{eq:vturb}. This is shown in Figure~\ref{fig:turbvshear}. At high shear (high $v_{\rm vel}$), there is a strong scaling of $v_{\rm turb}$ with $v_{\rm rel}$ shown by the orange dashed line, consistent with $v_{\rm turb} \sim f_{\rm rel}v_{\rm rel}$ with $f_{\rm rel} \sim 1/15$. However, as $ v_{\rm rel}$ gets smaller, $v_{\rm turb}$ levels off at roughly the sound speed of the cool gas, shown by the teal dashed line. This supports the picture that the KH instability is but one way of generating turbulence in the mixing layer, and that turbulence can continue to persist and drive mixing even when clouds become entrained in the wind (see discussion above equation~\eqref{eq:vturb}).

\subsection{Non-thermal Pressure Support}
Figure~\ref{fig:pressure_balance} illustrates the lack of thermal pressure balance between the clouds and the background wind and the contributions of the two main sources of non-thermal pressure support in the clouds: magnetic pressure and turbulent pressure. The top left panel shows a slice of thermal pressure normalized by the mean over the entire slice. The clouds appear generally as under-pressured regions relative to their surroundings. The top right panel shows this more clearly, where the orange dashed line shows the mean pressure of cool gas in the orange shaded region, and the red dashed line represents the mean pressure in the hot gas marked by the red shaded region. The strong dip in the intermediate region where the cooling time is short indicates that the mixing layer itself is not well resolved \citep{fielding20}. From left to right, the bottom row of histograms show ratios of (i) thermal pressures in the cloud and the surrounding wind, (ii) magnetic pressure and thermal pressure in the cloud, and (iii) turbulent pressure and thermal pressure in the cloud. From the upper right and lower left panels, we see that just looking at thermal pressures, clouds are under-pressured relative to their environment by a factor of 2. Magnetic pressure support is not large enough to make up the different, with the magnetic plasma beta $\beta$ only being $\sim 4$. This is despite having $\beta \sim 1$ in the ISM prior to the onset of the SN driven wind. Instead, the missing pressure support is provided by turbulent pressure, which we define as $\rho_{\rm cl} \langle v^2 \rangle_{\rm cl}$. The turbulent pressure is roughly equal too the thermal pressure (which is equivalent to having a turbulent velocity equal to the sound speed of the cool gas). Hence, pressure support in the cloud is provided mostly in equal parts by thermal and turbulent pressure support, with a minor contribution from magnetic pressure. Taken together the \textit{total} pressure of the clouds is, on average, equal to the hot wind pressure.
\begin{figure}
    \centering
    \includegraphics[width=\columnwidth]{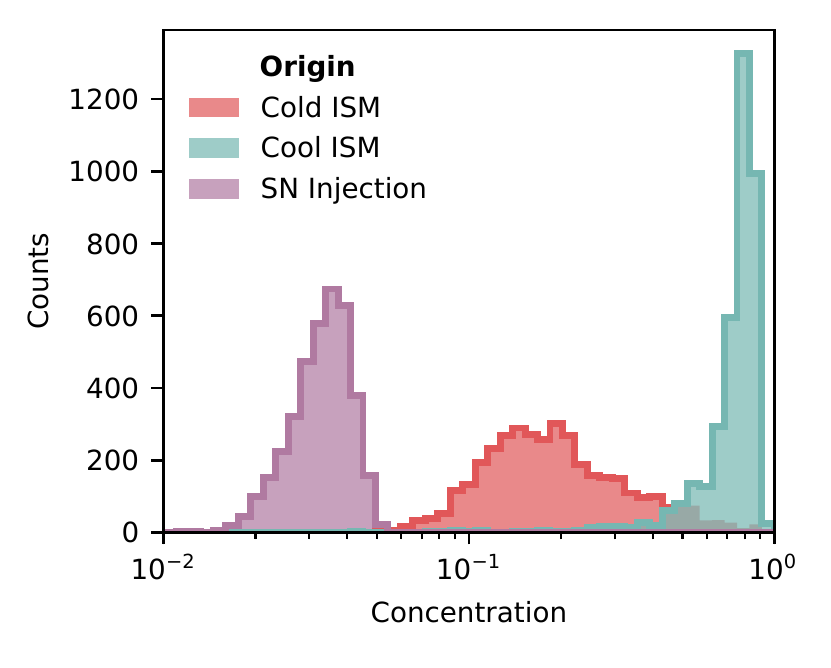}
    \caption{Histogram showing that most clouds (for the time window spanned, see Figure~\ref{fig:mass_flux_time_profile}) are comprised of gas that was originally part of the cool ISM pre-SN ($s_1$), with a small amount from the cold ISM ($s_0$). The contribution from SN injected mass ($s_4$) is negligible.}
    \label{fig:cloud_scalar}
\end{figure}

\subsection{Cloud Origins: Passive Scalars}
The passive scalars we have employed reveal some interesting points about the origins of these clouds. Figure~\ref{fig:cloud_scalar} shows histograms of concentrations of passive scalars in our clouds which track the amount of cloud material that was originally (right before the first SN) cold ISM gas ($s_0$), cool ISM gas ($s_1$), or mass injected by SN events ($s_4$). These histograms are constructed from all the clouds in the catalog which spans a range of times as denoted in Figure~\ref{fig:mass_flux_time_profile}. We do not see any time dependence of the histograms within this window. We find that most of the clouds are comprised of gas that was originally part of the cool ISM, supporting the fragmentation origin of the clouds, with only a small fraction originating from the cold ISM gas. It should also be noted that in both cases, the gas likely mixed and cooled to different temperatures, either in the wind or cloud, as evidenced by the other passive scalars ($s_2$ and $s_3$) in Section~\ref{sect:results_wind}. The close to unity values of $s_2$ and $s_3$ show that almost all gas had at different points in time been both cold and hot.

\subsection{Cloud-Wind Alignment}
\begin{figure}
    \centering
    \includegraphics{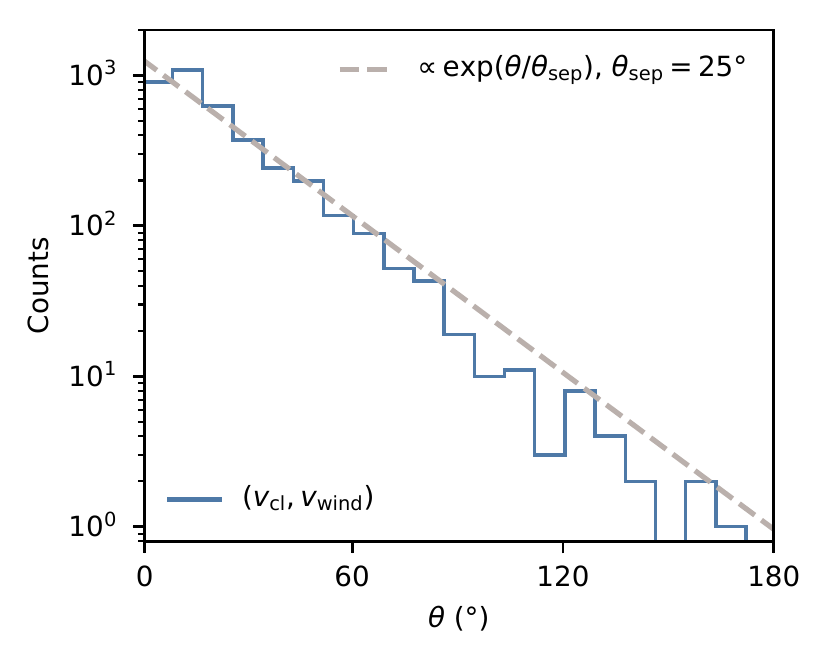}
    \caption{The angular separation between cloud and wind velocities is well described by an exponential distribution with a scale angle of $\sim 25\degree$.}
    \label{fig:cloud_alignment}
\end{figure}
Idealized setups involving clouds accelerated in a laminar wind or infalling under gravity often show an initial spherical cloud developing a pronounced head-tail structure, a morphological prediction that observations have widely been compared to \citep[e.g.,][]{putman12}. Clouds being accelerated in a highly turbulent wind exhibit far more complex structures, with the turbulence generating `tails' that result in a rich diversity in cloud morphologies, as can be seen in the lower panel of Figure~\ref{fig:clouds_3D}, where all clouds are pictured with the bulk flow in the upwards direction. Only a minority show similarities to classic head tail structures. In general, the head-tail paradigm serves only as a first order indicator of the direction of bulk velocities. We can quantify the alignment between clouds and a turbulent wind by looking at the angular separation between cloud and local wind velocities. This is shown in Figure~\ref{fig:cloud_alignment}, where we find that this angular separation follows an approximately exponential distribution with a scale angle of $\sim 25\degree$. While most clouds are well-aligned with the wind, a handful are significantly misaligned. The existence of such misaligned clouds is consistent with observations of clouds in the Fermi Bubbles, but to our knowledge there have not been any detailed studies of such clouds \citep[in contrast with aligned clouds, e.g.,][]{teodoro20,noon23}.

\dobib

\newpage

\section{Discussion} \label{sect:discussion}

\subsection{Connection to Small Scale Simulations}
It is exciting that the physical insights and analytic scalings derived from small scale idealized simulations translate well to larger scales. This was demonstrated in \citet{tan23} for individual infalling clouds, and now in this work for a population of clouds in turbulent magnetized outflows. In particular, these models build upon the work on the physics of interfacial turbulent radiative mixing layers \citep[e.g.,][]{ji19,fielding20,tan21}, minimizing the need for empirical tuning of parameterized models. This success is promising for future use of physically informed subgrid models in larger scale simulations. Nonetheless, some nebulous points remain---we briefly discuss some of these issues, connections to other work, and limitations/caveats in this section.

What is the physical process that determines the initial cloud distribution? Inverse power law distributions, where the probability of a quantity taking on some value varies inversely with the power of that value, are widely observed in both the physical and social sciences, with a large existing body of empirical evidence supporting their existence \citep{newman05}. They are commonly referred to as Zipf's law in the discrete case, or Pareto distributions in the continuous case. Their theoretical origins are, however, much less certain. In particular, it is difficult to explain their seeming universality. A $\sim$\,$-2$ exponent, which indicates that there is a constant contribution from each logarithmic bin, is ubiquitous, even within astronomy \citep{guszejnov18}. Take, for example, the stellar initial mass function, which has a roughly $dN(m)/dm \propto m^{-2.35}$ scaling at higher masses \citep{krumholz14}. This has been attributed to turbulent fragmentation \citep{padoan02,hopkins13}, where initially turbulent gas fragments into clumps. In this case, fragmentation is driven primarily by gravitational collapse. \citet{gronke22} find in simulations of turbulent multiphase boxes that the cool gas breaks up into droplets which follow a similar distribution. This setup is akin to a population of clouds fully entrained in a turbulent background wind. They argue that this power law, and in particular the $-2$ exponent, arises from competition between fragmentation and growth, where growth is provided from both a multiplicative source (cooling induced growth) as well as an additive source (coagulation of multiple smaller clumps with a larger clump).\footnote[1]{\citet{gronke22b} explore the effects of coagulation further and identify a critical Mach number below which coagulation dominates. This critical Mach number is small in our simulations, suggesting that coagulation should not dominate.} Fragmentation, on the other hand is driven solely by the turbulent field. Likewise, \citet{fielding23} find the same mass distribution of cool clumps in turbulent magnetized boxes where the multiphase medium is formed via thermal instability. While our setup differs from these more idealized ones, many features are similar. Instead of a single large cloud in a turbulent velocity field, we have an ISM that is broken up by an expanding hot superbubble. More work is needed to understand this process of turbulent driven fragmentation in detail.

In systems where the density contrast $\chi$ is high ($\geq 10^3$), there is still remaining uncertainty around the cloud survival criterion. In this work, we have presented a criterion based on the idea that the characteristic cooling time of the mixing layers must be significantly shorter than the shear timescale that is motivated by and consistent with results of wind tunnel simulations at high $\chi$ \citep{abruzzo22}. This criterion predicts clouds must be larger than the \citet{gronke22} criteria by a factor of $\sqrt{\chi}$ in order to grow, and is consistent with phenomenological criteria \citep[e.g.,][]{sparre20, li20}. However, there remains further investigation required to pin down the relevant physical processes that determine cloud survival in high $\chi$ environments like multiphase outflows. 

We do not investigate cloud mass loss rates in this work. The methods we use to estimate cloud growth rates break down for clouds that are getting destroyed (see Appendix~\ref{sect:apdx_e}). Incorporating cloud destruction rates into this model would improve the coupling between the phases, and is especially important in parameter regimes where clouds are not expected to grow. Similarly, this also requires more work on the small scale simulation side. 

A final point on the topic of cloud survival, while not directly relevant to this work but important to keep in mind, is that \citet{farber22} show that the story is more complicated when $T_{\rm cloud} < 10^4$\,K. This is mainly due to the shape of the cooling curve, which \citet{abruzzo22a} also showed plays an important role. However, the survival criteria they find is equivalent to the \citet{gronke18} criterion when $T_{\rm cloud} \gtrsim 10^4$\,K as it is in our simulations.

What about magnetic fields? This perennial question has been the subject of many wind tunnel simulations of cool clouds \citep[e.g., ][]{gregori99,mccourt15,gronnow18,hidalgo23}. Despite that, their effect remains unclear. For example, \citet{gregori99} find that destruction can be enhanced by more rapid acceleration while \citet{mccourt15} find that they aid survival via magnetic draping inhibiting shear instabilities \citep[see also ][]{banda18,ji19,gronnow22}. \citet{sparre20} find some enhancement in cloud survival for $\beta \sim 10$, while \citet{li20} find no effect for $\beta \sim 10^6$. \citet{hidalgo23} find that for small $\beta$ ($\sim 1$), much smaller clouds are able to survive. However, $\beta$ is large in our winds and only significant in the disc---we hence conclude that it is unlikely that magnetic fields play a significant role in influencing cloud acceleration or survival in multiphase outflows. Magnetic fields don't seem to inhibit mass growth rates either \citep{gronke20,hidalgo23}, although morphologically clouds are reported to have very different filamentary structures as compared to their hydrodynamical counterparts \citep[e.g., ][]{tonnesen14,gronke20,jung22}. Our model does not include magnetic fields, supporting their lack of impact on growth rates. Understanding why this is so is an interesting avenue for future work. While we do not compare to a hydrodynamical run without magnetic fields, we do not observe clear filamentary structures---which we attribute to a combination of a weakly magnetized wind and turbulence. In general, turbulence in the wind is the main generator of complex morphologies seen in the clouds.

\subsection{Connection to Galaxy/Cosmological Scale Simulations}
Having demonstrated that much of the insight garnered from small-scale simulations translates to more realistic larger-scale systems we can now address how these processes fit into the overall landscape of galaxy formation and global-scale simulations. The impact of capturing these multi-scale multiphase effects has been seen in isolated galactic scale simulations with self-consistently generated multiphase winds, which find that properties of the hot wind including temperature, density, and pressure fall off slower than expected with distance, and also travel slower than single-phase adiabatic winds \citep{fielding17,schneider20}. These effects are consistent with expectations of mass, momentum and energy exchange between cool clouds and the surrounding wind \citep{fielding22}. The total cool gas mass flux in \citet{schneider20} decreases with distance, suggesting that the cool gas is being destroyed and mass loading the hot phase, even for largest clouds. This may be understood by applying the $r_{\rm crit,shear}$ criterion for cloud growth as opposed to $r_{\rm crit,cc}$. In addition, they find that cool clouds are under-pressurized by up to a factor of 10. While we find that turbulent pressure support is significant, this cannot fully account for the large factor. Given that we also find that regions of phase space where the cooling time is shortest are the most underpressurized, this can likely be attributed to lower resolutions \citep{fielding20}. Additional detailed studies of multiphase winds at even higher resolutions (or with conditions in which it is easier to resolve $r_{\rm crit,shear}$) will be helpful in shedding further light on how the multiphase interactions shape the overall evolution of the winds.

Recent large cosmological simulations also exhibit galactic winds with self-consistently generated cool phases. While most have insufficient resolution to resolve any clouds, certain zoom-in and high resolution simulations can do so, albeit marginally. For example, multiphase galactic winds are seen in TNG50 \citep{nelson19} and FIRE-2 \citep{pandya21}. The latter characterized the multiphase nature of the outflow by analyzing the contribution to fluxes from each phase and found results broadly consistent with our findings and past tall-box ISM patch simulations \citep[e.g.,][]{fielding18,kim20a}. 

Correctly capturing the multiphase nature of galactic winds is not only essentially for accurately modeling the winds themselves, but as recent work has shown, may also be essential for capturing the correct regulation of star formation and thus galaxy evolution. When winds are able to separate into multiple phases, the hot phase, which has very high specific energy, heats and stirs the CGM, which prevents new star forming material from entering the galaxy \citep{fielding17a}, and the cold phase ejects material directly out of the ISM. Standard galactic feedback models cannot capture the high specific energy phase in particular and as a result may be missing important regulation mechanisms, as was recently shown using regulator and semi-analytic models \citep{pandya22,carr22}, as well as isolated galaxy simulations \citep{smith23}. It is uncertain if cosmological simulations will ever achieve the resolutions necessary to properly capture the physical scales relevant to multiphase wind launching and dynamics. In which case, simulations and models such as those presented in this work are important in being able to bridge the gap towards achieving the capability to include realistic multiphase outflows using subgrid techniques.

\subsection{Implications for Observations}
A key missing piece in our understanding of galaxy evolution is the amount of mass and energy carried by winds from the ISM into the CGM and beyond. Most observations of galactic winds come from probes that are sensitive to cool gas with $T \sim 10^4$\,K \citep[e.g.,][]{heckman90,martin99,rubin14}, although in some rare nearby cases the hot phase is observable in X-ray \citep[e.g.,][]{Lopez:2020,Lopez:2023}. Translating from observed quantities to an inferred mass flux is a difficult and uncertain process, however in almost all cases the inferred mass flux is up to orders of magnitude smaller than what standard theoretical models predict. For example, \citet{McQuinn:2019} analyzed a sample of 12 nearby starburst dwarfs from the STARBIRDS survey and found mass loading factors of $0.2 - 7$ compared to the $1 - 100$ predicted from simulations. \citet{Concas:2022} studied ionized gas outflows in 141 main-sequence star-forming galaxies at $1.2 < z < 2.6$ from the KLEVER survey, and found that the ionized gas mass only accounts for less than 2 percent of predicted values. Resolving this profound conflict between theory and observations requires a better treatment and understanding of multiphase outflows on both sides. Here, we have shown that the nature of feedback is likely to be dramatically different from standard single-phase galactic feedback models, which in part helps to relieve this tension. We can, however, also use this multiphase wind picture to help refine our understanding and modeling of observations.

In order to make the most of galactic wind observations, particularly new and planned spatially resolved emission observations \citep[e.g.,][]{ReichardtChu:2022}, new observational modeling paradigms that take the multiphase nature of galactic winds into account are required. In particular, our finding is that the vast majority of the readily observable cool gas is in the form of clouds with a fairly well-understood size distribution and a relatively small volume-filling fraction. Furthermore, because we have shown the properties of these cool clouds are closely coupled to the energy containing hot phase, with the two phases shaping each other's properties, future multiphase models may be able to constrain not only the mass flux (cool phase) but also the energy flux (hot phase). 

Our findings also provide insight into the nature of the CGM. Observations in the CGM suggest that the cool phase is not in pressure equilibrium with the hot phase \citep{werk14}. One way of accounting for this discrepancy is the addition of non-thermal components. As previously discussed, our results suggest that turbulence within clouds is a significant contributor to pressure support, while magnetic fields are a minor actor, although other candidates such as cosmic rays may also provide further cold phase pressure support \citep{butsky20}. 

\subsection{Further Considerations}
Despite modelling and simulating a more realistic turbulent magnetized multiphase system, we ultimately still make simplifying assumptions. The following are some such limitations and caveats, many of which are each interesting enough in their own right to warrant exploration in future work.

\subsubsection{Additional Physics}
The most obvious and direct of these is the inclusion of additional underlying physics which were not incorporated into this work, but could possibly affect such systems by significantly modifying the dynamics or thermodynamics of the outflow. Detailed investigations and treatments of these processes are thus needed to accurately access their impact and importance, in order that they can be properly accounted for when modelling the behavior of multiphase winds. One such source of uncertainty that could potentially have a large impact is cosmic rays, which have been shown to be able to modify the properties of multiphase winds \citep{huang22,armillotta22}. They can provide an additional source of non-thermal pressure support and affect cooling \citep{butsky20}, or accelerate clouds directly \citep{wiener19}. Explicit viscosity \citep[e.g., ][]{li20, jennings21} and thermal conduction \citep[e.g., ][]{bruggen16,li20} have also not been included here (although we expect that in turbulent mixing layers, mixing is generally dominated by turbulent diffusion \citep{tan21}, explicit conduction can affect observables \citep{tan21b}). The inclusion of non-equilibrium cooling as well as more sophisticated non-equilibrium chemical evolution models could potentially be important, especially for predictions of observables such as ion column densities \citep{Sarkar:2022}. We also assume solar metallicity and abundances everywhere. This is clearly an oversimplification. When metallicities of clouds and their environment differ, the mixing can have significant implications for observables \citep{gritton14,heitsch22}. Dust survival or depletion in these systems is also important to understand, in particular at lower temperatures \citep{farber22}. Closely tied to this is self-shielding. While we assume that the whole box is optically thin, leading to clouds with $T \sim 10^4$\,K, more massive clouds with $N_{\rm HI} > 10^{17}$~cm$^{-2}$ are likely able to self-shield and hence posses cold cores \citep{farber22}. In fact, molecular gas is observed both in the Milky Way \citep{teodoro20,noon23} and other systems such as M82 \citep{walter02}.

\subsubsection{Geometry}
Besides being limited in terms of spatial extent, the nature of our setup lacks the correct geometry to track the outflowing wind further into the halo. While wind properties are often converged with resolution, local box simulations like this are limited by the Cartesian geometry of the setup which restrict the reliability of quantitative predictions of wind properties, which are often not converged with respect to box height instead. In particular, such a geometry which lacks the inverse squared distance scaling does not allow the adiabatic expansion and corresponding subsonic to supersonic transition of steady state winds that are a hallmark of predictions from analytic galactic wind models like \citet{chevalier85} \citep{martizzi16, fielding17}. This in part motivated our focus on a cloud-centric analysis, rather than an extended look at associated wind properties such as mass and energy loading that are typically done in such setups.

\subsection{Looking Forward}
A natural next step is to apply and test cloud models further downstream in the wind. Analytic models of cloud growth/destruction as they are carried out further in the wind such as the framework outlined in \citet{fielding22} can easily be extended and applied to populations of clouds in the manner we have done here. While we have determined a physically motivated evolution and initial distribution, validating the time evolution of such a distribution coupled  to an expanding wind is the natural next step (Anthony Chow et al., in prep). Additionally, the scope of the simulations can be expanded by exploring the effects of the different mechanisms listed above (such as cosmic rays) on the outflow. This will provide us with a more comprehensive understanding of the physical processes that drive multiphase galactic winds. Finally, understanding the impact of varying the properties of the galactic environment, such as gas surface density and metallicity, on the outflow will allow us to test the robustness of our results and determine the degree to which they are dependent on specific initial assumptions. In the long term, such models can inform subgrid approaches to modelling unresolved multiphase outflows in large scale galactic and cosmological simulations, allowing us to study the macro impacts on galaxy formation and evolution.

\dobib

\newpage

\section{Conclusions}  \label{sect:conclusions}
At the frontier of the field of galaxy formation and evolution lies the challenge to understand the multiphase nature of the environment and its implications. Galactic winds are a key component of these complex ecosystems and can be driven by stellar feedback channels. Observations reveal them to be common and multiphase in structure. Theoretical modeling and simulations of galactic winds have become increasingly sophisticated, aiming to reproduce and understand these outflows. In this work, we have built on turbulent radiative mixing layer theory and applied this to understanding and modelling the formation of multiphase outflows. We have run 3D MHD tall box patch simulations with a clustered SN driven wind outflow, with a focus on analyzing and modelling the properties of the resulting seeded cloud population. The main findings are summarized as follows:

\begin{itemize}
  \setlength{\itemsep}{5pt}
  \item {\it Seeded by Fragmentation}: During the breakout phase, the hot expanding bubble propagates through a multiphase ISM. The multiphase nature of the ISM tends to lead to asymmetric breakouts, and causes the outflows to fluctuate in power and direction. More importantly, it also leads to the fragmentation of the ISM during the breakout, which seeds the resulting hot outflowing wind with a population of cool clouds. We leave a detailed study of this process to future work. Consistent with this formation history, we find that clouds are mostly comprised of gas that was originally part of the cool ($T\sim10^4$\,K) ISM, rather than from cold ($T\lesssim10^2$\,K) or hot material ($T\gtrsim10^6$\,K).

  \item {\it Turbulent Clouds \& Winds}: The uneven breakout and the presence of these clouds induces large scale turbulence in the wind. The clouds gradually get entrained via mixing induced accretion of momentum. The turbulent Mach number in the hot phase of the wind is $\sim 0.3$, and magnetic pressure is extremely weak, meaning thermal pressure is dominant. However, clouds have turbulent Mach numbers $\sim 1$ internally and at their surfaces where mixing occurs. 

  \item {\it Clouds Exhibit Complex Morphologies}: This turbulent environment naturally leads to complex cloud morphologies that do not always conform to the head-tail description. In some cases, clouds can appear to be extremely misaligned with the bulk flow.
  
  \item {\it Cloud Size Distribution}: Cloud sizes are well described by a power law distribution of $dN/dm \propto -2$. While this has been observed in previous works, we find that this scaling seems to hold even with the inclusion of magnetic fields in the ISM. The lower and upper scale cutoffs are consistent with estimates of the cloud survival radius and the scale height of the disc, respectively. We find that this scaling appears early on during the SN breakout stage---consistent with the source of clouds being the process of fragmentation of the ISM. 

  \item {\it Cloud Survival}: The critical radius below which clouds can survive in a turbulent wind are consistent with $r_{\rm crit,shear}$ which is given by equation~\eqref{eq:rcritshear} as proposed in \citet{abruzzo23}. This criterion is motivated by a combination of turbulent box simulations from \citet{gronke22} and cloud simulations with high $\chi$ in \citet{abruzzo22}, who proposed that the survival criteria is set by $t_{\rm cool,mix} < t_{\rm shear}$.

  \item{\it Cloud Growth}: By combining analytic models for the surface area to volume relationship of clouds ($A_{\rm cl} \propto V^{5/6}$) and mixing layer theory for the mass inflow velocity $v_{\rm in}$, we can model the growth rate of clouds. We find that the model predictions are a good match to what we observe in the simulations, including predicted scalings and estimated $\Dot{m}$.

  \item{\it Role of Wind Shear}: At high relative shear velocities between clouds and the surrounding wind, there is a strong relationship between the shear and the turbulent velocities which drive mixing and hence growth. However, as clouds get entrained, the turbulence becomes independent of the shear and is roughly the sound speed of the cool gas. The KHI is thus not the sole driver of turbulent mixing. Post entrainment, turbulence must instead be sustained by cooling or cooling-induced pulsations.
  
  \item {\it Non-Thermal Pressure Support}: Because turbulent velocities within these clouds are high (roughly the sound speed within the cloud), turbulent pressure provides as much support as thermal pressure. Surprisingly, even though the initial ISM had $\beta \sim 1$, clouds are much more weakly magnetized, likely due to significant mixing with the hot high beta wind. These sources of non-thermal pressure support bring the clouds into \emph{total} pressure equilibrium with their surroundings, despite having a factor of 2 lower thermal pressure.

\end{itemize}

In summary, we have shown that many of the physical insights and analytic scalings from idealized small scale simulations of radiative turbulent mixing layers and singular clouds in wind tunnels translate well to larger scale, more realistic turbulent magnetized winds. The multiphase component of these winds (a population of cool embedded clouds) can hence be well modelled, allowing for physics informed subgrid prescriptions which account for unresolved coupling between the various phases to be utilized in galactic and cosmological simulations where resolution limits are prohibitive. While refinements are required (e.g., the survival of molecular gas), moving forward, proper treatment of the small scales in this manner promises to pave the way towards tackling burning questions that remain regarding the role of multiphase feedback in galaxy formation and evolution.

\section*{Acknowledgements}
We thank Siang Peng Oh, Greg Bryan, Yan-Fei Jiang, and Matthew Abruzzo for helpful discussions, as well as the anonymous referee for an extremely helpful and constructive report which greatly improved the structure of this work. We have made use of the yt astrophysics analysis software suite \citep{yt}, matplotlib \citep{Hunter2007}, numpy \citep{van2011numpy}, scipy \citep{2020SciPy-NMeth}, CMasher \citep{cmasher} and Blender \citep{blender} whose communities we thank for continued development and support. 
BT acknowledges support from NASA grant 19-ATP19-0205 and NSF grant AST-1911198. BT was supported partly by the Simons Foundation through the Flatiron Institute’s Predoctoral Research Fellowship. DBF is supported by the Simons Foundation through the Flatiron Institute. 

\section*{Data Availability}
The data underlying this article will be shared on reasonable request
to the corresponding author.

\dobib

\newpage
\bibliography{references}
\clearpage

\appendix

\section{Hydrostatic Equilibrium Test} \label{sect:apdx_c}
\begin{figure}
    \centering
    \includegraphics[width=\columnwidth]{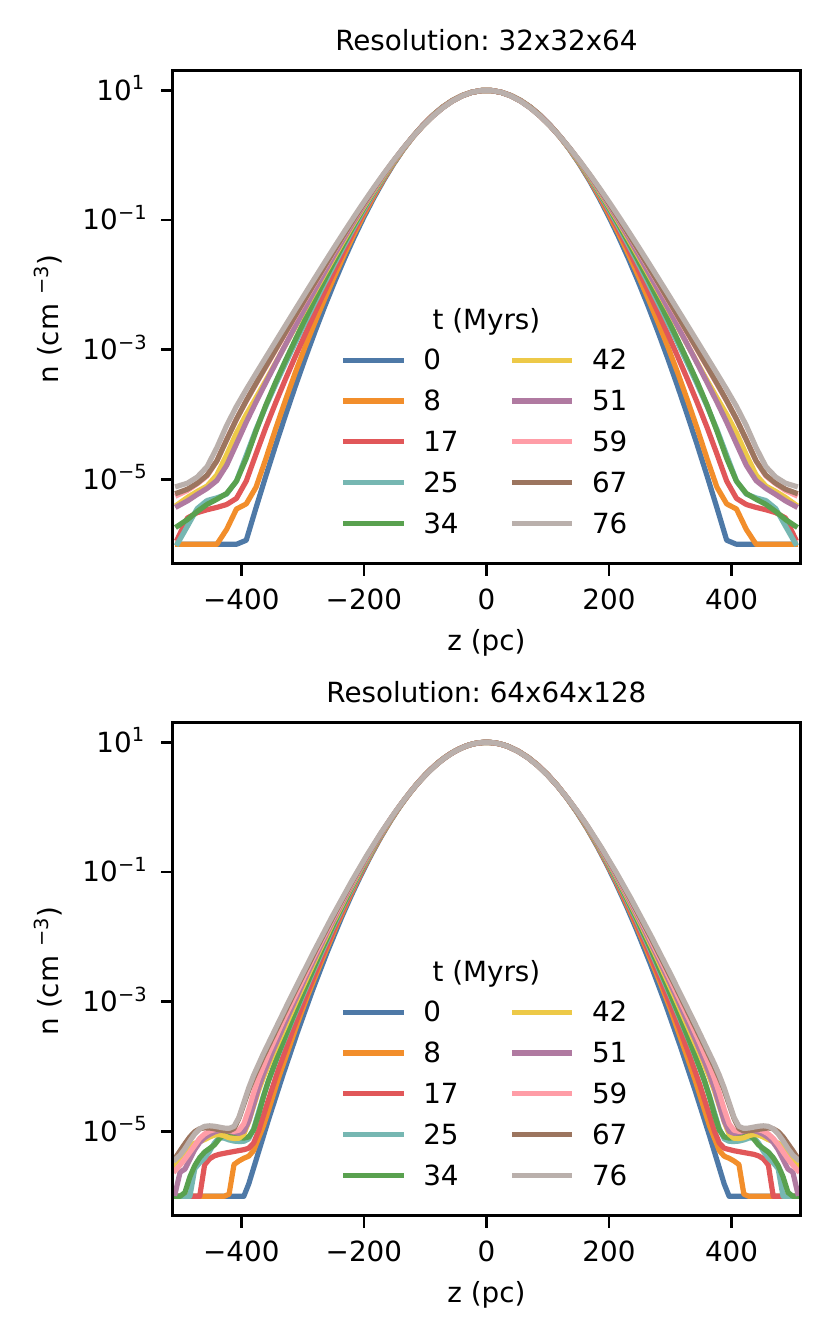}
    \caption{A test of the initial hydrostatic profile for two low resolution boxes. We initialize 3D boxes with the analytical profiles at $T=10^4$\,K and see how well hydrostatic equilibrium is maintained. The error is smaller at higher resolutions and is negligible. There is a density floor at $10^{-6}$\,cm$^{-3}$.}
    \label{fig:test_HE}
\end{figure}

Figure~\ref{fig:test_HE} shows a simple test setup in hydrostatic equilibrium for two boxes at different resolutions (low; below fiducial). The box is at $T=10^4$\,K throughout with a mid-plane density of $10$\,cm$^{-3}$. There is a density floor set at $10^{-6}$\,cm$^{-3}$. In both cases, the error over time gets smaller, and is consistently smaller in the higher resolution box. These tests provide simple sanity checks of the initial conditions. In practice, outflowing gas quickly becomes the dominant effect.

\section{Constrained Turbulence Test} \label{sect:apdx_d}
\begin{figure}
    \centering
    \includegraphics[width=\columnwidth]{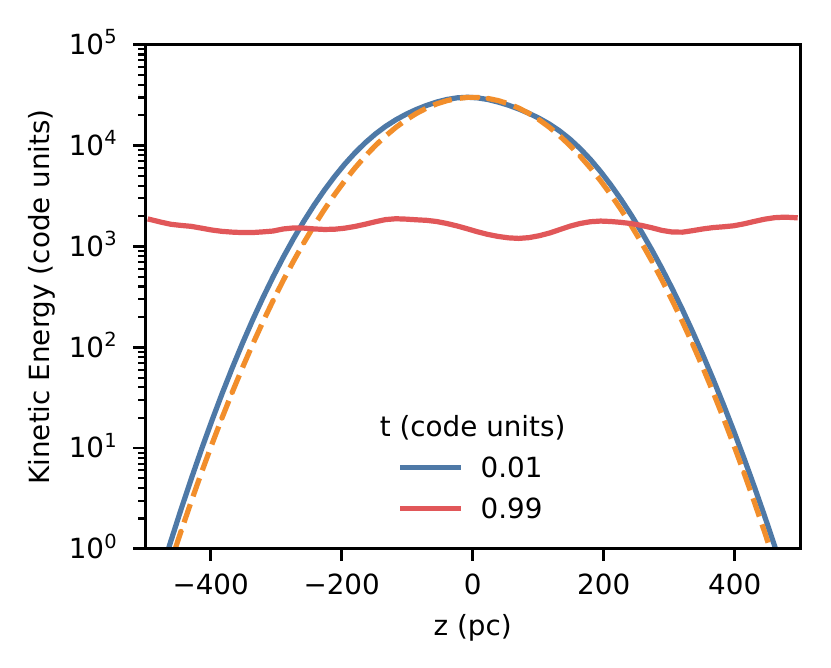}
    \caption{A test of constrained turbulence. Turbulence is injected at the start of the simulation into a uniform box with periodic boundary conditions. The orange dashed line marks the expected kinetic energy as a function of height in the box at the time of injection, while the blue and red lines show early and late time values from the simulation respectively.}
    \label{fig:test_turb_constraint}
\end{figure}

For our implementation of turbulence, we include a turbulent scale height constrain $z_{\rm turb}$. This is done by applying a Gaussian weight factor $\exp{(-z/z_{\rm turb})^2}$ to the velocity perturbations in the turbulent driver. We test our implementation by injecting decaying turbulence with some scale height into a uniform periodic box, as shown in Fig.~\ref{fig:test_turb_constraint}. The orange dashed line marks the expected kinetic energy as a function of height in the box at the time of injection centered at zero (in practice, due to the zeroing of the total momentum, this has some shift), while the blue and red lines show early and late time values from the simulation. At late times, there is some decay in kinetic energy since we do not continuously drive. The turbulence also becomes uniform throughout the box as expected.

\section{Extended Townsend Algorithm} \label{sect:apdx_b}
\subsection{Exact Integration Scheme}
We first review the original Townsend algorithm \citep{townsend}. This is an algorithm for computing the change in temperature due to isochoric radiative cooling over some given time interval. It is based on an exact solution by noting that if one knows the cooling function, one can in principle simply integrate this function over the given time interval. The algorithm is as follows. We want to solve the following cooling equation:
\begin{align}
    \frac{dT}{dt} = - \frac{(\gamma-1)\rho\mu}{k_{\rm B}\mu_{\rm e}\mu_{\rm H}}\Lambda(T).
\end{align}
We can define the dimensionless Temporal Evolution Function (TEF) as follows:
\begin{align}
    Y(T) \equiv \frac{\Lambda(T_{\rm ref})}{T_{\rm ref}} 
    \int_{T}^{T_{\rm ref}} \frac{dT}{\Lambda(T)}.
\end{align}
In principle, this only requires that $1/\Lambda(T)$ be analytically integrable. This is basically a normalized measure of the time taken to cool/heat from $T_{\rm ref}$ to $T$. We can then integrate the cooling equation over a timestep such that the integrated cooling function becomes:
\begin{align}
    \frac{T_{\rm ref}}{T^{\rm n}}\frac{\Lambda(T_{\rm n})}{\Lambda(T^{\rm ref})}
    \left[ Y(T^{\rm n} - Y(T^{\rm n+1}) \right] = -\frac{\Delta t}{t_{\rm cool}},
\end{align}
where $t_{\rm cool}$ is the single point cooling time defined as
\begin{align}
    t_{\rm cool} = \frac{k_{\rm B} \mu_{\rm e}\mu_{\rm H} T}{(\gamma-1)\rho \mu \Lambda(T)},
\end{align}
and hence we can update the temperature as
\begin{align}
    T^{\rm n+1} = Y^{-1} \left[Y(T^{\rm n}) + 
    \frac{T_{\rm n}}{T^{\rm ref}}\frac{\Lambda(T_{\rm ref})}{\Lambda(T^{\rm n})}
    \frac{\Delta t}{t_{\rm cool}}
    \right].
\end{align}
In practice, this is done in 3 steps - computing the TEF, evolving it over the required timestep, and then transforming back to the new updated temperature. In the next section, we compute the TEF for the two most useful cases, piecewise power laws and piecewise linear functions.

\subsection{Temporal Evolution Functions}
\subsubsection{Piecewise Power Laws}
Physically realistic cooling function often come in the form of piecewise power laws that have been fitted to more complicated underlying models. The construction of the TEF and its inverse for piecewise power laws is given in the appendix of \citet{townsend}, which are as follows. We assume the following functional form:
\begin{align}
    \Lambda(T) = \Lambda_{\rm k} \left( \frac{T}{T_{\rm k}} \right)^{\alpha_{\rm k}},
\end{align}
for some temperature bin $T_{\rm k} \leq T \leq T_{\rm k+1}$. The TEF is then
\begin{align}
    Y(T) = Y_{\rm k} + 
    \begin{cases}
    \frac{1}{1-\alpha_{\rm k}} \frac{\Lambda_{\rm ref}}{\Lambda_{\rm k}} \frac{T_{\rm k}}{T_{\rm ref}} \left[ 1 - \frac{T_{\rm k}}{T}^{\alpha_{\rm k} - 1} \right]    &    \alpha_{\rm k} \neq 1 \\
    \frac{\Lambda_{\rm ref}}{\Lambda_{\rm k}} \frac{T_{\rm k}}{T_{\rm ref}} \ln \left( \frac{T_{\rm k}}{T} \right) & \alpha_{\rm k} = 1
    \end{cases}.
\end{align}
The constraint that $Y(T)$ is continuous leads to the recurrence relation
\begin{align}
    Y_{\rm k} = Y_{\rm k+1} - 
    \begin{cases}
    \frac{1}{1-\alpha_{\rm k}} \frac{\Lambda_{\rm ref}}{\Lambda_{\rm k}} \frac{T_{\rm k}}{T_{\rm ref}} \left[ 1 - \frac{T_{\rm k}}{T_{\rm k+1}}^{\alpha_{\rm k} - 1} \right]    &    \alpha_{\rm k} \neq 1 \\
    \frac{\Lambda_{\rm ref}}{\Lambda_{\rm k}} \frac{T_{\rm k}}{T_{\rm ref}} \ln \left( \frac{T_{\rm k}}{T_{\rm k+1}} \right) & \alpha_{\rm k} = 1
    \end{cases},
\end{align}
with $Y_{\rm ref} = Y(T_{\rm ref}) = 0$. For cooling(heating), $T_{\rm ref}$ is any temperature higher(lower) than the current temperature and we construct the TEF for decreasing(increasing) temperatures. Hence for heating, we express the above recurrence relation as $Y_{\rm k+1} = Y_{\rm k}+\dots$, starting from $T_{\rm ref} < T$. The inverse TEF is thus
\begin{align}
    Y^{-1}(Y) = 
    \begin{cases}
    T_{\rm k} \left[1-(1-\alpha_{\rm k})\frac{\Lambda_{\rm k}}{\Lambda_{\rm ref}} \frac{T_{\rm ref}}{T_{\rm k}} (Y-Y_{\rm k}) \right]^{1/(1-\alpha_{\rm k})}    &    \alpha_{\rm k} \neq 1 \\
    T_{\rm k} \exp \left[- \frac{\Lambda_{\rm k}}{\Lambda_{\rm ref}} \frac{T_{\rm ref}}{T_{\rm k}} (Y-Y_{\rm k})\right] & \alpha_{\rm k} = 1
    \end{cases}.
\end{align}

\subsubsection{Piecewise Linear Function}
We can likewise compute the TEF and its inverse for piecewise linear functions of the form
\begin{align}
    \Lambda(T) = \Lambda_{\rm k} + \alpha (T-T_{\rm k}), \;\;\;  \alpha = \frac{\Lambda_{k+1}-\Lambda_{\rm k}}{T_{\rm k+1}-T_{\rm k}},
\end{align}
where $\alpha$ is the slope in the temperature bin $T_{\rm k} \leq T \leq T_{\rm k+1}$. The TEF is then:
\begin{align}
    Y(T) =
    \begin{cases}
    Y_{\rm k} + \frac{\Lambda_{\rm ref}}{T_{\rm ref}}  \frac{1}{\alpha} \log\left( \frac{\Lambda_{\rm k}}{\Lambda_{\rm k} + \alpha(T-T_{\rm k})} \right) &   \alpha \neq 0 \\
    Y_{\rm k} + \frac{\Lambda_{\rm ref}}{T_{\rm ref}} \frac{T_{\rm k} - T}{\Lambda_{\rm k}} & \alpha = 0
    \end{cases},
\end{align}
and and its inverse is:
\begin{align}
    Y^{-1}(Y) = 
    \begin{cases}
    T_{\rm k} + \frac{\Lambda_{\rm k}}{\alpha} \left(\left(\exp \left[ \frac{\alpha T_{\rm ref}}{\Lambda_{\rm ref}} (Y-Y_{\rm k}) \right]\right)^{-1} - 1 \right)  &   \alpha \neq 0 \\
    T_{\rm k} + \frac{T_{\rm ref}}{\Lambda_{\rm ref}} \Lambda_{\rm k} Y-Y_{\rm k} & \alpha = 0
    \end{cases}.
\end{align}

\subsection{The Extended Algorithm}
The main benefit of the Townsend Algorithm is naturally that it is based on an exact solution and hence not sensitive to errors associated with temporal resolution. This makes it the best choice when using a simple treatment of cooling/heating that is only a function of temperature and does not have a functional dependence on quantities such as ionic abundances. One caveat is that thermal and hydrodynamical evolution are decoupled over the course of a single timestep. It is hence still recommended to include a further constraint on the timestep past the CFL condition that addresses this issue (for example requiring that the timestep be below some fraction of the cooling time).

In \citet{townsend}, the algorithm was only outlined for cooling. Including heating is non-trivial because heating scales differently with density compared to cooling. This means that the net cooling/heating function becomes density dependent, which means that we can no longer optimize by pre-computing all TEFs. Another challenge is the behavior of power laws near equilibrium points. As far as we know, the only attempt to include heating and tackle this problem is in \citet{zhu17}. In their implementation, they assume piecewise linear functions and only one equilibrium point. We generalize this further to generally work with piecewise power laws, as most cooling functions are represented as such, and an arbitrary number of equilibrium points. 

When calculating the net cooling/heating at a given density, we assume that the resulting table itself represents a piecewise power law, instead of being a linear combination of two piecewise power laws. The exception is that for bins that have a zero crossing, we assume they are piecewise linear functions. One advantage of the piecewise functions above is that we can use linear functions only for bins with equilibrium temperatures to interpolate smoothly across the bin, where a power law formulation breaks down. This is fine as long as we dot our i's and cross our t's when computing the TEF and its inverse.
Note that by construction, the TEF for any point with net zero cooling/heating will have an infinite TEF. Hence in practice one should set the cooling/heating to some non-zero tiny number with the appropriate sign. Furthermore, for cooling/heating, we use the next higher/lower bin as the reference temperature and only calculate the TEF down/up to the next equilibrium temperature, as we must always remain within a cooling/heating region if we begin there.

More specifically, we implement a new cooling module as a class that is initialized at the beginning of the simulation with some specified cooling and heating table. It is assumed that both cooling and heating are represented as piecewise power laws. The class gives the user the ability to call a cooling function for some input temperature, density and timestep and returns the new temperature in cgs units. The class also allows the user to query the single point cooling time at some given temperature and density, along with the minimum/maximum temperatures. For the cooling implementation, we use two preallocated scratch arrays, one to hold temperatures and one to hold the net cooling. At the start of the function, we first check that the temperatures are within the bounds of the provided table. If it is not, we return the lower/upper bound instead. We then populate the scratch arrays with a copy of the temperature table and the net cooling table for the given density. Next, we figure out which temperature bin we are in and check if there is net cooling or heating in that bin. If the net value is zero, we return the current temperature. We now case on whether there is net cooling or heating. For either case, the steps are similar, but the directions in which we compute quantities are opposite, along with the choice of some indexes. We first use the next bin over as the reference bin, but check for the edge case where we are in an equilibrium bin. In that case we modify the next bin to be at the equilibrium temperature and set the net cooling in that bin to be zero. If we assume in a bin that cooling and heating have the following power law forms:
\begin{align}
    \Lambda = \Lambda\left(\frac{T}{T_i}\right)^{\alpha_i} && \Gamma = \Gamma_i\left(\frac{T}{T_i}\right)^{\beta_i},
\end{align}
then the equilibrium temperature in a bin that transitions from heating to cooling or vice versa is
\begin{align}
    T_{\rm eq} = T_i \left( \frac{\Lambda_i}{n\Gamma_i} \right)^{1/{(\alpha_i - \beta_i})}.
\end{align}
We also flag the bin to be linear. We next calculate the TEF recursively downwards, until either we reach the bounds or we reach another equilibrium temperature, in which case we again modify the bin and flag it as a linear bin. We then compute the current TEF followed by the TEF after the timestep. Since the bins with equilibriums are linear, there is some finite value of $Y$ that if exceeded immediately returns the equilibrium temperature there. Lastly, we compute the inverse TEF, accounting for the fact that we might have crossed several bins. This gives us the new temperature.

\subsection{Testing}
\begin{figure}
    \centering
	\includegraphics[width=\columnwidth]{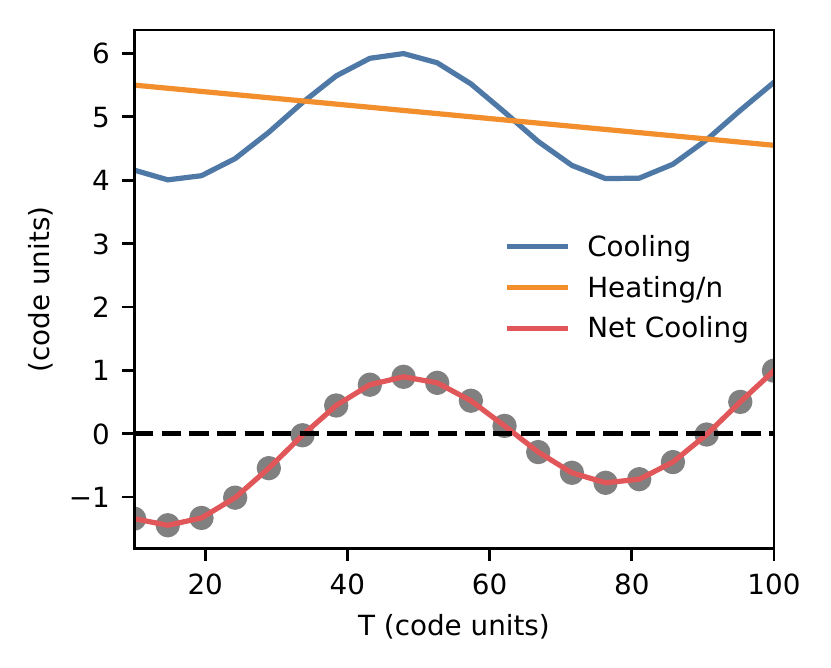}
	\caption{The cooling and heating functions used in our test case, with $n=2$. The grey points show the values in each bin.}
	\label{fig:test_cooling_curve}
\end{figure}
\begin{figure}
    \centering
	\includegraphics[width=\columnwidth]{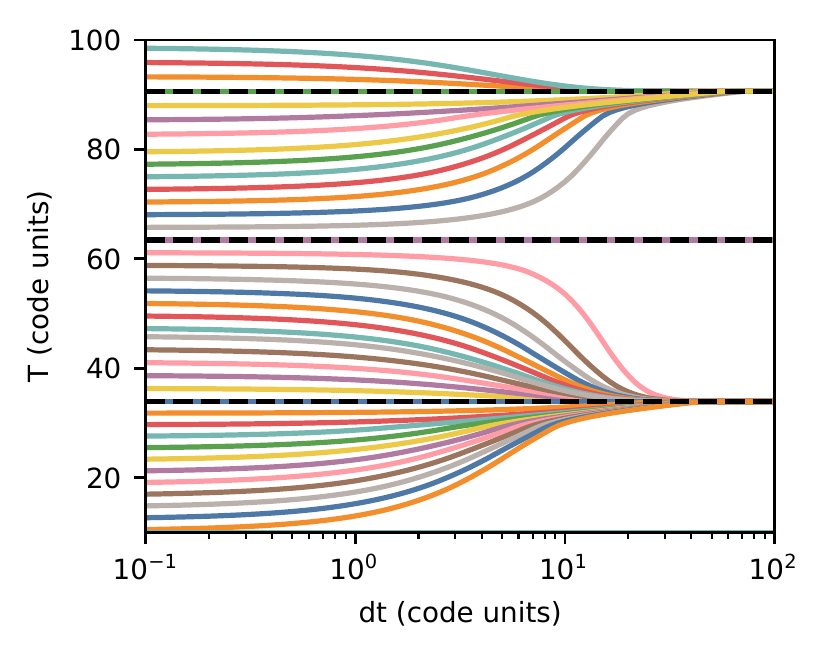}
	\caption{The computed heating/cooling over time for a range of starting temperatures. Black dashed lines show the expected equilibrium temperatures.}
	\label{fig:test_townsend}
\end{figure}
Figures~\ref{fig:test_cooling_curve} and \ref{fig:test_townsend} show a test of the above algorithm, where we have set up a simple cooling and heating curve with $n=2$ and computed the calculate new temperature as a function of timestep for a range of starting temperatures. The black dashed lines in Fig.~\ref{fig:test_townsend} mark the expected equilibrium temperatures.

\section{Estimating Cloud Growth Rates} \label{sect:apdx_e}
\begin{figure}
    \centering
	\includegraphics[width=\columnwidth]{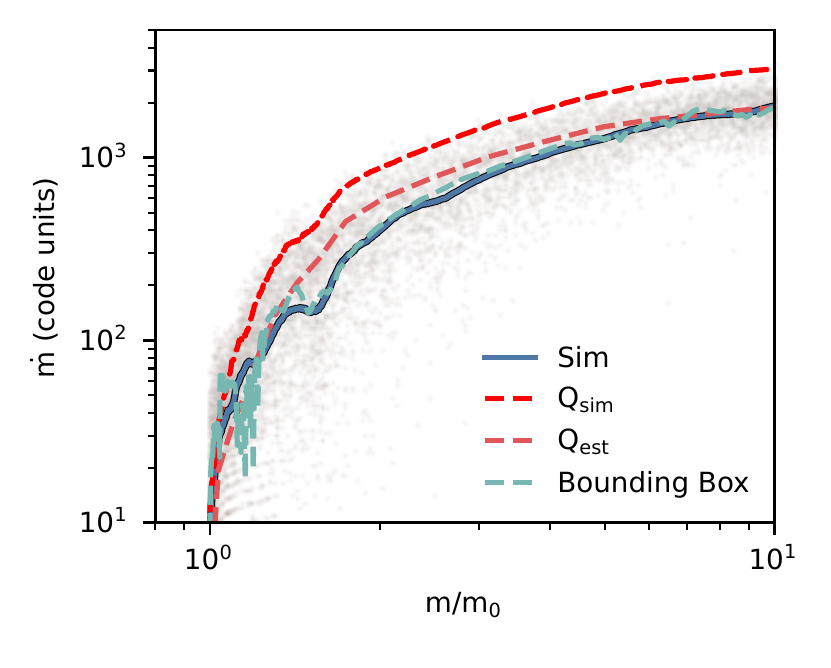}
	\caption{The mass growth rate as a function of mass from the simulations(grey points; smoothed result in blue) and from estimates using (i) total cooling luminosity $Q$ (both recorded and estimated) and (ii) mass flux through the bounding box of the cloud.}
	\label{fig:fc1}
\end{figure}
\begin{figure}
    \centering
    \includegraphics[width=\columnwidth]{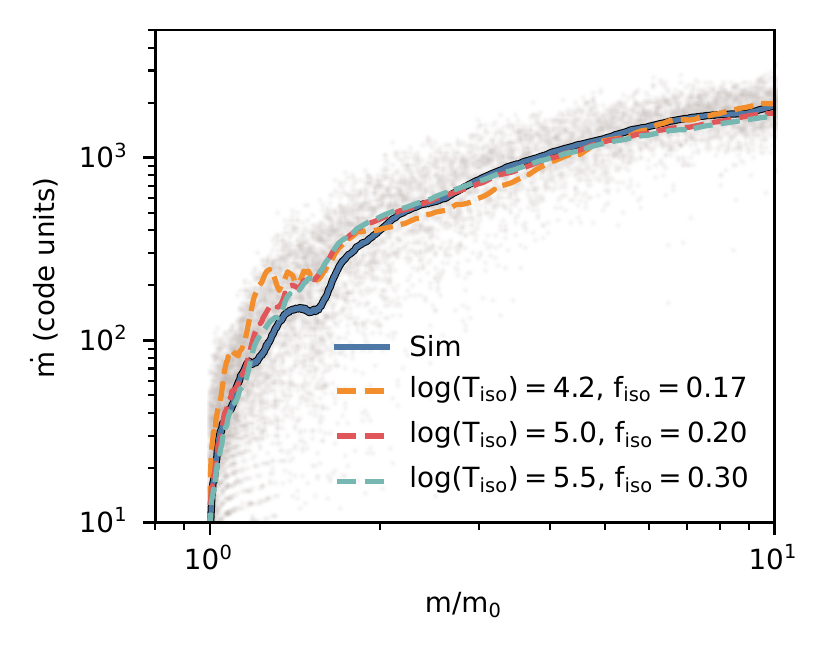}
    \caption{The model applied to different choices of temperature for the isosurface $T_{\rm iso}$. The scaling factor $f_{\rm iso}$ is seen to be dependent on $T_{\rm iso}$.}
    \label{fig:fc2}
\end{figure}
\begin{figure}
    \centering
    \includegraphics[width=\columnwidth]{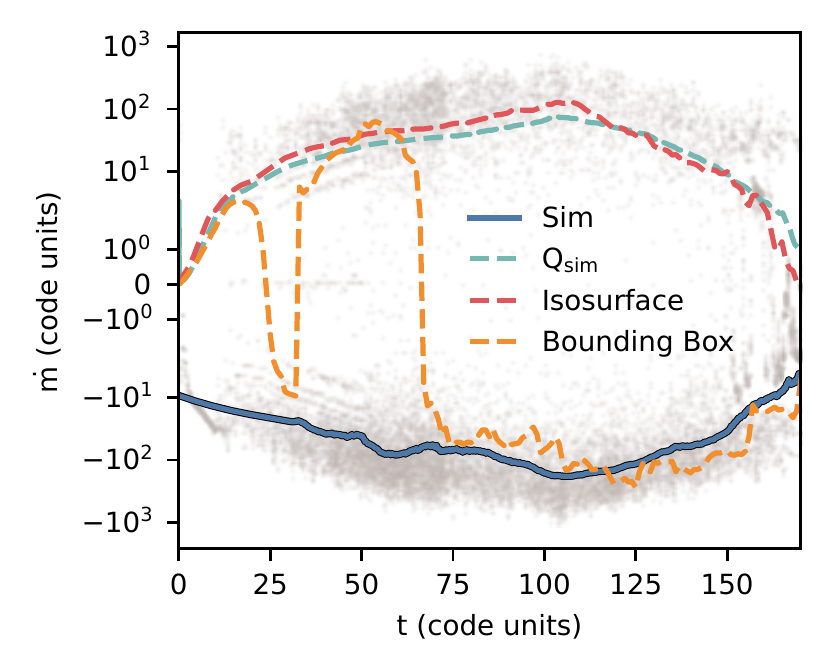}
    \caption{The mass loss rate for a cloud being destroyed, compared with the various methods to estimate mass growth rate. We find that these methods do poorly when the cloud is not actually growing.}
    \label{fig:fc3}
\end{figure}
\begin{figure}
    \centering
    \includegraphics[width=\columnwidth]{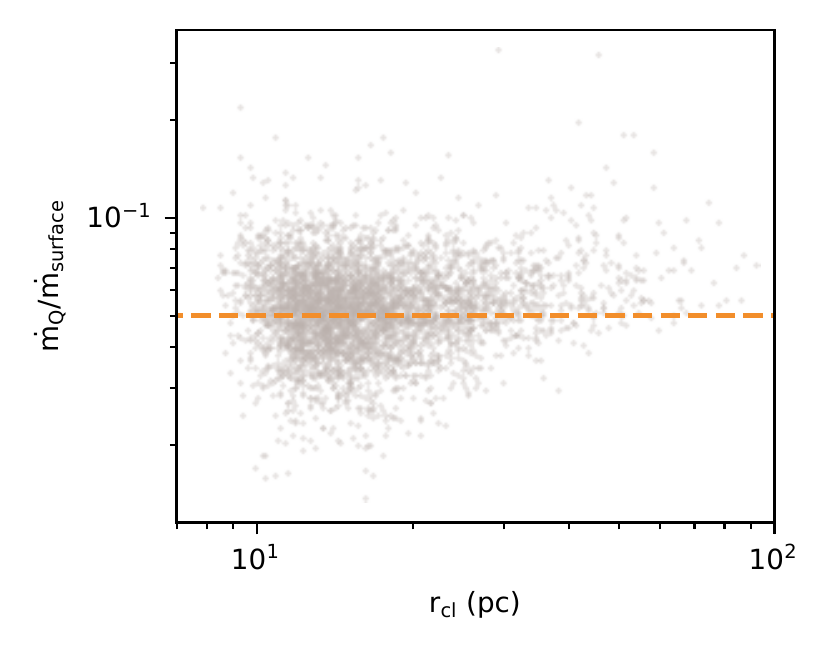}
    \caption{The ratio of $\Dot{m}_{\rm Q}$ and $\Dot{m}_{\rm surface}$ as a function of cloud     size. The value of 0.05 given by the orange dashed line is the final scaling value we use in our main analysis.}
    \label{fig:mdotcalibration}
\end{figure}
Given the large time interval between full data outputs, and the resulting difficulty in matching and tracking individual clouds across multiple snapshots, we would like to be able to estimate the mass growth rate of an individual cloud based solely on its properties from a single time snapshot alone. 

Here we compare several methods of doing this, which we test by applying to a controlled setup containing only a single cloud and where the mass growth rate of the cloud is tracked explicitly with high time resolution. This provides a `ground truth' which we can use to test our various estimators against. The setup consists of a cool cloud, initially at rest, falling under gravity in a constant hot background as detailed in Section~3 of \citet{tan23} (more specifically this setup corresponds to the $\Lambda_0 = 30$ and $r_{\rm cl} = 300$\,pc run). At fiducial resolution here, $r_{\rm cl}$ is resolved by $\sim 12$ cells.

We estimate the mass growth rate in three different ways. The first method is by estimating the total cooling luminosity $Q \equiv \int \rho \mathcal{L} dV$. If we then assume that radiative cooling balances enthalpy flux \citep[i.e., in the sub to transonic regimes;][]{ji19}, we can estimate the mass growth rate as $\Dot{m}_{\rm Q} \sim Q/c_pT_{\rm hot}$. This also assumes that the bulk of the cooling luminosity comes from the mixed gas in the turbulent mixing layer and that the contribution from the cool and hot gas components is negligible. In Figure~\ref{fig:fc1}, we show how this stacks up against the actual mass growth rate. The mass growth from the simulation is represented by the solid blue line, which is obtained by smoothing the instantaneous values of $\Dot{m}$ represented by the grey points. In this simulation, the cooling in each cell per timestep is explicitly tracked, which using the method above corresponds to a mass growth rate given by the red dashed line. However, this quantity is not similarly tracked in our main simulations, and hence must be estimated instead by computing $Q_{\rm est}$ using the density and temperature of each cell, which gives us an estimate shown by the orange dashed line. This does even better than $Q_{\rm sim}$ because we explicitly avoid contributions from cells within a percent of the temperature floor, to account for the lack of explicit heating in this simulation. 

The second method involves computing the mass flux through the bounding box of the cool gas. By computing the net mass influx into the volume defined by the bounding box, and assuming that this balances the mass flux from the hot to cool phase, we can estimate the mass growth rate. This is shown by the teal dashed line in Figure~\ref{fig:fc1}. This estimate does surprisingly well despite its simplicity. While it works well in this controlled experiment, it fails in our main simulations due to the turbulent nature of the environment leading to fluctuating mass fluxes through the bounding box.

The third and last method is to compute the total mass flux through a temperature isosurface $\Dot{m}_{\rm surface}$. This is the isosurface we construct using the marching cubes algorithm as described in the previous section. In order to measure the mass flux through this surface, we interpolate the velocity and log density at the centroid of each triangular face in the isosurface mesh. The mass flux through each face is given by the component of this velocity that is normal to the face and in the inward mesh direction, multiplied by the density and the area of the face. This gives us a total mass influx rate over the entire isosurface. However, just this quantity alone significantly overestimates the mass growth rate. This is likely because translating this instantaneous quantity into a mass growth rate directly assumes that all the instantaneous flux becomes cool gas, which does not hold here because our isosurface is itself sensitive to the velocity field. To account for this, we can determine a normalization constant $f_{\rm iso}$ by calibrating to $Q_{\rm est}$.
The constant of proportionality $f_{\rm iso} = 0.2$ for $T_{\rm iso} = 5.0$, which is our default choice. Figure~\ref{fig:fc2} shows this model for isosurfaces of various temperatures, with log($T$) = 4.2, 5.0 and 5.5. We have used $f_{\rm mix} = 0.17$, 0.2 and 0.3 for them respectively. We expect $f_{\rm iso}$ to vary with the choice of isosurface temperature and the velocity field. 

What if the cloud is not growing? Figure~\ref{fig:fc3} shows a run with weaker cooling ($\Lambda_0 = 1$) where the cloud is getting destroyed. Here, we see a drawback of the various methods presented---they do not account for mass loss. Hence, these methods should not generally extend to clouds that do not survive. The bounding box approach appears to work well at later times, but this is because the accreting cloud is able to form a steady inflow towards the cloud surface in this setup. In more turbulent environments, the signal from this method is washed out by the much higher turbulent velocities in the background. 

Ultimately, the three different methods of estimating $\Dot{m}$ from instantaneous cloud properties presented above give similar results that are good estimates of the actual cloud mass growth rate, assuming that the clouds are growing. It should be stressed that these are estimates---having multiple methods provides a cross check that allows us to make this estimate of $\Dot{m}$ more reliably. 

When measuring mass growth rates of the clouds in our main simulations, we thus use the mass flux through a temperature isosurface $\Dot{m}_{\rm surface}$, but normalized to match mass growth rates estimated using the cooling luminosity $\Dot{\rm m}_{\rm Q}$. This choice allows for more consistency when discussing the scalings related to both the inflow velocity through this isosurface and its computed surface area. 

The normalization is done by finding the constant of proportionality $f_{\rm iso}$ discussed above. In other words, we calibrate a scaling factor $f_{\rm iso}$ such that $\Dot{m} \sim f_{\rm iso}\Dot{m}_{\rm surface}$, where $f_{\rm iso} \equiv \langle \Dot{m}_{\rm Q}/\Dot{m}_{\rm surface} \rangle$. Figure~\ref{fig:mdotcalibration} shows $\Dot{m}_{\rm Q}/\Dot{m}_{\rm surface}$ as a function of cloud size for the clouds used in the analysis in Section~\ref{sect:results_clouds}, with $f_{\rm iso} = 0.05$ represented by the orange dashed line being the scaling factor we adopt. 

\section{`Growth' in a Failed Wind} \label{sect:apdx_f}

\begin{figure}
    \centering
    \includegraphics[width=\columnwidth]{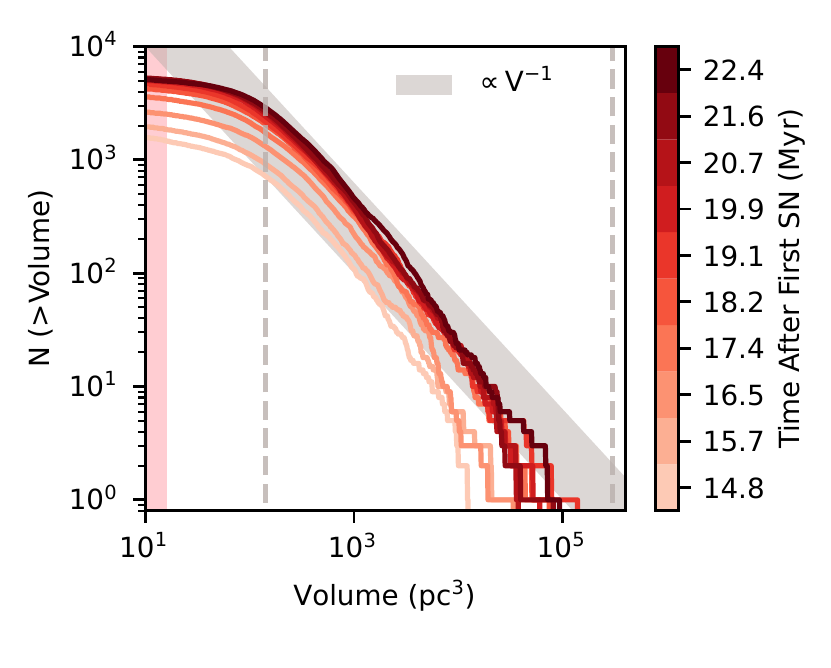}
    \caption{Cumulative cloud volume distribution at various time slices from the failed wind run. The pink region indicates where clouds are resolved by fewer than 16 cells. Grey regions show the expected power law scaling of $-1$. Grey dashed lines show expected lower (for survival) and upper bounds on cloud sizes.}
    \label{fig:cloud_hist_512}
\end{figure}

\begin{figure}
    \centering
    \includegraphics[width=\columnwidth]{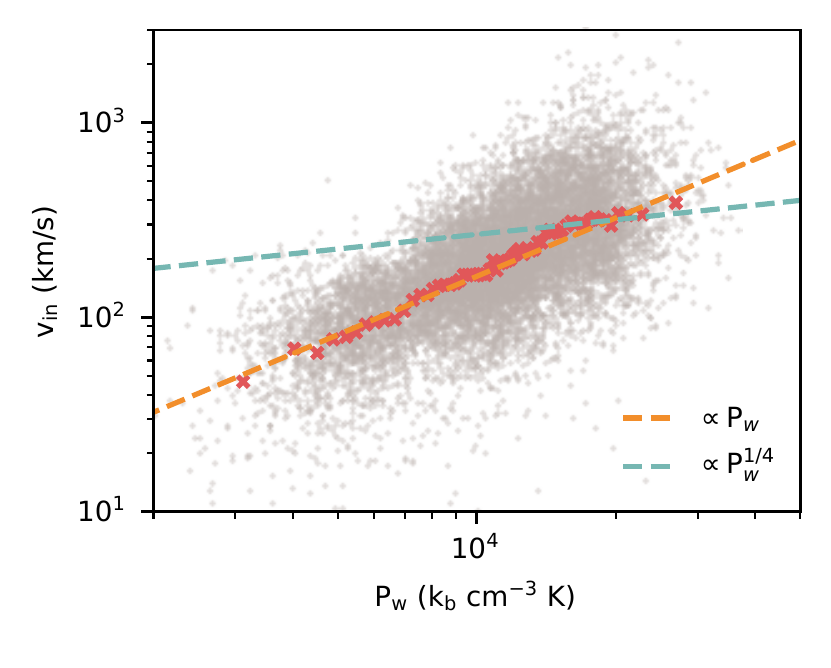}
    \caption{The inflow velocity here is shown to scale linearly with pressure in the simulation run which fails to generate a wind.}
    \label{fig:linear_cooling}
\end{figure}

\begin{figure*}
    \centering
    \includegraphics[width=\textwidth, valign=c]{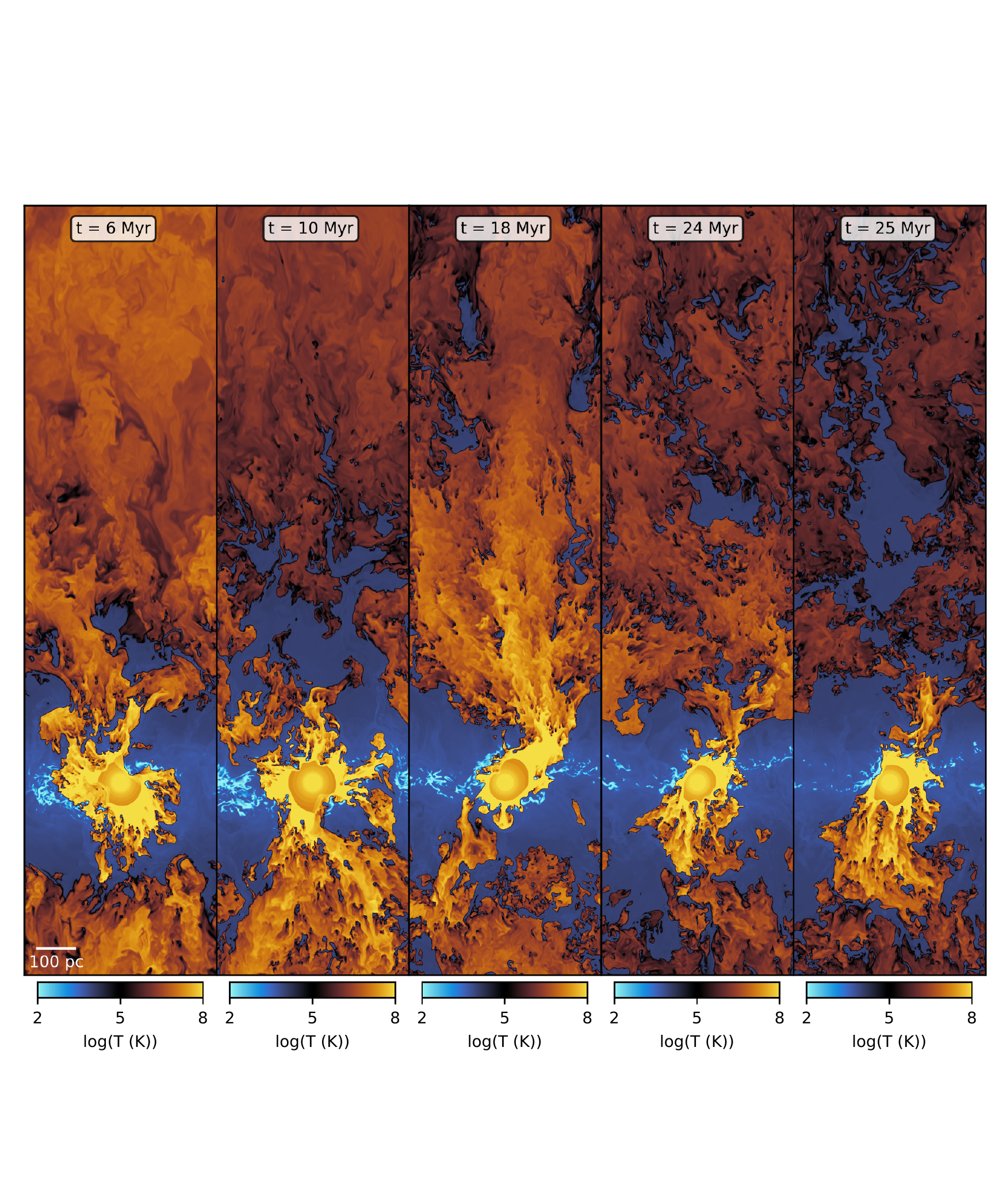}
    \caption{Slices of temperature at various times from a simulation where the breakout is not successful in launching a wind on the top side and is disrupted too early by cold gas in the ISM. This causes the breakout to cool and fall back towards the disc, a cycle which is repeated several times.}
    \label{fig:slices_progression_512}
\end{figure*}

\begin{figure*}
    \centering
    \vspace*{\fill}
    \includegraphics[width=\textwidth, valign=c]{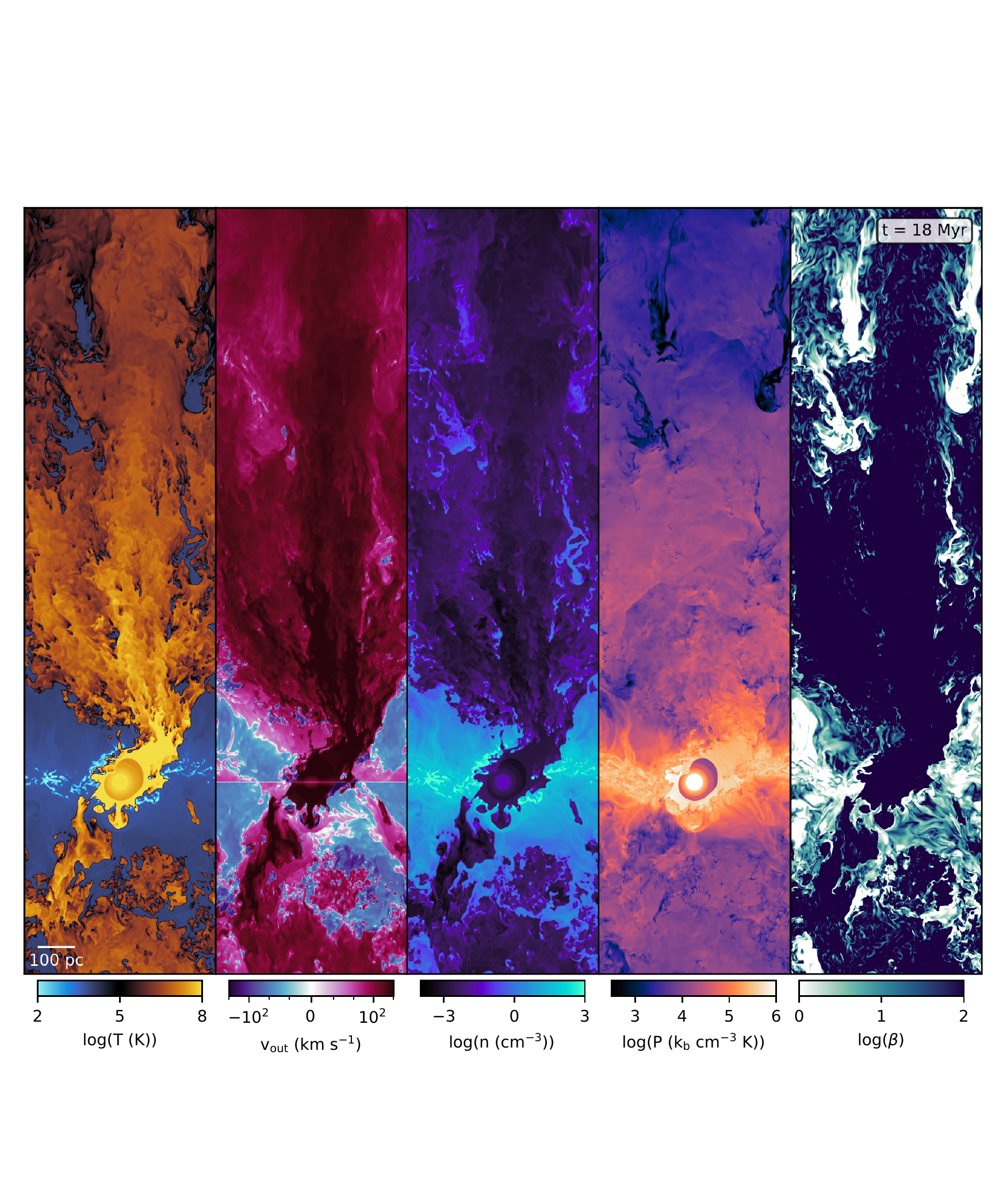}
    \vspace*{\fill}
    \caption{Slices of temperature, outflow velocity, number density, pressure and magnetic plasma beta for a single time snapshot 18 Myr after the first SN.}
    \label{fig:slices_prim_512}
\end{figure*}

It is interesting to present some results which show what happens when the superbubble breakout fails to generate a wind. In Figures~\ref{fig:cloud_hist_512} and \ref{fig:linear_cooling}, we repeat some of the analysis presented in the main text for one such simulation run so as to highlight several important differences. Figure~\ref{fig:slices_progression_512} shows temperature slices at various times in this simulation run. After the superbubble breaks out briefly and vents the hot gas, the hot material quickly begins to cool and falls back to the disc instead of sustaining a hot outflow. Figure~\ref{fig:slices_prim_512} shows various quantities such as outflow velocity, density, and pressure at 18~Myr.

Figure~\ref{fig:cloud_hist_512} shows the same analysis of the cumulative size distribution as Figure~\ref{fig:cloud_hist}. Here, the distribution appears slightly steeper for larger clouds and is more consistent with $N(>V) \propto V^{-1.2}$. We note that this simulation has a higher resolution of 1\,pc. It is possible that this slightly steeper slope is related to the higher resolution, but more likely it is due to the brief and weak nature of the wind---these clouds show very different growth scalings which we discuss next. We find that while the surface area to volume scaling above still holds, $v_{\rm in}$ in the cooling wind is independent of both cloud size and turbulent velocity, suggesting that the cool clouds are not growing via the same mechanism of turbulent mixing layers. In Figure~\ref{fig:linear_cooling}, we show that instead, $v_{\rm in}$ scales linearly with $P_{\rm w}$ (i.e., inversely with the cooling time), consistent with the picture above that the growth is instead directly driven by large scale cooling and condensation of the wind (much like a cooling flow).


\dobib

\bsp
\label{lastpage}
\end{document}